\documentclass[fleqn,usenatbib]{mnras}

\usepackage{newtxtext,newtxmath}
\usepackage[T1]{fontenc}

\DeclareRobustCommand{\VAN}[3]{#2}
\let\VANthebibliography\thebibliography
\def\thebibliography{\DeclareRobustCommand{\VAN}[3]{##3}\VANthebibliography}

\usepackage{graphicx}	% Including figure files
\usepackage{amsmath}	% Advanced maths commands
\usepackage{float,geometry,color,amsfonts,textcomp,upgreek,hyperref}
\usepackage{wasysym}
\usepackage{mathrsfs}
\usepackage[T1]{fontenc}
\usepackage{ae,aecompl} 
\usepackage{times}
\usepackage[pscoord]{eso-pic}
\usepackage[normalem]{ulem}
\usepackage[dvipsnames]{xcolor}

\usepackage{natbib}
\usepackage[super]{nth}
\usepackage{xspace}

\usepackage{verbatim}
\usepackage{pdflscape}

\newcommand{\decode}{\textsc{decode}\xspace}

\defcitealias{grylls_paper1}{G19}
\defcitealias{grylls_paper2}{G20}

\title[\decode and galaxy merger rates]{Testing the key role of the stellar mass-halo mass relation in galaxy merger rates and morphologies via DECODE, a novel Discrete statistical sEmi-empiriCal mODEl}

\author[H. Fu et al.]{
	Hao Fu$^{1}$\thanks{E-mail: \href{mailto:h.fu@soton.ac.uk}{h.fu@soton.ac.uk}},
	Francesco Shankar$^{1}$\thanks{E-mail: \href{mailto:f.shankar@soton.ac.uk}{f.shankar@soton.ac.uk}},
	Mohammadreza Ayromlou$^{2,3}$,
	Max Dickson$^{1}$,
	Ioanna Koutsouridou$^{4,5}$,
	\newauthor
	Yetli Rosas-Guevara$^{6}$,
	Christopher Marsden$^{1}$,
	Kristina Brocklebank$^{1}$,
	Mariangela Bernardi$^{7}$,
	Nikolaos \newauthor Shiamtanis$^{1}$,
	Joseph Williams$^{1}$,
	Lorenzo Zanisi$^{1}$,
	Viola Allevato$^{8}$,
	Lumen Boco$^{9}$,
	Silvia Bonoli$^{6}$,
	Andrea \newauthor Cattaneo$^{10}$,
	Paola Dimauro$^{11}$,
	Fangzhou Jiang$^{12}$,
	Andrea Lapi$^{9}$,
	Nicola Menci$^{13}$,
	Stefani Petropoulou$^{1}$,
	\newauthor
	Carolin Villforth$^{14}$
	\\\\
	$^{1}$School of Physics and Astronomy, University of Southampton, Highfield, SO17 1BJ, UK
	\\
	$^{2}$Universit{\"a}t Heidelberg, Zentrum f{\"u}r Astronomie, Institut f{\"u}r theoretische Astrophysik, Albert-Ueberle-Str. 2, 69120 Heidelberg, Germany
	\\
	$^{3}$Max Planck Institute for Astrophysics, Karl-Schwarzschild-Str. 1, 85741 Garching bei M\"{u}nchen, Germany
	\\
	$^{4}$Dipartimento di Fisica e Astronomia, Universita degli Studi di Firenze, Via G. Sansone 1, I-50019 Sesto Fiorentino, Italy
	\\
	$^{5}$INAF/Osservatorio Astrofisico di Arcetri, Largo E. Fermi 5, I-50125 Firenze, Italy
	\\
	$^{6}$Donostia International Physics Centre (DIPC), Paseo Manuel de Lardizabal 4, 20018 Donostia-San Sebastian, Spain
	\\
	$^{7}$Department of Physics and Astronomy, University of Pennsylvania, Philadelphia, PA 19104, USA
	\\
	$^{8}$INAF - Osservatorio di Astrofisica e Scienza delle Spazio di Bologna, OAS, Via Gobetti 93/3, 40129, Bologna, Italy
	\\
	$^{9}$SISSA, Via Bonomea 265, 34135 Trieste, Italy
	\\
	$^{10}$Observatoire de Paris/LERMA, PSL University, 61 av. de l’Observatoire, 75014 Paris, France
	\\
	$^{11}$Observat\'{o}rio Nacional, Minist\'{e}rio da Ciencia, Tecnologia, Inovação e Comunicações, São Crist\'{o}vão, 20921-400, Rio de Janeiro,Brazil
	\\
	$^{12}$TAPIR, California Institute of Technology, Pasadena, CA 91125
	\\
	$^{13}$INAF - Osservatorio Astronomico di Roma, via di Frascati 33, 00078 Monte Porzio Catone, Italy
	\\
	$^{14}$Department of Physics, University of Bath, Claverton Down, Bath, BA2 7AY, UK
}

\date{Accepted XXX. Received YYY; in original form ZZZ}

\pubyear{2015}

\begin{document}
\label{firstpage}
\pagerange{\pageref{firstpage}--\pageref{lastpage}}
\maketitle

\begin{abstract}
The relative roles of mergers and star formation in regulating galaxy growth are still a matter of intense debate. We here present our \decode, a new Discrete statistical sEmi-empiriCal mODEl specifically designed to predict rapidly and efficiently, in a full cosmological context, galaxy assembly and merger histories for any given input stellar mass-halo mass (SMHM) relation. \decode generates object-by-object dark matter merger trees (hence discrete) from accurate subhalo mass and infall redshift probability functions (hence statistical) for all subhaloes, including those residing within other subhaloes, with virtually no resolution limits on mass or volume. Merger trees are then converted into galaxy assembly histories via an input, redshift dependent SMHM relation, which is highly sensitive to the significant systematics in the galaxy stellar mass function and on its evolution with cosmic time. \decode can accurately reproduce the predicted \textit{mean} galaxy merger rates and assembly histories of hydrodynamic simulations and semi-analytic models, when adopting in input their SMHM relations. In the present work we use \decode to prove that only SMHM relations implied by stellar mass functions characterized by large abundances of massive galaxies and significant redshift evolution, at least at $M_\star \gtrsim 10^{11} \, M_\odot$, can simultaneously reproduce the local abundances of satellite galaxies, the galaxy (major merger) pairs since $z \sim 3$, and the growth of Brightest Cluster Galaxies. The same models can also reproduce the local fraction of elliptical galaxies, on the assumption that these are strictly formed by major mergers, but not the full bulge-to-disc ratio distributions, which require additional processes.%, such as disc instabilities at lower stellar masses.
\end{abstract}

\begin{keywords}
galaxies: general -- galaxies: evolution -- galaxies: abundances -- galaxies: haloes
\end{keywords}

%%%%%%%%%%%%%%%%%%%%%%%%%%%%%%%%%%%%%%%%%%%%%%%%%%

    %%%%%%%%%%%%%%%%% BODY OF PAPER %%%%%%%%%%%%%%%%%%

    \section{Introduction}\label{sec_intro}
	
	The field of galaxy formation and evolution is still far from settled, with several open and still hotly debated issues. For example, it is still unclear what are the relative amounts of stellar mass that galaxies grow "in situ", via star formation, and acquire "ex situ" from, e.g., mergers with other galaxies \citep[e.g.,][]{guo_2008, oser_2010, cattaneo_2011, lackner_2012, lee_2013, pillepich_2014, rodriguez_gomez_2016, qu_2017, clauwens_2018, pillepich_2018_Mstar_content, monachesi_2019, davison_2020}. In a $\Lambda$CDM Universe, galaxies are in fact believed to live at the centre of Dark Matter (DM) haloes, which grow their mass via mergers with other DM haloes along with smooth mass accretion from their environments \citep[e.g.,][]{murali_2002, conselice_2009, genel_2010, huillier_2012}. Each merger between two DM haloes could in principle trigger a merger between their central galaxies and, therefore, galaxy mergers should indeed be frequent in a DM-dominated Universe. However, cosmological models suggest that in many instances the dynamical friction timescale, i.e., the time that two galaxies take to merge, is longer than the age of the Universe, resulting in the smaller galaxy orbiting as an unmerged satellite of the most massive central galaxy \citep[e.g.,][]{khochfar_2006, fakhouri_2010, mccavana_2012}. A prominent case is the orphan satellite galaxies, whose DM subhaloes can no longer be resolved in the simulations, but they continue orbiting the central galaxy. It is thus clear that to impose more stringent constraints on the role of mergers in shaping galaxies in a $\Lambda$CDM Universe, it is first of all essential to correctly predict the merger histories of the host DM haloes. After this, the following vital step is to identify the correct mapping between galaxies and host DM haloes to translate DM merger trees into a galaxy assembly history, a task far from trivial \citep[e.g.,][]{hopkins_2010, grylls_paper1, grylls_paper2}.
	
    It has been noted that widely distinct merger histories could lead to similar morphologies and kinematic properties in the remnant galaxies \citep[][]{bournaud_2007}. Moreover, different hierarchical models often predict strongly divergent balances between the stellar mass formed in-situ during the early-epoch, highly star-forming and dust-enshrouded phase, and the fraction of stellar mass acquired ex-situ via mergers. For example, the SAM presented in \citet{gonzalez_2011} suggests that only a few percent of the final stellar mass is formed in a typical massive galaxy during its initial burst, while at the other extreme, several groups suggest that most of the stellar mass was acquired in a moderate-to-strong burst of star formation at high redshifts \citep[e.g.,][]{granato_2004, chiosi_2012, merlin_2012, lapi_2018}.
    
    Semi-empirical models (SEMs) have been introduced in the last decades as a powerful, \emph{complementary} tool to probe galaxy evolution \citep[see, e.g.,][]{hopkins_2009a, conroy_2009, cattaneo_2011, zavala_2012, shankar_2014, rodriguez_puebla_2017, moster_2018, grylls_paper1, behroozi_2019, drakos_2022}. By design, SEMs avoid the modelling of galaxy growth and assembly within DM haloes from first principles, unlike more traditional modelling approaches. In their simplest form, SEMs adopt abundance matching techniques \citep[e.g.,][]{kravtsov_2004, vale_2004, yang_2004, shankar_2006, moster_2010, behroozi_2010}, based on the matching between the cumulative number densities of the measured stellar mass functions (SMF) and the host DM halo mass functions (HMF), to generate a monotonic stellar mass-halo mass (SMHM) relation through which they statistically assign galaxies to host DM haloes at different redshifts. Starting from this mapping, SEMs can then focus on specific questions, such as the merger rates of galaxies implied by a specific SMHM relation, or the role played by mergers in forming bulges in galaxies \citep[e.g.,][]{hopkins_2010a, behroozi_2010, moster_2010, moster_2013, grylls_paper1}. SEMs are based on minimal input assumptions and associated parameters, allowing for a high degree of transparency in the results whilst avoiding degeneracies.
    %At the very least, SEMs only depend on the shape and dispersion of the SMHM relation implied by the input SMFs \citep{grylls_paper2}.
    Additional assumptions can be gradually included in the modelling, for example, varying the major merger mass ratio threshold for forming ellipticals (as further discussed below), but always allowing for extreme flexibility and transparency.

    In the STatistical sEmi-Empirical modeL \textsc{steel}, \citet[][hereafter referred to as G19 and G20, respectively]{grylls_paper1, grylls_paper2} showed how a mean SMHM relation can convert mean DM halo assembly histories into galaxy merger histories, characterized by a total mean accretion track and cumulative mass accreted by merging satellites. From these quantities, galaxy star formation histories can then be computed subtracting from the total galaxy growth the contribution via mergers. The resulting star formation histories can then be compared with independent observational data. The number of surviving satellites can also be compared with relevant data at different redshifts and host halo masses. By including an empirically motivated linear relation between galaxy size and host halo size, \citep[][]{kravtsov_2013}, \citet{stringer_2014} and \citet{zanisi_2020, zanisi_2021a, zanisi_2021b} were able to reproduce the strong size evolution of massive galaxies and their size functions up to redshift $z=0$. \citet{marsden_2021} \citep[see also][]{ricarte_2018} have then extended these SEMs by including empirical estimates of the evolution of the stellar mass profile of galaxies (e.g., \citealt{shankar_2018} and references therein) to predict the full velocity dispersion profiles of central galaxies via detailed Jeans modelling. SEMs are thus a powerful tool to explore mean trends in the assembly, structural, dynamical and star formation histories of galaxies. \citet{grylls_paper3} have however recently highlighted \citep[see also][]{oleary_2021} that even relatively moderate differences in the “mapping” between galaxy stellar mass and host halo mass, i.e., in the input SMHM relations, can generate significantly distinct galaxy pairs and ultimately galaxy merger rates (along with their associated star formation histories). In a SEM framework, differences in the SMHM relation are mostly induced by systematics in the input galaxy SMFs (e.g., \citetalias{grylls_paper2}), but the loophole identified by \citetalias{grylls_paper2} can in fact be extended to all theoretical models predicting different SMFs and thus SMHM relations. This strong dependence of the merger rates on the underlying SMHM relation severely limits the effectiveness of the comparison between data on merger rates and hierarchical models developed largely independently of the fitted data used to measure the merger rates (or pair fractions). For example, the answer to the (still open) question whether galaxy major mergers with a mass ratio above, say, 1/4, can generate the right abundances of ellipticals at different epochs \citep[see, e.g.,][]{hopkins_2009a, hopkins_2010a, shankar_2013, grylls_paper2}, will strongly depend on which SMHM relation is employed in the hierarchical model at hand (either semi-empirical or not), as we will further prove in this work.

    The aim of the present work is twofold. 1) We first present our new Discrete statistical sEmi-empiriCal mODEl, \decode, specifically designed to efficiently and rapidly predict the merger histories, star formation histories, and satellite abundances of galaxies of any stellar mass at any redshift $z<3$, for a given set of input SMF. \decode, as detailed in Section \ref{sec_dream}, further improves on its predecessor \textsc{steel} by replacing \textit{statistical} distributions with catalogues of \textit{distinct} objects, similarly to an N-body simulation, and by a more accurate treatment of the subhaloes. Nevertheless, it still retains the flexibility and  higher computational performance of \textsc{steel}, severely reducing (and in some cases completely eliminating) the limitations imposed by resolution problems in mass and volume, which can heavily impact the modelling of galaxies in a full cosmological setting (see, e.g., discussion in \citealt{vdb_2014}). 2) We then use \decode to study how different renditions of the measured SMF at different epochs impacts the number of galaxy mergers, the formation of ellipticals, and the mean bulge fraction of galaxies in the local Universe. We will show that major mergers may be sufficient to account for all local ellipticals, but additional processes such as disc instabilities and disc regrowth mechanisms must be invoked to simultaneously explain the mean bulge-to-total distributions of local galaxies.
	
	This paper is organized as follows. In Section \ref{sec_data} we describe the data set we use in our work. In Section \ref{sec_dream} we introduce our model \decode, provide an overview on its numerical implementation, and test the performance of our model against available observational data sets, SAMs and hydrodynamic simulations. In Section \ref{sec_results} we present our model's predictions for the satellite abundances, merger histories, morphology and Bulge-To-Total (B/T) ratios of central galaxies, as well as the mean growth history of Brightest Cluster Galaxies (BCGs). Finally, in Sections \ref{sec_discussion} and \ref{sec_conclusions} we discuss our results and draw our conclusions. In this paper we adopt the $\Lambda$CDM cosmological model with parameters from \cite{planck_2018_cosmo_params} best fit values, i.e. $(\Omega_{\rm m}, \Omega_{\Lambda}, \Omega_{\rm b}, h, n_{\rm S}, \sigma_8) = (0.31, 0.69, 0.049, 0.68, 0.97, 0.81)$, and use throughout a Chabrier \citep{chabrier_2003} stellar initial mass function.

	\section{Data}\label{sec_data}
	
	In this work we make heavy use of both numerical/theoretical and observational data sets. We use the former mostly for validation tests of \decode, while the latter are used both as inputs for \decode, as well as outputs to test \decode's predictions. More specifically, we make use of: 1) the Millennium simulation to test the accuracy of the abundances in the surviving/unmerged subhaloes in \decode; 2) the TNG100 simulation, to compare the performance of \decode to a hydrodynamic simulation in terms of galaxy properties (in this work mostly fraction of ellipticals and B/T mass ratios); 3) the \textsc{GalICS} SAM, to compare how \decode compares to a full ab-initio analytic model of galaxy formation; 4) SDSS and MaNGA to compare \decode's predictions on satellite abundances, fraction of ellipticals, and B/T ratios with large data sets of local galaxies. Below we provide relevant details on each of these comparison data sets. In Appendix \ref{app_mass_growth} we discuss how \decode can faithfully reproduce the galaxy assembly histories of other cosmological models when it receives in input their mean SMHM relations. We will also compare with another cosmological SEM (EMERGE, see \citealt{moster_2018}).%, the \textsc{GalICS} SAM and the TNG simulation.

	    \subsection{The Millennium simulation}\label{sec_valid_millennium}

        We use the Millennium DM-only simulation \citep{springel_2005} scaled to the Planck cosmology \citep{planck_2016_cosmo_params} employing the method of \citet{angulo_2010} and \citet{angulo_2015}. All DM haloes are identified using a Friends-Of-Friends (FOF) algorithm \citep{davis_1985}. Furthermore, all DM subhaloes are detected using the \textsc{Subfind} algorithm \citep{springel_2001}, based on which each FOF halo has one central subhalo, and the rest of the subhaloes are labeled as satellite subhaloes. The \textsc{Subfind} algorithm considers a minimum number of particles $n_{\rm min}= 20$ for identifying subhaloes. We consider the infall time of each satellite subhalo as the time when it last changes its type from central to satellite. This information is taken from the publicly available L-Galaxies semi-analytical model \citep[][]{ayromlou2021}\footnote{\href{https://lgalaxiespublicrelease.github.io}{https://lgalaxiespublicrelease.github.io}} that runs on top of the Millennium simulation.
        
        We note that although the halo virial mass $M_{200}$ is reported as the mass within the halo virial radius $R_{200}$, the FOF halo could extend beyond this scale. Therefore, satellite subhaloes of a FOF halo can exist beyond the $R_{200}$ as well. Nevertheless, this does not constitute a limitation in the comparison of the subhalo mass function (SHMF) with \decode since we base our subhaloes generation on the unevolved total SHMF from \citet{jiang_vdb_2014, jiang_vdb_2016}, which has already been shown to be in good agreement with the one from the Millennium simulation as presented in \citet{li_mo_2009}. Furthermore, recently \citet{green_2021} presented a more accurate SHMF computed with an updated version of their model SatGen, where they are able to follow subhalo orbits and effects of numerical disruption. The SHMF that we use slightly differs with the one from \citet{green_2021} by a factor of $\sim 0.1$ dex only at $M_{\rm h,sub} / M_{\rm h,par} \lesssim 10^{-3}$, where their contribution to the galaxy mergers is completely negligible. Nevertheless, we checked that the unevolved total subhalo distribution given by the updated \citet{green_2021} version of SatGen is statistically unchanged with respect to that presented in \citet{jiang_vdb_2016}. This is also guaranteed by the fact that the unevolved SHMF is a manifestation of the progenitor mass function used in the extended-Press-Schechter formalism, which depends only on the cosmological parameters.
        
        \subsection{The TNG simulation}
        
        We make use of the public data release\footnote{\href{https://www.tng-project.org/data}{https://www.tng-project.org/data}} from the TNG100 hydrodynamical simulation of the IllustrisTNG project \citep{nelson_2019}. The IllustrisTNG simulations \citep[][]{nelson_2018, pillepich_2018_Mstar_content, springel_2018, marinacci_2018, naiman_2018} is a set of cosmological hydrodynamical simulations performed using the code \textsc{Arepo} \citep{springel_2010}. Employing subgrid physics, TNG implements astrophysical processes relevant to galaxy evolution, such as the cooling of the hot gas, star formation, the evolution of stars, supernova feedback, supermassive black hole formation (seeding), and AGN feedback \citep[see][for full model description]{pillepich_2018, weinberger_2017}. The TNG model has been presented in three different cosmological boxes so far ($l_{\rm box}\sim 50, 100, 300 \, {\rm Mpc}$). Here, we take the 100 Mpc box (TNG100) for our analysis because this is the simulation box used to calibrate the TNG model against observations. Therefore, TNG100 outputs the most reliable results among the other TNG simulations.
        
        The galaxy morphologies are calculated as in \citet{genel_2015} and \citet{marinacci_2013}. The kinematic decomposition of galaxies is based on the distribution of the circular parameter of individual stellar particles, which is defined as $ \rm \epsilon = J_z/J(E)$ \citep[][]{du_2019}. Here, $\rm J_z$ corresponds to the specific angular momentum in the symmetric axis of the galaxy and $ \rm J(E)$ the maximum specific angular momentum possible at the specific binding energy (E) of the stellar particle. The bulge component of each galaxy comprises the stellar particles with the $\epsilon < 0$ and a fraction of stellar particles with a positive circular parameter that mirrors around zero. The mass of the bulge is twice the mass of the stellar particles, with a circular parameter $\epsilon <0$. The elliptical galaxies are defined as the galaxies with bulge-to-total stellar mass ratio $B/T > 0.7$.

        \subsection{\textsc{GalICS}}
        
        We make use of the data computed via \textsc{GalICS} 2.2 \citep[][]{Koutsouridou22}. \textsc{GalICS} 2.2 is the latest version of the \textsc{GalICS} (Galaxies  In Cosmological Simulations) SAM of galaxy formation \citep{hatton_2003, cattaneo_2006, cattaneo_2017, cattaneo_2020}. The main differences between the GalICS 2.2 version used for this article and the latest published version \textsc{GalICS} 2.1 \citep{cattaneo_2020} are the presence of feedback from active galactic nuclei (AGN) in \textsc{GalICS} 2.2 (\textsc{GalICS 2.1} did not include AGN feedback) and the model for morphological transformations in galaxy mergers (in \textsc{GalICS} 2.1, major mergers destroyed the disc component completely; \textsc{GalICS} 2.2 adopts a more realistic model based on the numerical results of \citealt{kannan_2015}). Since \citet{cattaneo_2020}, there have also been small improvements in the modelling of supernova feedback and disc instabilities.
        
        The implementation of the disc instabilities is based on the results of \citet{devergne_2020} who studied the growth of pseudobulges in isolated thin exponential stellar discs embedded in static spherical haloes. They found that discs with $v_{\rm d} / v_{\rm c} > 0.6$ are unstable \citep[see also][]{efstathiou_1982}, where $v_{\rm c}$ is the circular velocity of the galaxy and $v_{\rm d}$ is the circular velocity considering only the disc's gravity. In what follows, we adopt the following fitting formula \citep{devergne_2020} for the B/T mass ratio
        \begin{equation}
            B/T = 0.5 f_{\rm d}^{1.8} (3.2 r_{\rm d}) \; ,
        \end{equation}
        where $f_{\rm d} = (v_{\rm d} / v_{\rm c})^2$ is the contribution of the disc to the total gravitational acceleration, and $r_{\rm d} = R_{\rm d} / R_{\rm vir}$ is the dimensionless exponential scale-length in units of the virial radius $R_{\rm vir}$.
        
        The mass ratio of the mergers is defined as $\mu = M_2 / M_1$, where $M_1$ and $M_2$ are the total (baryonic and DM) matter within the half-mass radii of the primary and secondary galaxies, respectively. We base our assumptions for the mergers on \citet{kannan_2015} which can be summarized as follows:
        \begin{itemize}
            \item a fraction $\mu$ of the stars in the disc of the primary galaxy is transferred to the central bulge;
            \item another fraction $0.2 \mu$ of the same stars is scattered into the DM halo, which acquires a stellar component in \textsc{GalICS};
            \item a fraction $\mu (1-f_{\rm gas-disc})$ of the gas in the disc of the primary galaxy is transferred to the central cusp, where $f_{\rm gas-disc}$ is the gas fraction on the primary galaxy's disc;
            \item even if the gas that remains in the disc undergoes a starburst in the case of major merger $\mu > 0.25$, we assume that this gas has the same star formation timescale as that in the central cusp (that is $t_{\rm SF} = M_{\rm gas}/{\rm SFR} = 0.2 \, {\rm Gyr}$), see also \citet{powell_2013} who showed that most major mergers exhibit extended SF in their early stages;
            \item the stars and the gas in the bar of the primary galaxy are transferred to the bulge and cusp of the merger remnant, respectively \citep{bournaud_2002, berentzen_2007};
            \item a fraction $\mu$ of all the stars of the secondary galaxy ends up in the bulge of the merger remnant, while the rest is added to the disc;
            \item all the gas of the secondary galaxy ends up in the central cusp.
        \end{itemize}
	    
    	\subsection{Sloan Digital Sky Survey and MaNGA}\label{sec_data_sdss}
    	
    	Our reference data from SDSS is the Data Release 7 \citep[DR7,][]{Abazajian+09} as presented in \citet{Meert+15,Meert+16}, which has a median redshift of $z \sim 0.1$. Stellar masses are computed using the best-fitting \texttt{S\'ersic-Exponential} or \texttt{S\'ersic} photometry of r-band observations, and by adopting the mass-to-light ratios by \citet{Mendel+14}. Furthermore, we adopt the truncation of the light profile as prescribed in \citet{Fischer+17_truncation}. The Meert et al. catalogues are matched with the \citet{Yang+07,Yang+12_groups} group catalogues, which allow us to identify central and satellite galaxies and provide an estimate of the group halo mass. Neural-network based morphologies from \citet{Dominguez-Sanchez+18} are adopted in the following.
    	
    	The first observable against which we test our model is the stellar mass function of satellites, which is computed using standard $V_{\rm max}$ weighting. We further test the model against the fraction of elliptical galaxies, $f_{\rm ellipticals}$ as a function of stellar mass. The \citet{Dominguez-Sanchez+18} catalogue provides estimates of T-Types, as well as the probability for Early Type Galaxies (i.e., $\text{T-Type} \leq 0$) of being S0, $P_{\rm S0}$. We compute $f_{\rm ellipticals}$ by considering only Early Type Galaxies for which $P_{\rm S0}$ falls below a certain threshold. We have accurately tested that the dependence of the ellipticals fraction on such threshold is extremely mild, showing variation within $10\%$ even for $P_{\rm S0} \lesssim 0.3$. In this work we adopt $P_{\rm S0}<0.5$ according to the results from \citet{Dominguez-Sanchez+18}.  We estimate the error bars on the satellite stellar mass functions and on $f_{\rm ellipticals}$ using the Poisson statistics.
    	
    	Later in this work we will be interested in bulge-to-total ratios (B/T) of galaxies.  Determining B/T values from observed surface brightness profiles is slightly involved, since not all objects are well fit by two-component (SerExp) profiles.  While a careful analysis of B/T values for all SDSS galaxies is not yet available, reliable values of B/T have recently been provided by Bernardi et al. (in prep.) for the objects in the MaNGA survey (\citealt{bundy_2015, drory_2015, law_2015}).  MaNGA is a component of the Sloan Digital Sky Survey IV (\citealt{gunn_2006, smee_2013, blanton_2017}; hereafter SDSS IV) and uses integral field units (IFUs) to measure spectra across nearby galaxies. The MaNGA final data release (DR17 -- \citealt{sdssDR17}) includes observations of about 10000 galaxies: the {\tt DR17 MaNGA Morphology Deep Learning} Value Added catalog (DR17-MMDL-VAC) provides morphological classifications and the {\tt DR17 MaNGA PyMorph photometric} Value Added Catalog (DR17-MPP-VAC) provides S\'ersic (Ser) and S\'ersic + Exponential (SerExp) fits to the 2D surface brightness profiles of these objects, along with a detailed flagging system for using the fits (see \citealt{dominguez_sanchez_2022} for details). Bernardi et al. (in prep.) describe how to combine the photometric and flagging information to determine reliable B/T values for these objects, and how the B/T correlate with morphology. Because the MaNGA selection function is complicated, and because the MaNGA sample is much smaller than the SDSS, we use the SDSS to determine the shape of the stellar mass function and how $f_{\rm ellipticals}$ varies with stellar mass, but MaNGA to determine how B/T correlates with stellar mass.

	\begin{figure*}
        \includegraphics[width=\textwidth]{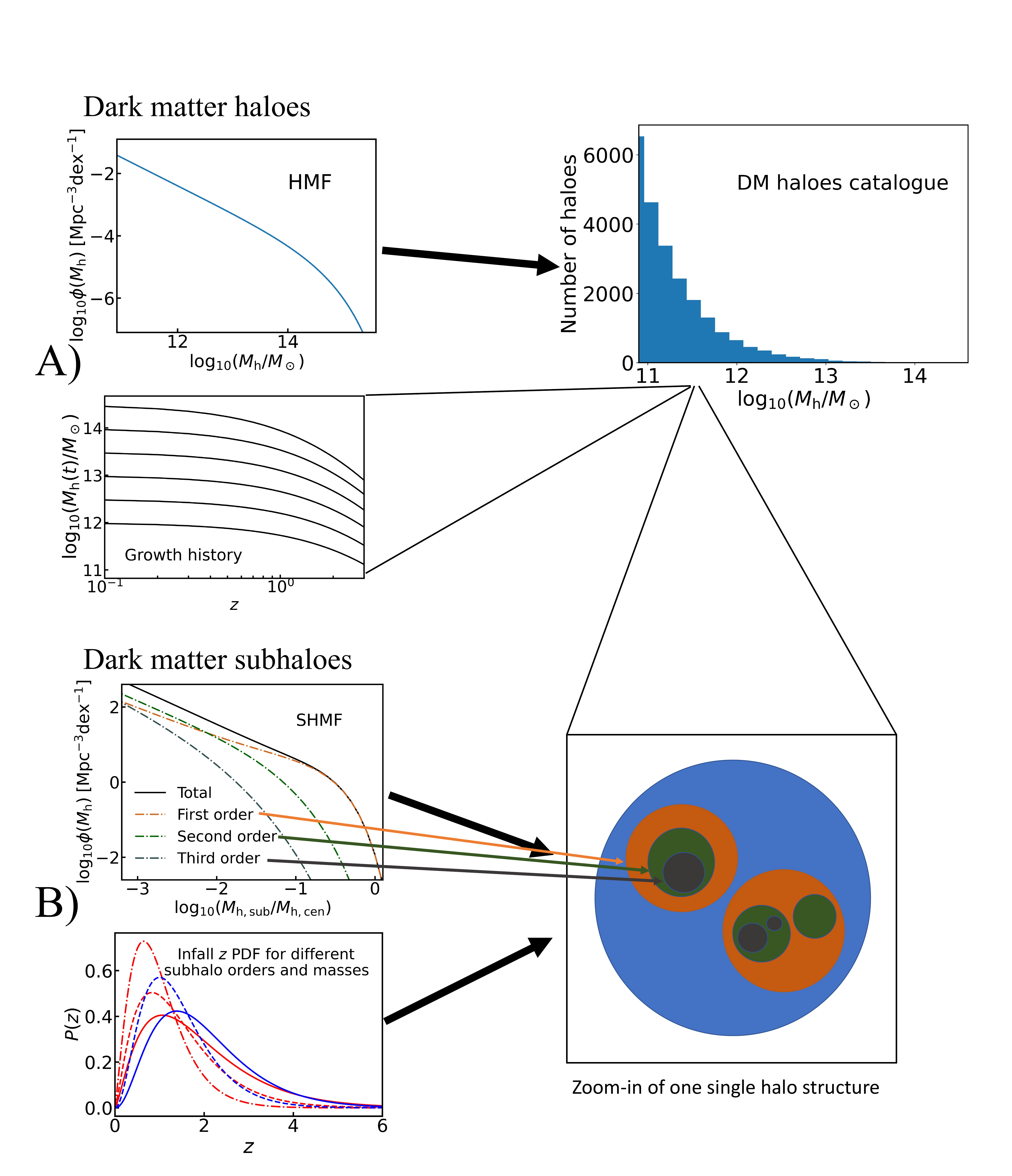}
        \caption{Scheme of the backbone of \decode for the dark matter side as described in Section \ref{sec_dream}. Panel A: functions used to generate the parent haloes catalogue (represented by the histogram in mass bins). The HMF is used as probability density distribution to generate the masses of the dark matter haloes, and a mean accretion history is assigned to each of them through an analytic fit. This contributes to build a set of main progenitors discretely, each of them characterized by a mean accretion history. The histogram in the top right panel represents a stochastic realization of the HMF. Panel B: statistical functions used to create the dark matter subhaloes. For each parent halo, we compute the SHMF for all subhaloes as well as for each order, and use it as probability density distribution for generating the subhalo population. The order of the subhaloes in the merger tree is assigned using the SHMFs distinguished by order (coloured dash-dotted lines), and in this work we limit our attention up to the second-order. Finally, the redshift of infall is assigned to subhaloes via fitted analytic equations, depending on their order and mass (see left panel of Figure \ref{fg_z_pdf_comparison} for the distinction for different orders and masses). In this way, the merging structure of each halo is known, i.e., the infalling subhaloes' order, mass and time at infall.}
        \label{fg_cartoon_haloes}
    \end{figure*}

    \section{The DECODE implementation}\label{sec_dream}
	
	In this Section we present our state-of-the-art discrete semi-empirical model, \decode. \decode is designed as a flexible, fast and accurate tool to predict the average merger and star formation histories of central galaxies at any given epoch without the need of resorting to a full SAM, hydrodynamical simulation, or even a complex, multi-parameter cosmological SEM. In this work we will mostly focus on the mean mass assembly and merger histories of central galaxies, along with their satellite abundances. We will leave the study of star formation histories to a separate study.
	
	The main steps of \decode can be summarized as follows:
    \begin{itemize}
        \item   generation of the central DM halo population (Section \ref{sec_num_density_haloes});
        \item   generation of the DM subhalo population (Section \ref{sec_dream_sub_pop});
        \item   evolution of subhaloes after infall (Sections \ref{sec_dream_z_inf} and \ref{sec_dream_evo_surv});
        %\item   generation of the galaxy population (Section \ref{sec_dream_gal_halo_connect}).
        \item   populating haloes with galaxies (Section \ref{sec_dream_gal_halo_connect}).
    \end{itemize}
	
    Figure \ref{fg_cartoon_haloes} depicts our general framework to generate the population of DM haloes and subhaloes which we further detail in the subsections below. In brief, our methodology relies on generating large catalogues of parent haloes extracted from the HMF and endowed with a mean halo growth history as derived from N-body numerical simulations and analytic models \citep[e.g.,][]{vdb_2005}. Subhaloes are subsequently extracted from the unevolved SHMFs, also based on accurate studies performed on N-body simulations \citep[e.g.,][]{jiang_vdb_2016}.
    In its implementation of the evolutionary tracks for central galaxies, \decode is similar in concept to its predecessor statistical semi-empirical model \textsc{steel} (\citetalias{grylls_paper1}), although it crucially differs from it in its implementation, beyond the performance itself. \decode in fact avoids continuous statistical weights of central and satellite galaxies, but it directly works on discrete objects, similarly to an N-body simulation, making it easier to handle object-by-object variations, but still preserving \textsc{steel}'s extreme flexibility and broad independence on volume and/or mass resolution limitations.
    
    In addition, as detailed below, \decode distinguishes between satellites and satellites of satellites, a key feature that was absent in \textsc{steel}. \decode's main objective is to still predict \textit{mean} galaxy growth histories, as in \citetalias{grylls_paper1}'s statistical model \textsc{steel}, but instead of using statistical weights, it relies on the generation of stochastic samples of haloes and galaxies that, on average, grow in mass as predicted by \textsc{steel}, as further detailed and demonstrated in Appendix \ref{app_weighted_discrete_consistency}. By working with discrete sources, \decode avoids the need to assign weights to each evolutionary step, which is a far from trivial task when propagated through different layers of complexities in the galaxy modelling.
    We also note that, although we adopt a Planck cosmology throughout, as specified in Section \ref{sec_intro}, our results are insentitive to the exact choice of input cosmological parameters within reasonable ranges. This independence on cosmological parameters is mostly induced by the heavy use of the SHMF, which has been shown several times in the literature to be of very similar shape in different simulations \citep[e.g.,][]{jiang_vdb_2016, green_2021}. We will further reiterate on this point in Section \ref{sec_validation}.
    
    \subsection{Generating the population of parent haloes}\label{sec_num_density_haloes}
    
    Our first step consists in generating a large catalogue of parent DM haloes extracted from the HMF. In this work we adopt the definition and parameterization of the HMF according to \cite{tinker_2008}, which accounts only for central haloes\footnote{We make use of the \texttt{lss.mass\_function} module in the Python \texttt{colossus} package (\citealt{diemer_2018}).}. We extract haloes and their masses in our catalogue from the cumulative HMF multiplied by an input cosmological volume. We choose here to use a reference box of 250 Mpc on a side, which allows throughout to balance speed with accuracy, although we stress that \decode is flexible enough to generate even larger volumes with ease. This method is extremely rapid and allows to simulate even large samples of massive cluster-sized haloes. 
    
    The average mass accretion history of each central halo is then computed using the methodology described in \cite{vdb_2014}. Instead of computing the growth of each single halo, we predefine a fine halo grid in mass of 0.1 dex width and assign the same mean history to all the DM haloes contributing to the same cell in the grid. This initial step provides the average growth history of the ``main progenitor'' (when compared to a traditional merger tree).
    
    \subsection{Generating the population of subhaloes of different orders}\label{sec_dream_sub_pop}
    
    The average number of subhaloes of a given mass that falls onto the parent halo at any given time is given by the SHMF. Here, we adopt the definition of unevolved SHMF distinguished by ``order'' of accretion in the merger tree \citep{jiang_vdb_2014, jiang_vdb_2016}, with first-order subhaloes being the ones falling directly onto the main branch, second-order subhaloes the ones already satellites in first-order subhaloes at the time of accretion onto the main branch, and so on (Panel B in Figure \ref{fg_cartoon_haloes}). 
    
    The same methodology applied to parent haloes is used to calculate the number and mass of all the subhaloes that have ever fallen onto each given parent halo, by making use of the cumulative total unevolved SHMF. Once the number and mass of the subhaloes are known, the order of the subhaloes are assigned by considering the SHMF distinguished by order, for which we use the recipe from \citet{jiang_vdb_2016} for the first-order SHMF and Equation (17) of \citet{jiang_vdb_2014} for higher-orders. The probability that a given subhalo of a given mass is of first- or higher-order, is then simply given by the relative ratio between first/higher-order SHMF and the total SHMF computed in the chosen bin of (sub)halo mass.
    
    We stress that \decode is a statistical SEM, where the mock catalogues of DM haloes are stochastic realizations of the input HMF and SHMF, taken from analytic fits to N-body simulation. The only free parameter in our model is the scatter in stellar mass at given halo mass, which is included in the abundance matching relation in Equation (\ref{eq_aversa_AM}), and it only contributes to the shape of the mean SMHM relation, as we will further detail in Section \ref{sec_dream_gal_halo_connect}.
    
    \subsection{Infall redshifts}\label{sec_dream_z_inf}
    
    To predict a robust merger history of central galaxies, clear knowledge of infall redshifts ($z_{\rm inf}$) of their satellite galaxies is required. In our model, we adopt the definition of infall redshift as the time when a DM halo was accreted for the \textit{first} time as subhalo, or in other words when it entered the virial radius of another halo. We assign infall redshifts to the subhaloes in a statistical way distinguishing between first-order and higher-order subhaloes. For first-order subhaloes, we apply the redshift probability distribution dictated directly by the SHMF, as follows. In practice, at any given redshift $z$, for a parent halo growing by $\mathrm dM_{\rm h,par} (z)$ at any redshift interval $z + \mathrm dz$ to $z$, the probability density function (PDF) of infall redshifts of subhaloes of infall mass $M_{\rm h,sub}$ is given by the derivative the SHMF with respect to the redshift, formally
    \begin{equation}\label{eq_pdf_z_inf}
        {\rm PDF}(z_{\rm inf}) \propto \frac{\mathrm d}{\mathrm dz} \phi(M_{\rm h,sub}, M_{\rm h,par}(z)) \; .
    \end{equation}
    As mentioned above, the unevolved SHMF $\phi (M_{\rm h,sub})$ provides the total number and mass of the subhaloes that have ever merged with the parent halo at any epoch. The mass of the parent halo $M_{\rm h,par}(z)$ will grow with cosmic time when moving from $z+ \mathrm dz$ to $z$. The change in the associated SHMF $\mathrm{d} \phi(M_{\rm h,sub}, M_{\rm h,par}(z))$, will provide the number and mass of the subhaloes of mass within $M_{\rm h,sub}$ and $M_{\rm h,sub} + \mathrm dM_{\rm h,sub}$, that have merged with the parent halo and contributed to its mass growth in the redshift interval $\mathrm dz$. Thus, the (normalized) derivative with redshift of the SHMF at a fixed subhalo mass $M_{\rm h,sub}$ given in Equation (\ref{eq_pdf_z_inf}), will provide the PDF for the subhaloes of mass within $M_{\rm h,sub}$ and $M_{\rm h,sub} + \mathrm d M_{\rm h,sub}$, merging with the parent halo $M_{\rm h,par}(z)$ at any given redshift $z$. Redshifts of infall for subhaloes of any given mass $M_{\rm h,sub}$ are then generated by randomly extracting them from the PDF given in Equation (\ref{eq_pdf_z_inf}).
    
    The methodology described so far to assign redshifts of infall to subhaloes can only be applied to first-order subhaloes, as the SHMF only provides information on the subhaloes merging with the parent halo. For second-order subhaloes, we adopt a similar, but not identical, recipe. We first generate the full merger tree associated to a given parent halo P0 by following its mass accretion history backwards in time and, using the recipe described above, computing the population of first-order subhaloes S1 and their redshifts of infall from the first-order SHMF \citep[][]{jiang_vdb_2014, jiang_vdb_2016}. For each S1 we then follow its mass and satellite accretion history using again the first-order SHMF. The satellites of S1 will be second-order, S2, with respect to P0. We repeat this loop up to the third-order\footnote{We explore and provide results for the distribution of infall redshifts up to the third-order for completeness. However, for the purposes of this paper discussed in Section \ref{sec_results} we limit our investigation to the second-order subhaloes. We have tested that orders higher than the second have negligible contribution to the amount of mergers.} as orders higher than the third have an insignificant contribution to the total number density of satellites \citep[][]{jiang_vdb_2014}. In order to speed up the computational time, we first run fine merger histories for all relevant subhaloes S2 and S3, and then compute analytic fits to the ${\rm PDF}(z)$ of their redshifts of infall. The parameterization for the infall redshift distribution we adopt is given by the following formula
    \begin{equation}\label{eq_z_infall_pdf}
        \mathcal P(z) = A z^\alpha \frac{1}{\delta e^{\beta z} - \gamma} \;,
    \end{equation}
    where $A$, $\alpha$, $\beta$, $\gamma$ and $\delta$ are dimensionless free fitting parameters, with best-fit values reported in Section \ref{sec_validation}. We use the analytic PDF of Equation (\ref{eq_z_infall_pdf}) to assign the infall redshifts statistically \footnote{In reality, for each single parent halo, a generic subhalo of the $i$-th order must have fallen at a redshift higher than the infall redshift of the $(i-1)$-th subhalo. We therefore set the infall redshift of the $(i-1)$-th order subhalo as a lower bound for the $i$-th order subhalo redshift.} to all second- and third-order subhaloes, especially when simulating large boxes and cluster-sized parent haloes.
    
    The procedure described above generates a stochastic merger tree of subhaloes for a given \emph{mean} parent halo mass accretion track $M_{\rm h,par} (z)$. In other words, \decode produces a stochastic distribution of subhaloes merging on a mean halo. As quantitatively proven and discussed in Appendix \ref{app_weighted_discrete_consistency}, when averaged over a large population of subhaloes, this approach is equivalent to an average one in which discrete subhaloes are replaced by statistical weights given by the SHMF, as carried out in \textsc{steel}. We stress that the main advantage of building halo assembly histories via discrete sources resides in the extreme flexibility of working with discrete objects and not with statistical weights, especially when transitioning to galaxies and the modelling of their evolutionary properties.
    
    \subsection{Merging timescales and surviving subhaloes}\label{sec_dream_evo_surv}
    
    Once a subhalo first falls into its host halo, it is affected by tidal stripping and dynamical friction, resulting in an overall net mass loss. Many works have carefully studied via numerical simulations these processes \citep[see, e.g.,][]{vdb_2005, giocoli_2008, angulo_2009, jiang_vdb_2016, vdb_2016, vdb_2017, green_2019}, and have found that the average mass loss rate of satellite subhaloes can be analytically expressed as
    \begin{equation}\label{eq_subhalo_MLR}
        \dot M_{\rm h, sub} = - \mathcal A \frac{M_{\rm h, sub}}{\uptau_{\rm dyn}} \bigg( \frac{M_{\rm h, sub}}{M_{\rm h, host}} \bigg)^\zeta \; ,
    \end{equation}
    where $\mathcal A = 1.54$, $\zeta = 0.07$, and $M_{\rm h, sub}$ and $M_{\rm h, host}$ are, respectively, the masses of the subhalo and halo that hosts the former subhalo. $\uptau_{\rm dyn}$ is the halo dynamical timescale given by
    \begin{equation}
        \uptau_{\rm dyn} (z) = 1.628 h^{-1} {\rm Gyr} \bigg[ \frac{\Delta_{\rm vir} (z)}{178} \bigg]^{-1/2} \bigg[ \frac{H(z)}{H_0} \bigg]^{-1} \; ,
    \end{equation}
    with $H(z)$ being the Hubble's parameter at redshift $z$ and $\Delta_{\rm vir}$ the virial parameter taken from Equation (6) of \cite{bryan_norman_1998}. The typical timescale that a subhalo needs in order to fully merge with its progenitor from the time of first accretion is well described by the merging timescale formula given by Equation (5) of \cite{boylan_kolchin_2008}, which we report below,
    \begin{equation}\label{eq_tau_merge}
        \uptau_{\rm merge} = \uptau_{\rm dyn} A \frac{(M_{\rm h, host} / M_{\rm h, sub})^b}{\ln (1 + M_{\rm h, host} / M_{\rm h, sub})} \exp \bigg[ c \frac{J}{J_c(E)} \bigg] \bigg[ \frac{r_c(E)}{R_{\rm vir}} \bigg]^d \; ,
    \end{equation}
    where $J/J_c(E) \def \xi$ represents the orbital circularity, $r_c / R_{\rm vir}$ the orbital energy and $(A, b, c, d)$ are free parameters that govern the dependence of the merging timescale on the mass ratio. Here we adopt the fitting parameters provided by \cite{mccavana_2012}. In particular, in order to apply Equation (\ref{eq_tau_merge}), we assign an orbital circularity $\xi$ to galaxies according to \cite{khochfar_2006}, by extracting a random value from a Gaussian distribution centered in $\bar \xi = 0.5$ and with standard deviation $\sigma_\xi = 0.23$, and compute the ratio between the average radius of the orbit $r_c$ and the host halo virial radius $R_{\rm vir}$
    \begin{equation}\label{eq_orb_energy}
        \frac{r_c}{R_{\rm vir}} = \frac{\xi^{2.17}}{1 - \sqrt{1- \xi^2}} \; .
    \end{equation}
    The analytic recipes described above are an approximation to the complex dynamics of DM subhaloes, and also the numerical simulations from which they are extracted may themselves suffer from resolution and/or incompleteness effects. To allow for some flexibility in the merging timescales, following \citetalias{grylls_paper1}, we also include a fudge factor $f_{\rm dyn}$ in Equation (\ref{eq_tau_merge}), $\uptau_{\rm merge} \rightarrow f_{\rm dyn} \uptau_{\rm merge}$, which we assume to be slightly dependent on parent mass, as detailed in Section \ref{sec_validation}.
    
    To test the validity of the methodology described in the previous paragraph, we analyze the population of the surviving subhaloes at present day via the unevolved surviving SHMF. For the latter we adopt the definition of \cite{jiang_vdb_2016}, which is the number density of the surviving subhaloes still present today as a function of their mass at the time of first accretion. We again assume that subhaloes of third-order or higher are not statistically significant to the surviving population \citep{jiang_vdb_2014}.
    
    The next step is to assign merging timescales to all subhaloes of different ranking, which we implement in \decode in the following way. For any parent halo of mass $M_{\rm P0}$, with a first-order subhalo of mass $M_{\rm S1}$ and a second-order subhalo of mass $M_{\rm S2}$:
    \begin{enumerate}
        \item we first calculate the merging timescale of the first-order subhalo which depends on the ratio $M_{\rm P0} / M_{\rm S1}$;
        \item depending on the first accretion epoch and the merging timescale, we consider and implement in \decode the three following possibilities: 1) the first-order subhalo has survived today and we assume at this step that its higher-order subhaloes inside still exist; 2) the first-order subhalo has not survived and it releases all its higher-order subhaloes to the parent\footnote{We investigate also different ways of treating the evolution of subhaloes after the time of infall. In particular, we explored two additional possibilities: 1) every higher-order subhaloes merge together with their host first-order subhalo, 2) higher-order subhaloes have a dynamical friction longer than the age of the Universe and never merge. In both cases, there is not any appreciable difference in terms of satellite abundances and mergers.} \citep[][]{jiang_vdb_2016}; 3) the higher-order subhaloes have been tidally disrupted before their first-order subhalo has merged;
        \item we assign the merging timescale to the second-order subhalo which depends on the ratio $M_{\rm S1} / M_{\rm S2}$;
        \item finally, to the second-order subhaloes that are released from a first-order to the parent we assign a new merging timescale using the ratio $M_{\rm P0} / M_{\rm S2}$\footnote{Here, $M_{\rm S2}$ is the evolved mass of the second-order subhalo S2, that following a mass evolution according to Equation (\ref{eq_subhalo_MLR})}, when released to the parent halo P0.
    \end{enumerate}
    In Section \ref{sec_validation} we will compare \decode's predicted abundances of local unmerged subhaloes, described in terms of the surviving SHMF, with the analytic model of \citet{jiang_vdb_2016} and with the resolved SHMF of the Millennium simulation, showing very good agreement when adopting a fudge factor of $f_{\rm dyn} \sim 0.64$. We stress already here that simply adopting $f_{\rm dyn} \sim 1$ does not alter any of our main results.
    
    \subsection{Building the mapping between galaxy stellar mass and host dark matter halo mass}\label{sec_dream_gal_halo_connect}
    
    One of the key components of our modelling is the relation between stellar mass and host halo mass, the SMHM relation. The latter is computed via the formalism put forward in \citet[][see Equation 37 therein]{aversa_2015}, which allows to calculate the mean stellar mass at given halo mass
    \begin{equation}\label{eq_aversa_AM}
    \begin{split}
        \int_{\log M_*}^{+\infty} \phi(M_*', z) & \mathrm d \log M_*' = \\ 
        \int_{-\infty}^{+\infty} & \frac{1}{2} \mathrm{erfc} \Bigg\{ \frac{\log M_h (M_*) - \log M_h' }{\sqrt{2} \Tilde{\sigma}_{\log M_*}} \Bigg\} \\
        & \cdot \phi(M_h', z) \mathrm d \log M_h' \; ,
    \end{split}
    \end{equation}
    where $\Tilde{\sigma}_{\log M_*} = \sigma_{\log M_*} / \mu $, with $\sigma_{\log M_*}$ being the Gaussian scatter at fixed halo mass and $\mu = \mathrm d \log M_* / \mathrm d \log M_h$ the derivative of the SMHM relation. Equation (\ref{eq_aversa_AM})\footnote{We test that Equation (\ref{eq_aversa_AM}) provides accurate results for $\sigma_{\log M_*} \lesssim 0.3$.} provides a fast and flexible methodology to compute the SMHM relation numerically, without the need for a pre-defined analytic fit, but requiring in input only one parameter, the scatter in stellar mass at fixed halo mass \citep[see also][]{kravtsov_2018}. When applying Equation (\ref{eq_aversa_AM}), the other two main ingredients are the observational SMF and the HMF including the subhalo term. We make use of the total HMF which accounts for both parent haloes and subhaloes $\phi(M_{\rm h,tot}) = \phi (M_{\rm h,par}) + \phi (M_{\rm h,sub}) $, where $\phi (M_{\rm h,sub}) = k \cdot \phi (M_{\rm h,par})$ with $k$ being a correction factor (as described in Appendix \ref{app_corr_HMF}). The additional term $\phi (M_{\rm h,sub})$ allows to include all unstripped subhaloes and unmerged up to redshift $z$, as predicted by \decode following the recipes detailed in Section \ref{sec_dream_evo_surv}. We find that the new total HMF $\phi(M_{\rm h,tot})$, inclusive of the surviving satellites, is similar to the parent HMF $\phi (M_{\rm h,par})$ but, as expected, with a steeper low mass end. A similar approach was adopted by \cite{behroozi_2013} who, in their Appendix G provide a redshift-dependent analytic formula to correct the HMF for the abundances of surviving satellites. We adopt their analytic formula which we fit to reproduce our realizations of halo+subhalo mass functions with \decode. Our best-fit parameters are given in Appendix \ref{app_corr_HMF}. The SMHM relation generated at each redshift via Equation (\ref{eq_aversa_AM}) is then given as input in \decode  to assign galaxies to all parent haloes and to all subhaloes of any rank at the time of infall. We also test our SMHM relations computed via Equation (\ref{eq_aversa_AM}) with the SMHM relations computed using the halo peak velocity function and the same SMF as input (see Appendix \ref{app_sham}).
    
    \begin{figure*}
        \includegraphics[width=0.52\linewidth]{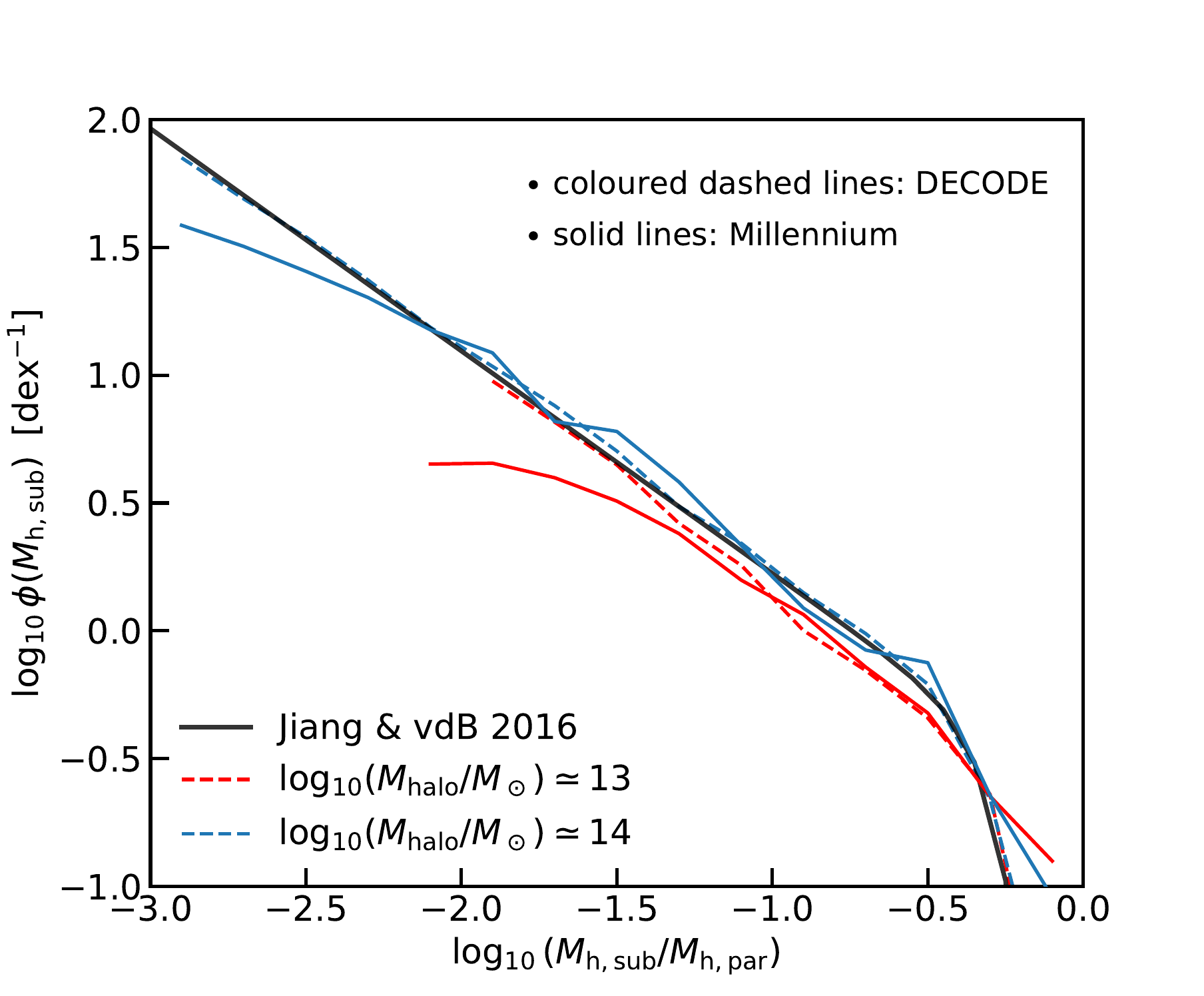}
	    \includegraphics[width=0.47\linewidth]{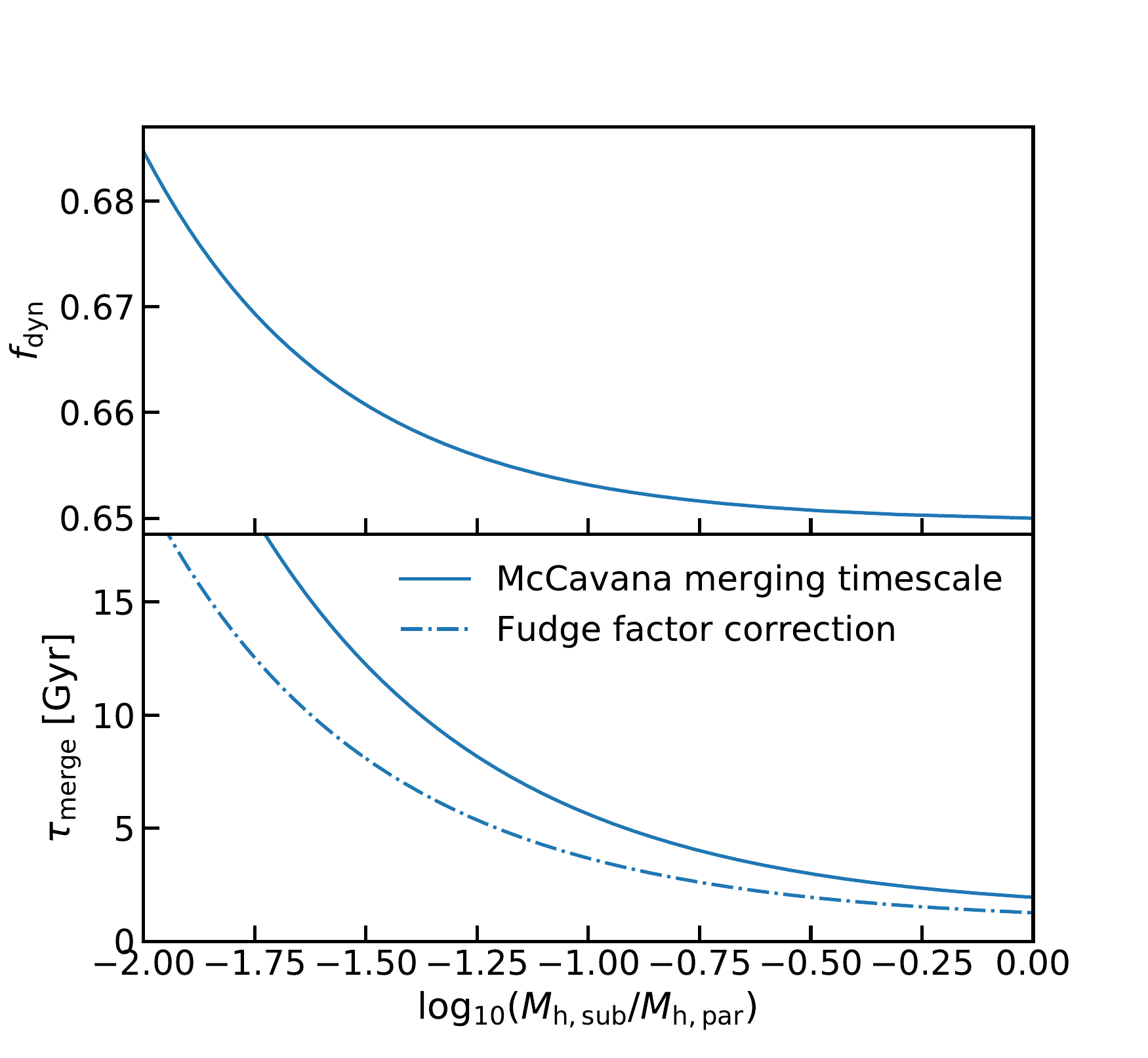}
	    \caption{Left panel: Comparison between the surviving unevolved subhalo mass function for two different parent halo masses at redshift $z = 0$. The coloured dashed lines are the results from \decode, the solid lines are the results extracted from the Millennium simulation (as described in Section \ref{sec_data}) and the solid black line the analytic form taken from \citet{jiang_vdb_2016}. Right upper panel: fudge factor as function of the mass ratio according to Equation (\ref{eq_fudge}). Right lower panel: merging timescale from \citet{mccavana_2012} (solid line) compared with that computed by applying the fudge factor correction (dash-dotted line).}
	    \label{fg_survSHMF_fudge_tau_merge}
    \end{figure*}
    
    Throughout we assume that satellites at infall follow the same SMHM relation of centrals at that epoch. We note that if we were to relax this assumption by allowing for a somewhat different SMHM relation for satellites, our main results would be unaltered in the stellar mass range of interest here $M_* \gtrsim 10^{10} M_\odot$, in line with the findings of other SEMs that suggest similar distributions of centrals and satellites at the high-mass end \citep[see, e.g.,][]{rodriguez_puebla_2012, dvornik_2020, engler_2021, contreras_2021}. Furthermore, we consider only "frozen" models in this work (i.e., the mass of the satellite is assumed to be constant after the infall). As also shown by \citetalias{grylls_paper1}, allowing for some star formation and stellar stripping in the satellites after infall, following standard recipes in the literature, does not alter any of our conclusions on the abundances of satellites, at least for galaxies of stellar mass above $10^{10} M_\odot$. We will study the full impact of stellar stripping and latent star formation in satellites in a separate work. We point out that galaxies are assigned to DM haloes via the \textit{mean} SMHM relation, derived via Equation (\ref{eq_aversa_AM}), because \decode at this level of development is mostly sensitive to the mean galaxy growth and mean merger histories. Therefore, in this paper we show only the mean predictions for the satellite abundances, ellipticals and B/T ratios.

	\subsection{Validating the dark sector}\label{sec_validation}
	
	\begin{table*}
        \begin{center}
    	\begin{tabular}{ |c@{\hskip 0.25in} c@{\hskip 0.25in} c c c c c| }
			\hline
    		subhalo order & mass range & $A$ & $\alpha$ & $\beta$ & $\gamma$ & $\delta$ \\
			\hline
			second & $10 < \log_{10} (M_{\rm h,sub} / M_\odot) < 11$ & $10.06_{-3.86}^{+3.51}$ & $1.07_{-0.14}^{+0.16}$ & $13.78_{-5.78}^{+3.51}$ & $1.86_{-7.80}^{+9.25}$ & $1.77_{-0.37}^{+0.44}$\\
			\hline
			second & $11 < \log_{10} (M_{\rm h,sub} / M_\odot) < 12$ & $24.93_{-10.31}^{+6.24}$ & $1.77_{-0.18}^{+0.22}$ & $8.87_{-3.64}^{+3.39}$ & $0.42_{-6.71}^{+6.22}$ & $1.94_{-0.35}^{+0.43}$\\
			\hline
			second & $\log_{10} (M_{\rm h,sub} / M_\odot) > 12$ & $24.81_{-9.19}^{+5.64}$ & $2.67_{-0.21}^{+0.26}$ & $2.31_{-0.87}^{+0.88}$ & $-2.60_{-4.08}^{+2.52}$ & $1.86_{-0.34}^{+0.45}$\\
			\hline
			third & $10 < \log_{10} (M_{\rm h,sub} / M_\odot) < 11$ & $8.37_{-3.36}^{+2.76}$ & $1.50_{-0.18}^{+0.21}$ & $13.50_{-5.79}^{+3.74}$ & $3.00_{-8.75}^{+10.33}$ & $3.29_{-0.51}^{+0.53}$\\
			\hline
			third & $\log_{10} (M_{\rm h,sub} / M_\odot) > 11$ & $23.71_{-12.63}^{+7.73}$ & $2.54_{-0.31}^{+0.31}$ & $3.41_{-1.70}^{+1.57}$ & $-3.56_{-4.65}^{+4.59}$ & $3.12_{-0.49}^{+0.70}$\\
			\hline
		\end{tabular}
	    \end{center}
	    \caption{Best fit parameters of the infall redshift distribution parameterization from Equation (\ref{eq_z_infall_pdf}), for different subhalo orders and mass intervals.}
	    \label{tb_z_pdf_comparison}
	\end{table*}
	    
	\begin{figure*}
        \includegraphics[width=0.485\linewidth]{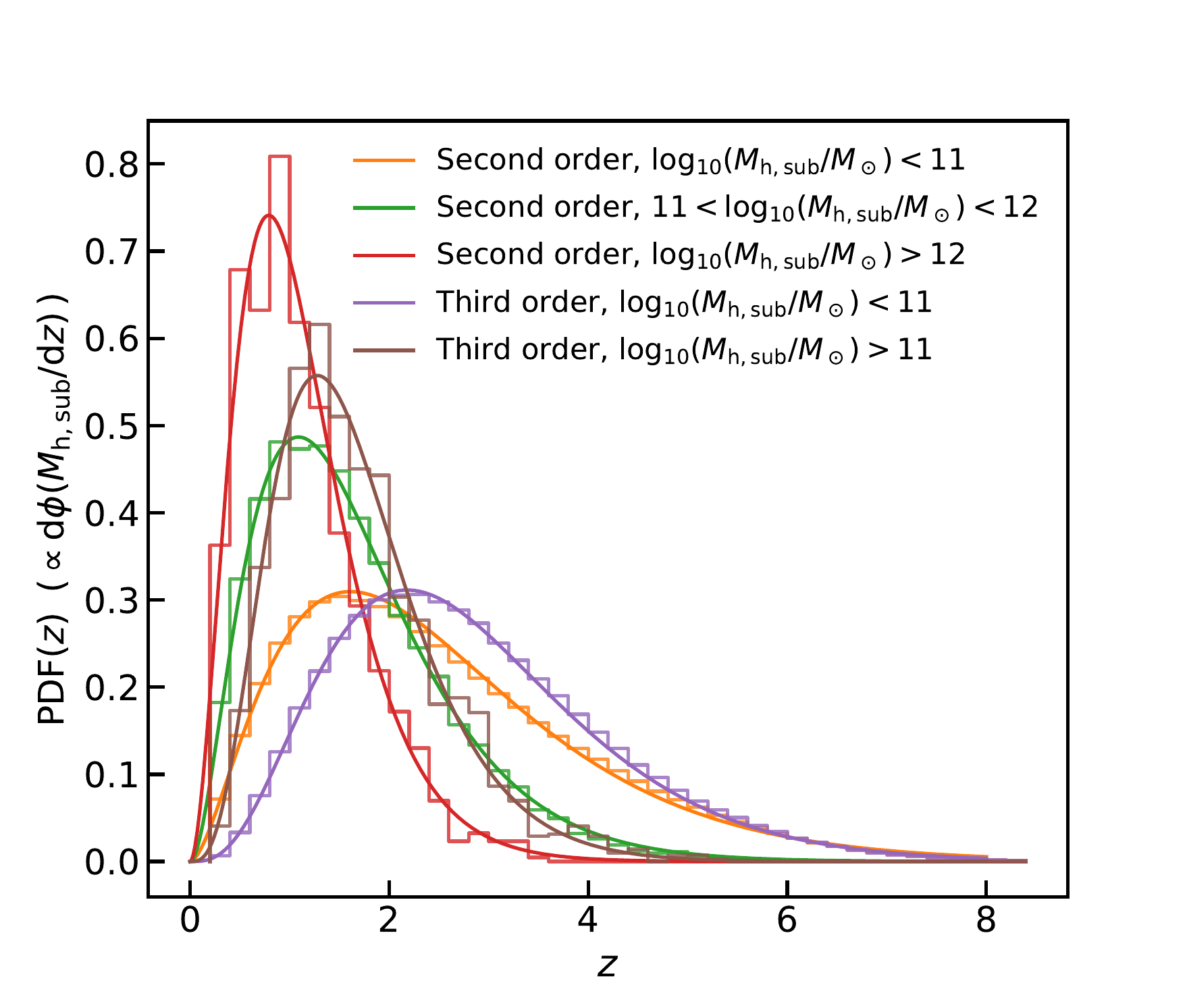}
	    \includegraphics[width=0.51\linewidth]{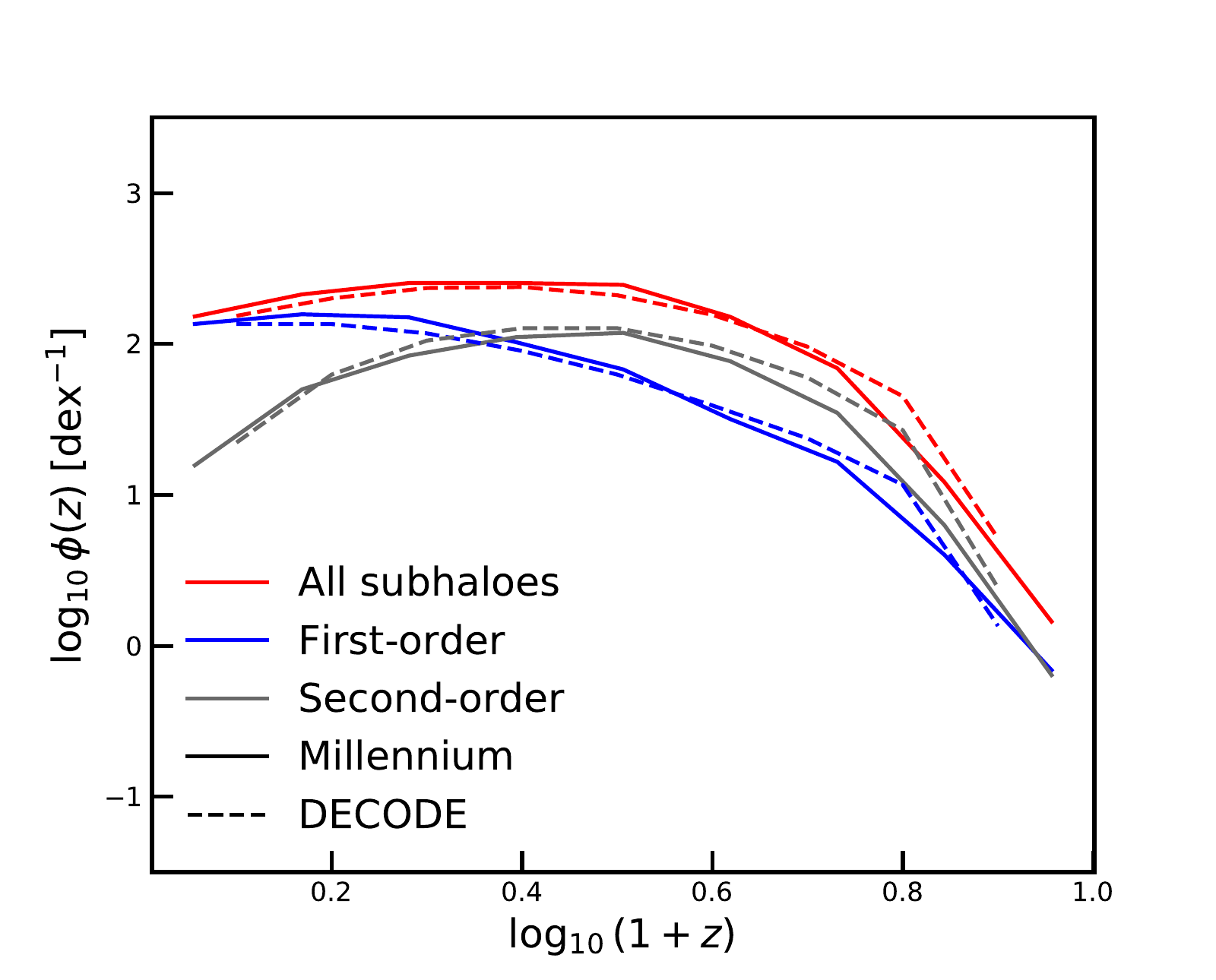}
        \caption{Left panel: analytic (normalized) probability distributions of the infall redshifts adopted in this work to generate the mock catalogues. The results are organized for different subhalo orders and mass ranges. The histograms show the results from the merger tree and the curves show the best fits. Right panel: comparison between number densities of the infall redshifts from our model \decode (dashed lines) and from Millennium simulation (solid lines) for parent haloes of mass selected between $10^{14} M_\odot$ and $10^{14.1} M_\odot$. Results are shown for subhaloes of all orders (red lines), first-order (blue lines) and second-order (grey lines). Similar results are found for other parent halo masses.}
	    \label{fg_z_pdf_comparison}
    \end{figure*}
	
	Before presenting the power and flexibility of \decode in efficiently probing crucial aspects of galaxy evolution, such as merger pairs and bulge formation, we test the accuracy of \decode in matching the number densities and $z_{\rm inf}$ distributions of unevolved and unmerged subhaloes of first- and second-order as predicted by N-body simulations and SAMs. We first note that the unevolved total SHMF fitted by \citet{jiang_vdb_2016} from the MultiDark simulation \citep{klypin_2011}, is well consistent, we verified, with the total unevolved SHMF extracted from the Millennium simulation, at least above the resolution limit of the latter. This is quite significant as it further proves the universality of the SHMF with respect to the underlying cosmological model and also other aspects of the simulations, such as the halo finder algorithm.
	    
	The left panel of Figure \ref{fg_survSHMF_fudge_tau_merge} shows the unevolved \emph{surviving} SHMF (i.e., composed by subhaloes not yet merged or completely disrupted) for two different values of parent halo masses, as labelled. The results are compared with those from the Millennium simulation and from the SAM of \cite{jiang_vdb_2016}, which reproduces well the results from the Millennium simulation. The predictions from \decode on the number of surviving, unstripped subhaloes, are plotted with dashed lines, and become indistinguishable from the grey solid line of \cite{jiang_vdb_2016}, when a small correction is applied to the merging timescales of \citet{mccavana_2012}, as also pointed out by \citetalias{grylls_paper1}.
	As shown in the right panel of Figure \ref{fg_survSHMF_fudge_tau_merge}, the fudge factor $f_{\rm dyn}$ in the dynamical friction timescale, we find, is well represented by the following linear relation with the halo-to-subhalo mass ratio $\psi = (M_{\rm h, host} / M_{\rm h, subhalo})$
    \begin{equation}\label{eq_fudge}
        f_{\rm dyn} = a \psi + b \; ,
    \end{equation}
    where the best-fit values for $a$ and $b$ are $0.00035$ and $0.65$ respectively.
	    
    In order to be used as a flexible tool to model, e.g., galaxy merger rates, it is essential for \decode to not only generate the correct abundances of subhaloes of different orders, but also to reproduce the correct probability distributions of their infall redshifts $z_{\rm inf}$. The left panel of Figure \ref{fg_z_pdf_comparison} shows the predicted $z_{\rm inf}$ PDFs predicted by \decode (histograms) for different subhalo masses and order, as labelled. As described in Section \ref{sec_dream_z_inf}, these PDFs have been calculated by generating full merger trees for each parent halo and subhalo. Such a procedure is of course not practical and time-consuming. For this reason, we fit analytic functions to the PDFs of all second- and third-order subhaloes of different masses of interest here, and we report the best fit parameters for Equation (\ref{eq_z_infall_pdf}) in Table \ref{tb_z_pdf_comparison}. We recall that the PDFs for first-order subhaloes are instead directly computed by the change of the SHMF along the parent halo mean mass accretion track (Section \ref{sec_dream_evo_surv}). The right panel of Figure \ref{fg_z_pdf_comparison} compares the number densities of the infall redshifts of first- and second-order subhaloes accreting onto a parent haloes of mass between $10^{14} M_\odot$ and $10^{14.1} M_\odot$ as predicted by \decode (dashed lines) and the Millennium simulation (solid lines). The agreement is good, further validating the accuracy of our modelling.

	\section{Results}\label{sec_results}
        
        As introduced in the previous Sections, \decode is a flexible statistical SEM that can predict, for a given input SMHM relation, the implied star formation and merger histories of central galaxies. It is in concept similar to its predecessor, \textsc{steel}, but it has relevant new features, including the separation in subhalo accretion order and a more refined treatment of infall timescales.

        In this work we will mostly focus on the ability of \decode to rapidly predict the mean merger histories of central galaxies of different stellar mass for different input SMHM relations. In a separate work, we will extend \decode to predict star formation histories and the amount of intracluster light generated from the stellar stripping of infalling satellites.

        \citetalias{grylls_paper2} identified a major role played by the SMHM relation in shaping the merger history of a central galaxy, especially in the case of more massive galaxies. In brief, if the high-mass end of the SMHM is flatter, it will imply that a halo increasing with mass via mergers will correspond to progressively lower growth in the stellar mass of its central galaxy. In the extreme condition of a perfectly flat SMHM relation, any increase in halo mass would not be followed by any increase in stellar mass, i.e., the (especially major) merger rate of those type of central galaxies will be drastically reduced. 

        In this Section, we further expand on \citetalias{grylls_paper2} taking into account the different SMHM relations as derived from abundance matching using the latest data on the SMF at low and high redshifts (Section \ref{sec_smhm_models}). We will then discuss the implied merger rates (Section \ref{sec_res_mergers_rate}), number densities of "unmerged galaxies", i.e., satellites, in the local Universe for different SMHM relations (Section \ref{sec_res_sate}), the implied fraction of ellipticals originating from major mergers (Section \ref{sec_res_morphology}), and the distribution of B/T stellar mass ratios of galaxies in the local Universe induced by major mergers and some models of disc instabilities (Section \ref{sec_res_bt_ratio}). We will also present the prediction of the growth histories of BCGs in Section \ref{sec_res_bcgs}. We will show that, as expected, the aforementioned quantities are highly dependent on the input SMHM relation in a hierarchical DM-dominated framework of galaxy evolution.
        
        \begin{figure*}
            \includegraphics[width=0.495\linewidth]{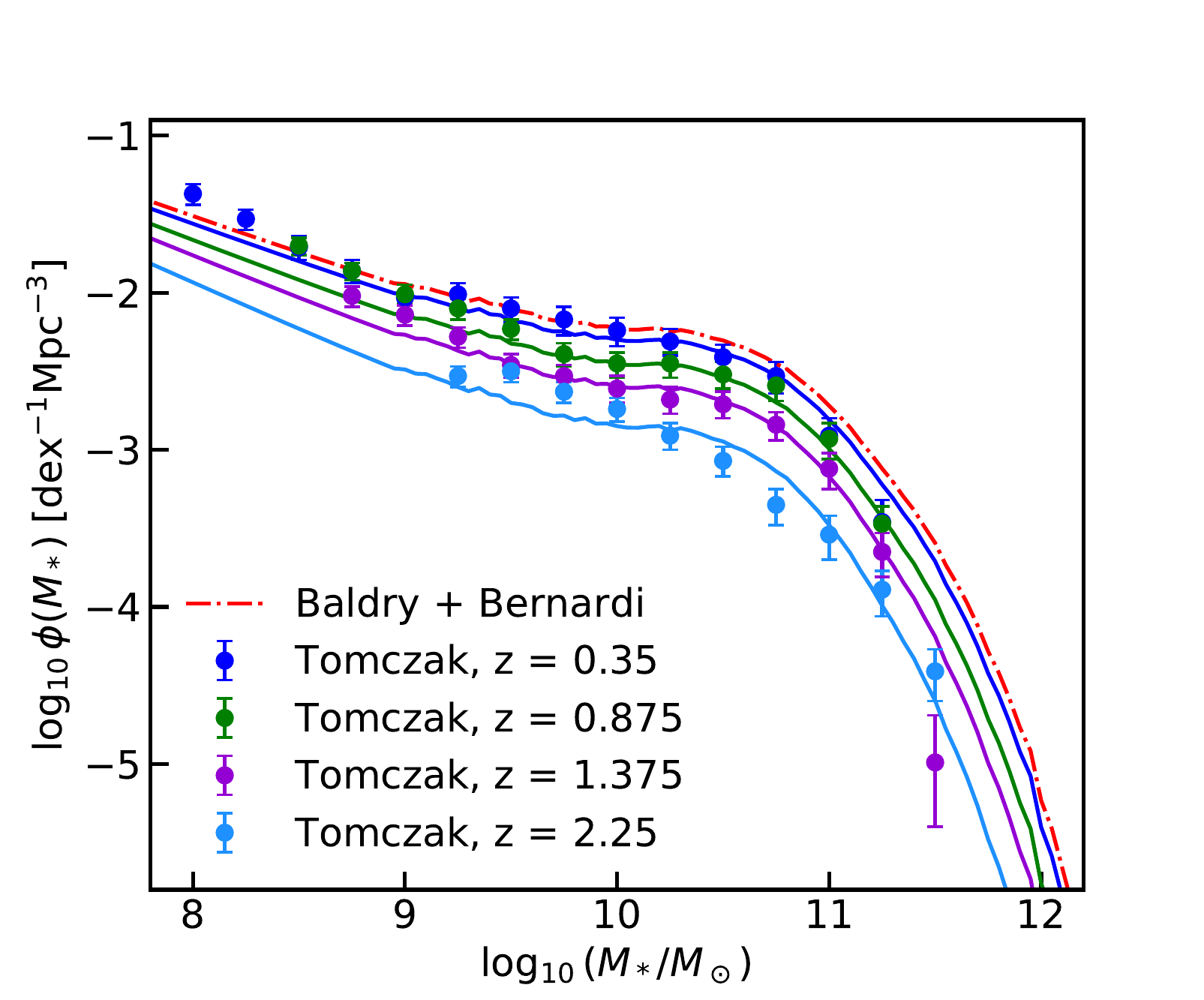}
            \includegraphics[width=0.495\linewidth]{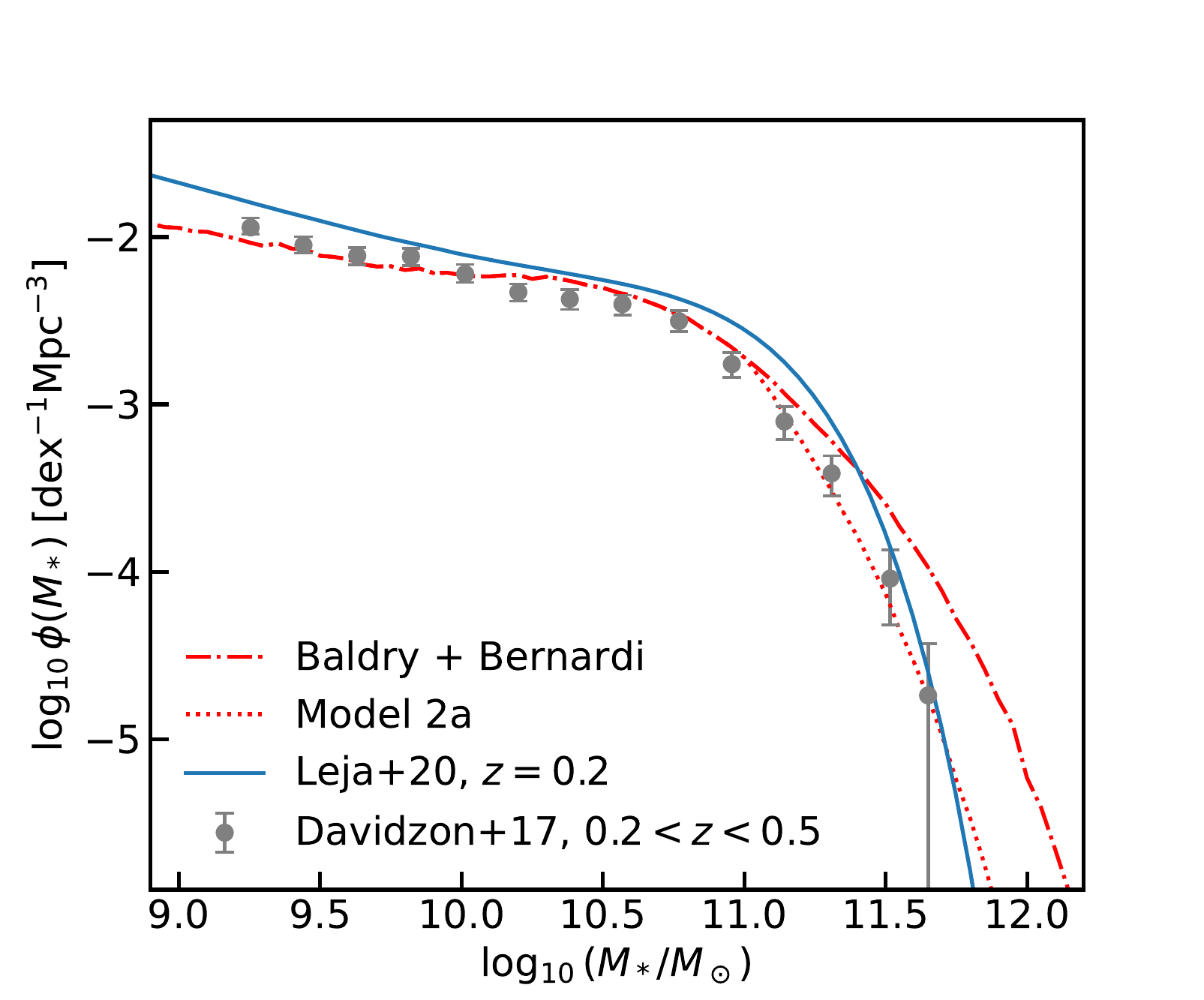}
            \caption{Stellar mass functions used for the abundance matching of the SMHM for the different models described in Section \ref{sec_smhm_models}. In both panels, the red dash-dotted line represents the combination of the SMF from \citet{baldry_2012} and \citet{bernardi_2017} at $z=0.1$. In the left panel, the dots with error bars represent the observational data from \citet{tomczak_2014} and the coloured solid lines the redshift-dependent normalization correction (Equation \ref{eq_tomczak_corr}) applied to the Baldry+Bernardi SMF to be consistent with the Tomczak+14 data at $z>0$. In the right panel, the dots with error bars show the data from \citet{davidzon_2017}, while the blue solid line and red dotted line show the SMF from \citet{leja_2020} and the correction (as described in Section \ref{sec_smhm_models}) to the Baldry+Bernardi SMF for being consistent with the Davidzon+17 data, respectively.}
            \label{fg_SMFs}
        \end{figure*}
        
        \subsection{Stellar mass-halo mass models}\label{sec_smhm_models}
        
        As anticipated above, and further discussed in detail by many groups \citep[e.g.,][]{guo_2011, wang_jing_2010, moster_2010, moster_2013}, the SMHM relation is strongly dependent on the shape and evolution of the measured SMF. Unfortunately, the latter is far from known with sufficient accuracy, not even in the local Universe. \cite{bernardi_2013, bernardi_2016, bernardi_2017}, for example, showed that the SMFs in SDSS and CMASS at $z=0.1$ and $z=0.5$, respectively, are highly dependent on the light profile chosen to fit the photometry. When the same methodology is consistently applied to infer stellar masses, no apparent evolution is detected in the high mass end of the SMF up to at least $z=0.5-0.8$ \citep[see also][]{shankar_2014}, whilst significant evolution at all masses is inferred when comparing \cite{bernardi_2017} and, e.g., \cite{davidzon_2017}, who derived stellar masses from SED fitting. Additional systematic differences in SMFs have been claimed by, e.g., \cite{leja_2020}, who support larger stellar masses at fixed SFR. The substantial systematic uncertainties in measured SMFs naturally propagate into the shape, scatter and evolution of the SMHM relation. In what follows, we present four models for the SMHM relation derived from direct abundance matching, as detailed in Section \ref{sec_dream_gal_halo_connect}, with different observed SMFs. Our aim is to estimate to what extent observationally informed systematic differences in the input SMHM relation impact the implied galaxy merger rates and bulge fractions, a task that is particularly suited to address with \decode. The four SMHM models considered in this work can be summarised as follows:
        \begin{itemize}
            \item   Model 1:
                    \begin{itemize}
                        \item   \cite{bernardi_2017} SMF, assumed to be constant up to $z \approx 1.5$. This is of course an extreme assumption but worth exploring as it is still unclear whether apparent evolution in the SMF at $z>0$ may be, at least in part, driven by non-ideal/inconsistent estimates of galaxy stellar masses, as mentioned above. Indeed, \cite{kawinwanichakij_2020} recently suggested that there is no measurable evolution in the SMF up to at least $z=1.5$.
                        \item   \cite{tinker_2008} HMF.
                        \item   0.15 dex constant scatter in stellar mass at fixed halo mass.
                    \end{itemize}
            \item   Model 2:
                    \begin{itemize}
                        \item   \cite{tomczak_2014} SMF
                        at $z>0$ and \cite{bernardi_2017} at $z=0$. This model is also somewhat extreme because, as detailed above, the $z>0$ SMF estimates may be affected by systematic measurement errors and/or incomplete. Nevertheless Model 2 and Model 1 should bracket the range of possible evolutionary patterns of the SMF, at least based on the present data and on the assumption of a constant IMF.
                        \item   \cite{tinker_2008} HMF.
                        \item   0.15 dex constant scatter in stellar mass at fixed halo mass.
                    \end{itemize}
            \item   Model 3:
                \begin{itemize}
                        \item  equivalent to Model 1 but assuming a linearly increasing scatter with redshift up to $z = 2$ as follows
                    \begin{equation}
                    \left \{ \begin{array}{l}
                    \sigma_{\log M_*} = 0.15 + 0.1 z \;\;\; {\rm for} \; z<2\\
                    \sigma_{\log M_*} = 0.25 \;\;\;\;\;\;\;\;\;\;\;\;\; {\rm for} \; z \geq 2
                    \end{array}
                    \right.
                    \end{equation}\label{eq_scatter_z}
                \end{itemize}
            \item   Model 4:
                \begin{itemize}
                    \item  equivalent to Model 2 but with the same $z$-dependent scatter as Model 3.
                \end{itemize}
        \end{itemize}
        
        All the reference Models listed above start from the same $z=0$ SMF. The latter is built joining the \cite{bernardi_2017} SMF, which is valid down to $M_* \sim 10^9 M_\odot$, and the \cite{baldry_2012} SMF which extends down to $M_* \sim 10^6 M_\odot$ \citep[see, e.g.,][]{shankar_2006, kravtsov_2018}. In Models 1 and 3, as detailed above, we strictly assume no evolution in the SMF up to $z=1.5$. Beyond this redshift it becomes unrealistic to assume no further evolution in the SMF and thus we extend the SMF at higher redshift with a toy model that smoothly decreases the normalization of the SMF from $z=1.5$ using the following log-linear correction, which simultaneously allows to track the \citet{tomczak_2014} SMF data at $z>1.5$ and to be consistent with the \citet{bernardi_2017} SMF up to $z=1.5$
        \begin{equation}\label{eq_bernardi_corr}
            \log_{10}\phi(M_*(z)) \simeq (0.99 + 0.13 (z - 1.5)) \cdot \log_{10} \phi(M_*(z=0.1)) \; .
        \end{equation}
        In Models 2 and 4 we instead assume the SMF to continuously evolve in normalization at $z>0$ in a way to be broadly consistent with the \cite{tomczak_2014} SMFs at $0<z<3$, as shown in the left panel of Figure \ref{fg_SMFs}. When applying the \cite{tomczak_2014} evolution to the SMF, we make use of the following correction to the \cite{bernardi_2017} + \cite{baldry_2012} SMF at $z=0$
        \begin{equation}\label{eq_tomczak_corr}
            \log_{10}\phi(M_*(z)) \simeq (0.99 + 0.13 z) \cdot \log_{10} \phi(M_*(z=0.1)) \; .
        \end{equation}
        Furthermore, we also explore a variant of Model 2, which we refer to as Model 2a, with a SMF matching the SMF calibrated by \citet{davidzon_2017} characterised by a less abundant number density of massive galaxies (dotted line in Figure \ref{fg_SMFs}). We will discuss in the following Sections how the latter SMF will modify the implied SMHM relation and in turn the number of merging pairs and fraction of ellipticals. It is relevant here to clarify that our method relies on direct abundance matching between the SMF and HMF at any given epoch. However, while for the latter, analytic fits extracted from N-body simulations are available at all redshifts, this is not the case for the SMF, for which analytic fits are provided only in some predefined redshift bins. In addition, some of the high redshift data of interest to our work lack or have a poor determination of the high mass-end of the SMF \citep[e.g.,][]{tomczak_2014}. These are the main reasons why we need to define a full shape for the input SMF at all redshifts. We reiterate here that the aim of our work is to explore how different shapes and evolutionary trends in the input SMF, within the range allowed by current observations, impact the implied SMHM relation and related quantities such as the galaxy merger rates.
        
        \begin{figure}
            \includegraphics[width=\columnwidth]{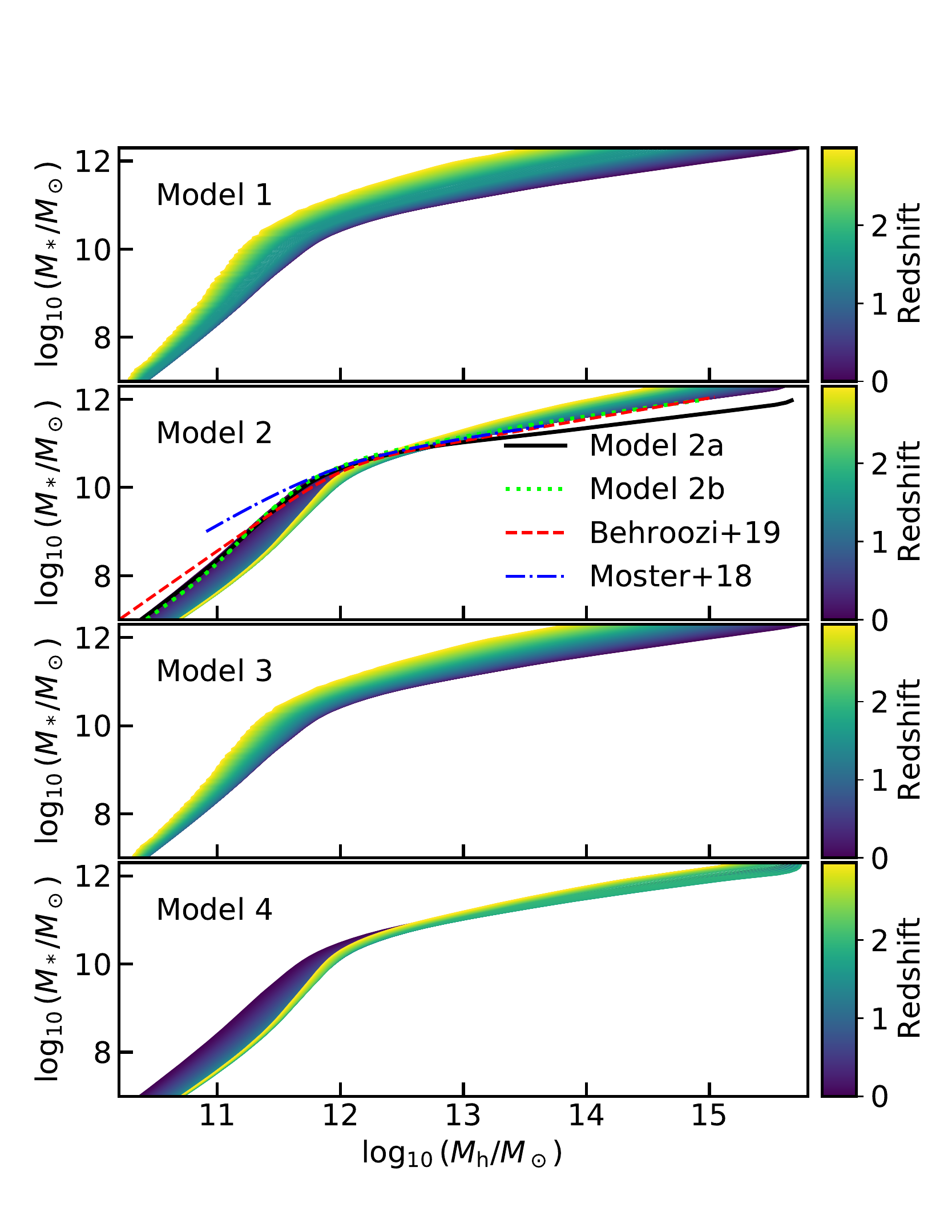}
            \caption{Stellar mass-halo mass relations computed via abundance matching. The four panels show the relations for the four models described in Section \ref{sec_smhm_models}, respectively, for the range of redshift denoted by the color code. In the second panel we show also Models 2a and 2b, along with two SMHM relations from other works in the literature \citep[][]{moster_2018, behroozi_2019} for comparison (black solid, green dotted, red dashed and blue dash-dotted lines, respectively).}
            \label{fg_SMHM_AM_scatter_comparison}
        \end{figure}
        
        Given the SMFs at any given epoch $z<4$ and for each Model, it is now necessary to specify the host halo mass functions to then apply the abundance matching routine given in Equation \ref{eq_aversa_AM}. We choose the \cite{tinker_2008} HMF, which provides the abundances of host parent haloes (we note that switching to other forms of the HMF would yield very similar results throughout). As the SMFs at $z>0$ contain both central and satellite galaxies (though the latter become progressively negligible in number densities at earlier epochs), we need to correct the \cite{tinker_2008} HMF by the abundances of surviving satellites at any epoch of interest, as specified in Section \ref{sec_dream_gal_halo_connect}.
        
        The SMHM relations, derived from our abundance matching algorithm for each Model, are shown in Figure \ref{fg_SMHM_AM_scatter_comparison} in the redshift range $0<z<3$. Models 1 and 3 are characterized by a high-mass slope $\mathrm d \log_{10} (M_* / M_\odot) / \mathrm d \log_{10} (M_{\rm h} / M_\odot)$ (between $M_{\rm h} = 10^{13} M_\odot$ and $M_{\rm h} = 10^{14} M_\odot$) of $0.550$, and a low-mass one (between $M_{\rm h} = 10^{11} M_\odot$ and $M_{\rm h} = 10^{11.5} M_\odot$) of $1.30$, both with a normalization of $10.509$ at $M_{\rm h} = 10^{12} M_\odot$, at $z = 1$. Models 2 and 4 instead have a high-mass slope of $0.588$ and $0.508$, a low mass slope of $1.15$, and a normalization of $10.138$ and $10.163$, respectively. As shown in Figure \ref{fg_SMHM_AM_scatter_comparison}, our Model 2 at $z=0$ is very close to both \citet{moster_2018} and \citet{behroozi_2019}, at least at high stellar masses. The main difference between Model 2a and the others is the significantly flatter slope of $0.414$ at the high-mass end. We will show below that such apparently small differences in the shape of the SMHM relations, especially the differences in the slopes at the high-mass end of the SMHM relations, are sufficiently large to generate significant systematic differences in, e.g., the major merger rates and implied elliptical fractions of up to a factor of 2-4 in some stellar mass bins. We notice that by keeping the SMF constant in time (Models 1 and 3) induces a weak evolution at low masses and a more pronounced one at larger masses. On the other hand, the models characterized by an evolving SMF (Models 2 and 4), generate a SMHM relation with evident redshift evolution at low stellar masses and a weak one at higher stellar masses (see also \citealt{shankar_2006, moster_2018}; \citetalias{grylls_paper2}).
        
        Finally, we also investigated the possibility of a mass-dependent scatter in stellar mass at fixed halo mass as input in our abundance matching. To this purpose, we assume a halo mass-dependent scatter, similarly to what suggested by other SEMs \citep[e.g.,][]{moster_2018, behroozi_2019}, constant at $M_{\rm h} > 10^{12} M_\odot$ and increasing linearly with $\log_{10} (M_{\rm h}/M_\odot)$ below that mass
        \begin{equation}\label{eq_scatter_Mhalo}
        \left \{ \begin{array}{l}
        \sigma_{\log M_*} = 2.95 - 0.23 \log_{10}(M_{\rm h}/M_\odot) \;\; {\rm for} \; M_{\rm h} < 10^{12} M_\odot\\
        \sigma_{\log M_*} = 0.15 \;\;\;\;\;\;\;\;\;\;\;\;\; \;\;\;\;\;\;\;\;\;\;\;\; \;\;\;\;\,\;\;\;\;\;\; {\rm for} \; M_{\rm h} \geq 10^{12} M_\odot \, .
        \end{array}
        \right.
        \end{equation}
        We show the SMHM relation at $z=0$ implied by the scatter in Equation (\ref{eq_scatter_Mhalo}) in the second panel of Figure \ref{fg_SMHM_AM_scatter_comparison}, labelled as Model 2b. The comparison shows that Model 2b is fully equivalent to Model 2 at high masses and slightly steeper at low masses, and these differences we checked are similar at all redshifts. We have tested that, since Model 2b is equivalent to Model 2 in the mass range where major mergers are significant, i.e., $M_\star \gtrsim 10^{11}\, M_\odot$, the predicted amount of major mergers remains unchanged and, hence, also all the other quantities analyzed in this work. Therefore, we do not show the results for Model 2b any further, and concentrate on the two models with redshift-dependent scatter (Models 3 and 4), which alters also the high-mass end of the SMHM relation and consequently the major merger rates.
        
        \begin{figure}
    	    \includegraphics[width=\columnwidth]{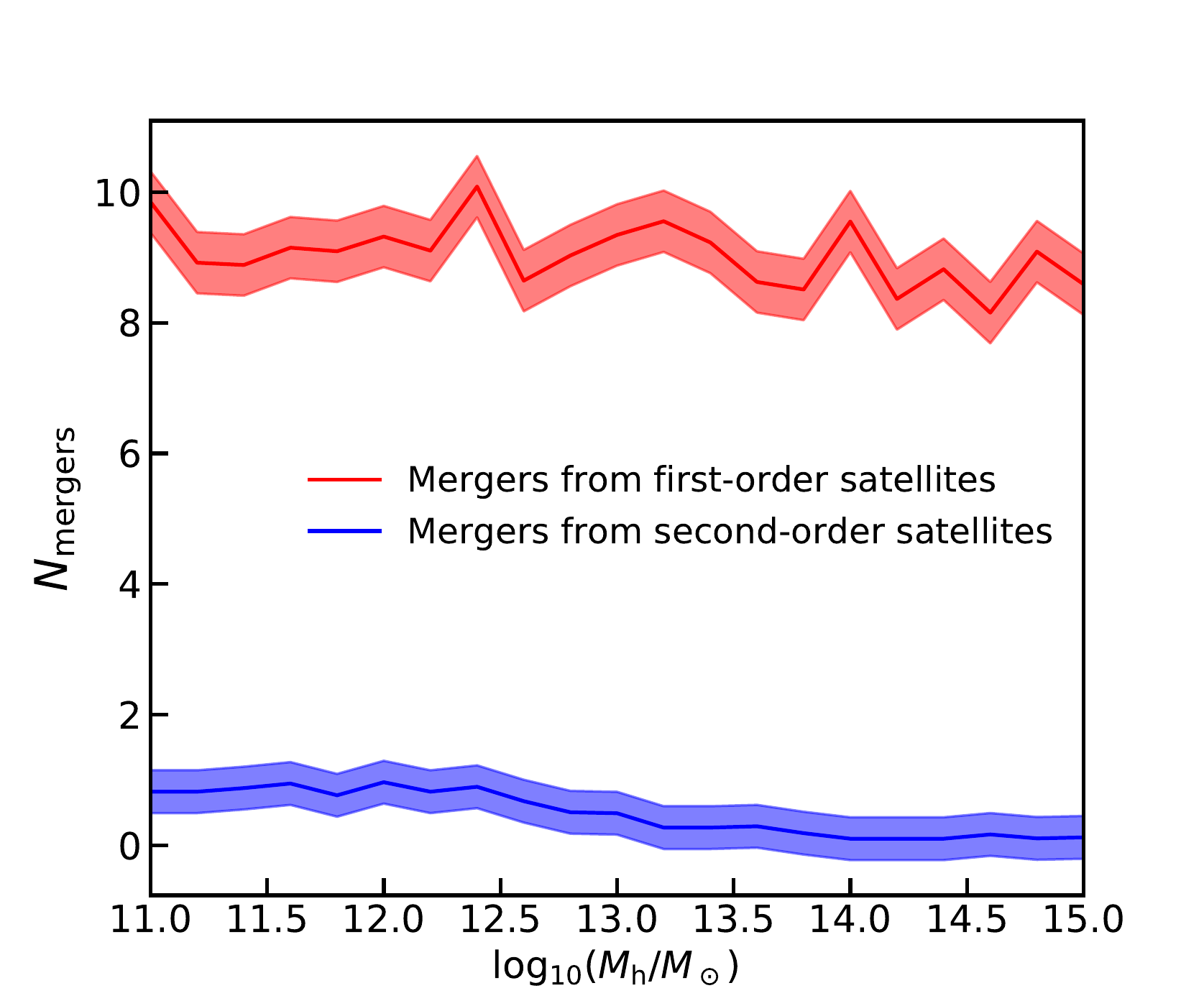}
            \includegraphics[width=\columnwidth]{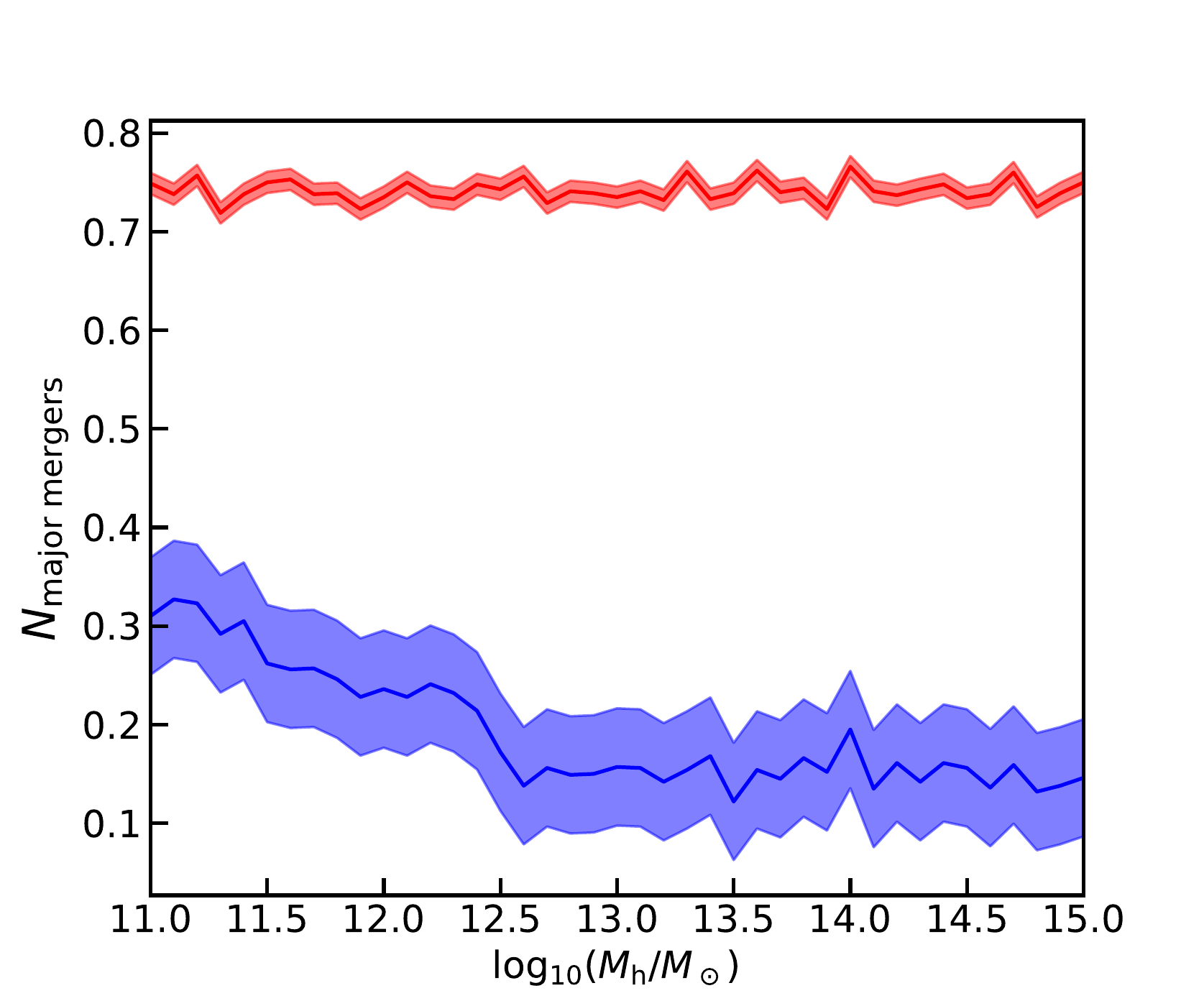}
    	    \caption{Upper panel: number of dark matter mergers from the contribution of first- and second-order subhaloes as function of the final parent halo mass at redshift $z=0$. The solid lines represent the mean value, while the shaded areas show the $1\sigma$ uncertainty. Lower panel: same as upper panel, but for major mergers, for which we assume a mass ratio $M_{\rm h, sub} / M_{\rm h, par} > 0.25$.}
    	    \label{fg_num_mergers}
        \end{figure}
        
        \begin{figure*}
	        \includegraphics[width=0.95\textwidth]{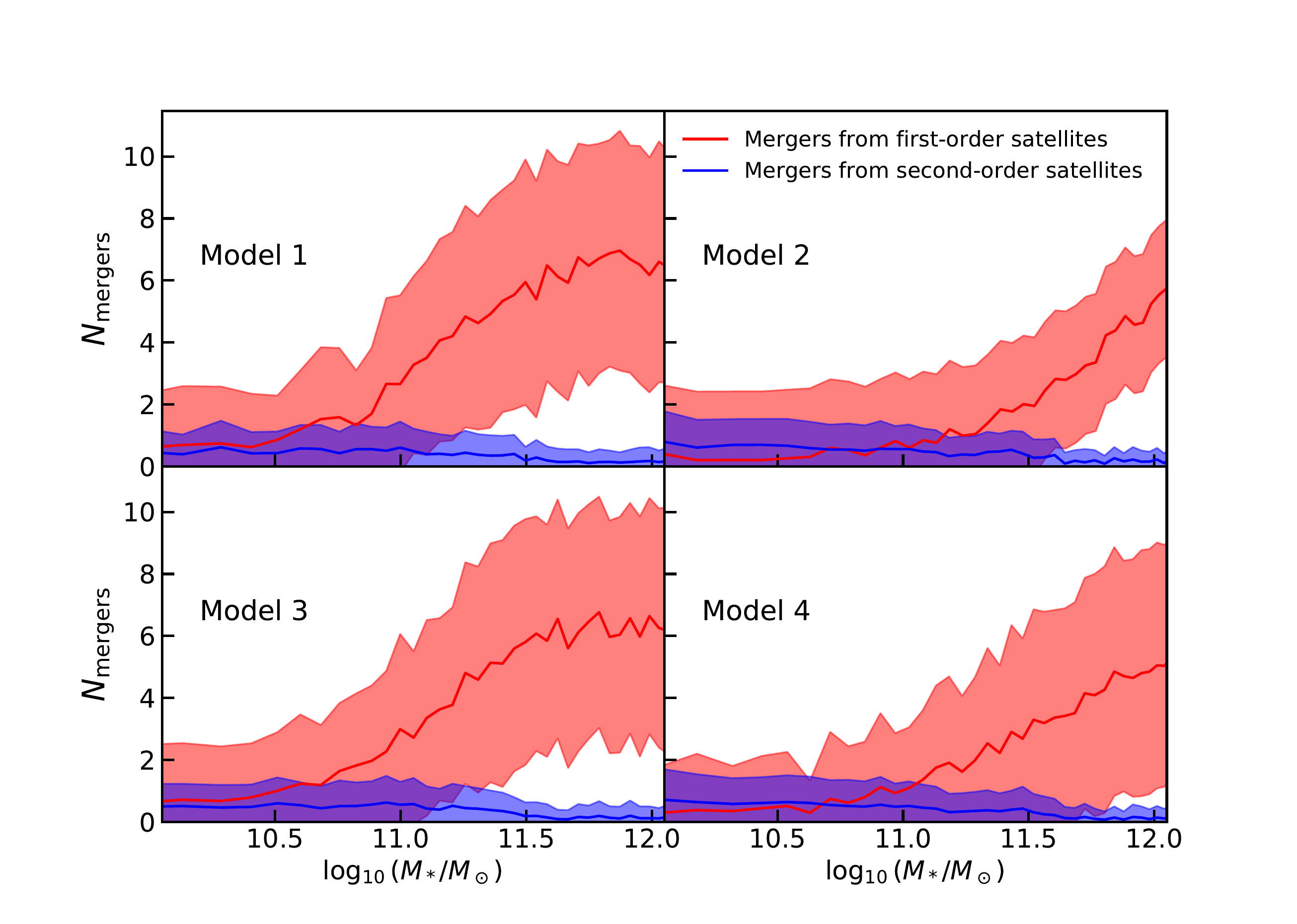}
	        \caption{Number of galaxy major mergers from first- and second-order satellites, with mass ratio $M_{\rm *, sat} / M_{\rm *, cen} > 0.25$, as a function of the central galaxy mass. The solid lines represent the mean value, while the shaded areas show the $1\sigma$ uncertainty. Results are shown for the four different models described in Section \ref{sec_smhm_models}, as labelled.}
	        \label{fg_num_mergers_Mstar}
        \end{figure*}
        
        \subsection{Merger rates}\label{sec_res_mergers_rate}
        
        The input SMHM relation has a direct impact on the merger rate of each galaxy that \decode produces \citep[e.g.,][]{stewart_2009a, hopkins_2010a, grylls_paper3, oleary_2021}, which in turn influences the implied satellite abundances, fraction of ellipticals, and B/T ratios. In this Section, we focus on galaxy merger rates and other predictions will be discussed in the following Sections.
        
        First of all, we checked that our halo-halo merger rates are consistent with the halo-halo merger rates derived by \citet{fakhouri_2010} from the Millennium simulation at $z\gtrsim 0.35$. We note that our fits drop slightly faster than those of \citet{fakhouri_2010} at $z < 0.35$, as also previously noted by \citetalias{grylls_paper2}, which we checked is mostly induced by our adopted halo mass mean accretion histories from \citet{vdb_2014} which are somewhat steeper than those presented in \citet{fakhouri_2010} at these redshifts. Figure \ref{fg_num_mergers} shows the cumulative number of total and major mergers from \decode (upper and lower panels, respectively) below $z<4$ predicted from \decode, expected in a hierarchical $\Lambda$CDM Universe, as a function of parent halo mass. We found that the number of mergers, with both first- and second-order subhaloes, is roughly constant with parent halo mass \citep[see also][]{shankar_2014}. However, such an invariance in halo mass is broken when mapping DM halo mergers to galaxy mergers via the double power-law shaped SMHM relation. We show this result in Figure \ref{fg_num_mergers_Mstar}, where we plot the average number of major mergers, along with its $1\sigma$ uncertainty, as a function of the (final) galaxy stellar mass. The red lines represent the contribution of mergers from first-order satellites and the blue lines the contribution of mergers from second-order. The plot clearly shows that, for all the four SMHM models, the contribution from the second-order satellites is in good approximation relatively negligible, at least for massive central galaxies. On the other hand, the number of mergers from the first-order satellites increases significantly when the central galaxy mass increases.
        
        The sharp increase of major mergers with galaxy stellar mass and the fact that major mergers for massive galaxies outnumbers the major mergers of the parent haloes can be both explained from the shape of the SMHM relation. Indeed, while the total number of mergers is the same for the DM halo and the host galaxy (and it is independent of the halo or stellar mass), the classification as major or minor merger depends on the mass ratio between halo or stellar mass ($M_{h,1}/M_{h,2}>0.25$ is the major merger threshold for halo mergers and $M_{\star,1}/M_{\star,2}>0.25$ for galaxy mergers). Since the SMHM relation sets $M_h\propto M_\star^\alpha$, the halo mass ratio is related to the stellar mass ratio as $M_{h,1}/M_{h,2}=(M_{\star,1}/M_{\star,2})^\alpha$. If the SMHM relation was linear ($\alpha$=1) the halo mass ratio would be equal to the stellar mass ratio and, consequently, a halo major merger would correspond to a galaxy major merger. On the other hand, a steep power law relation with $\alpha>1$ would decrease the stellar mass ratio with respect to the halo mass ratio, while a flat power law with $\alpha<1$ would increase it. As a consequence, the number of galaxy major mergers is reduced for a steep power law, while it is enhanced for a flat one. In other words it is less (more) likely to find similar mass galaxies for similar mass haloes for a steep (flat) SMHM. Since the SMHM relation in all the four considered models is a broken power law with a steep faint end and a flat bright end, the number of galaxy major mergers tends to naturally increase toward higher stellar masses. However the level of increase is different for the four models since even small variations in the SMHM relation produce strong effects in the merger rates. For example, Model 2 with a high-mass slope of $0.558$ predicts less than half of the mergers at $M_* \sim 1-3 \cdot 10^{11} M_\odot$ compared to Model 1 and Model 3 with a high-mass slope of $0.550$. On the other hand, Model 4, despite having the lowest high-mass end slope at $z=0$, produces slightly more mergers than Model 2 but less than Model 1 and Model 3, a trend that can be explained by the lower slope in the SMHM relation at higher redshifts in Model 4 than in Model 2, being for example $0.498$ and $0.551$ the average slope up to $z=1.5$, respectively. For more massive galaxies, Models 1 and 3 predict on average a higher number of major mergers compared to Models 2 and 4. This effect is also directly reflected in the fraction of ellipticals and B/T ratios, as we will see in the following Sections. We also note that, although the normalization of the SMHM does not directly impact the number of major mergers, it has an important role in determining the total merger history, which in turns will impact the star formation history, as we will discuss in future work. Furthermore, the amount of scatter in the SMHM relation instead seems to play a relatively less significant role in controlling the fraction of major mergers when assuming a constant SMF.
        
        \begin{figure}
    	    \includegraphics[width=\columnwidth]{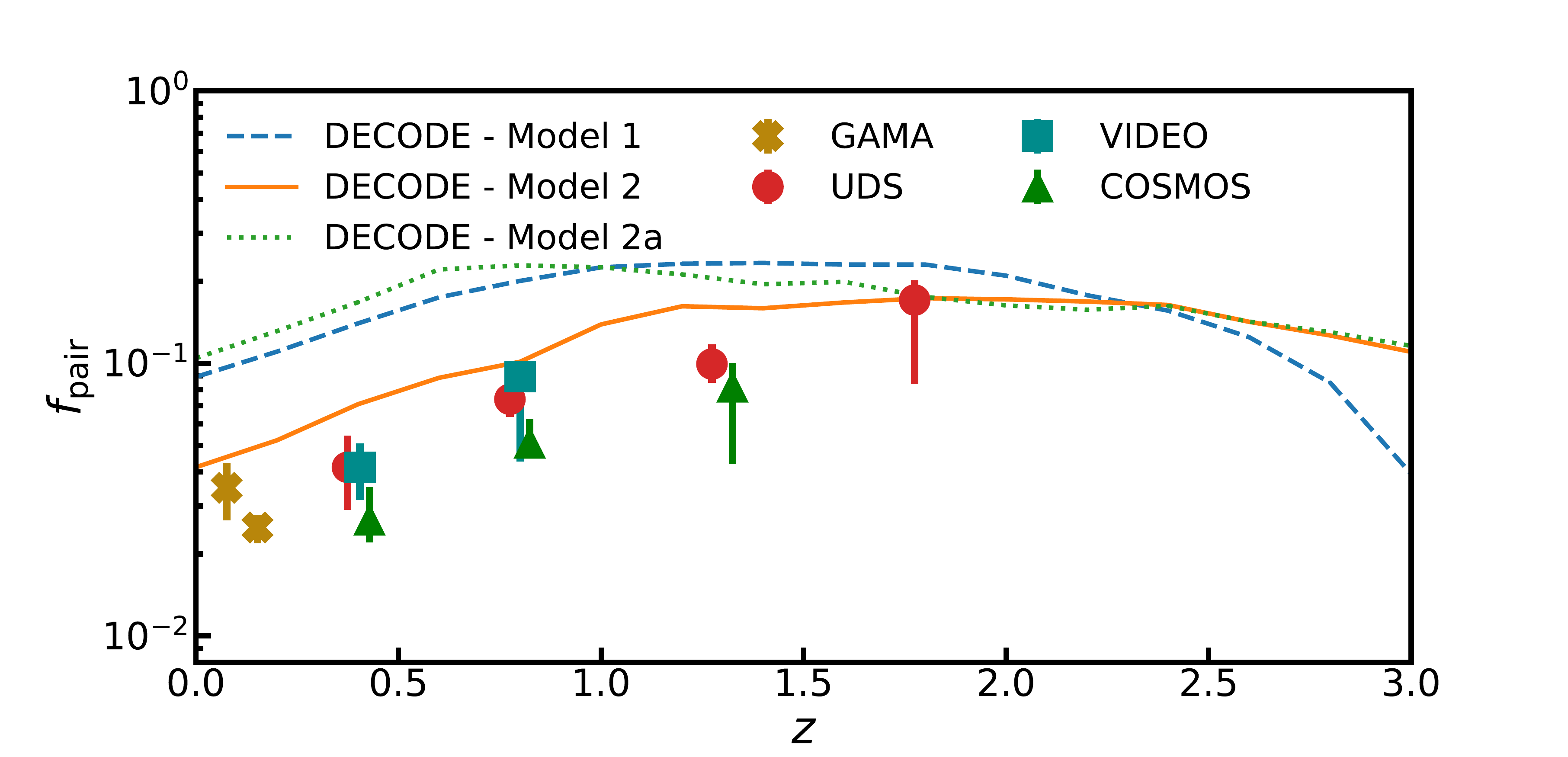}
    	    \includegraphics[width=\columnwidth]{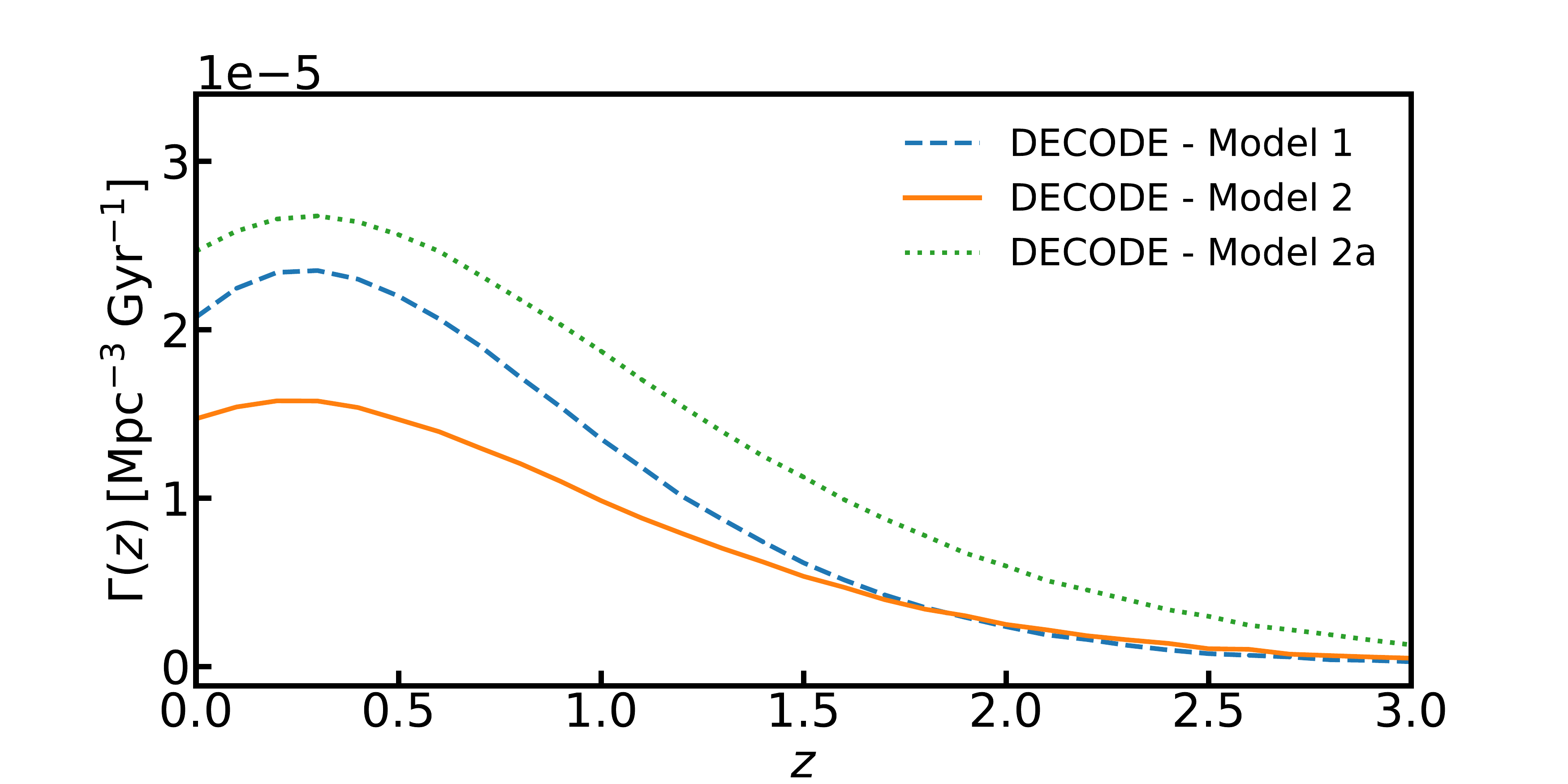}
    	    \caption{Upper panel: major merger pair fraction, with mass ratio $\mu \gtrsim 0.25$, for galaxies at $M_* \gtrsim 10^{11} M_\odot$ as predicted by \decode for Models 1, 2 and 2a (blue dashed, orange solid and green dotted lines, respectively). The points with error bars show the observational data of UDS, VIDEO, COSMOS and GAMA surveys, as presented in \citet{mundy_2017}. Lower panel: major merger rates as a function of redshift, as predicted by \decode's Models 1, 2 and 2a.}
    	    \label{fg_major_mergers_f_pair}
        \end{figure}
        
        Furthermore, we show in the upper panel of Figure \ref{fg_major_mergers_f_pair} the major merger pair fraction as predicted by \decode for Models 1, 2 and 2a. We compare our results with the data from UKIDSS UDS, VIDEO / CFHT-LS, UltraVISTA / COSMOS and GAMA surveys, presented in \citet{mundy_2017}. The pair fraction in \decode is calculated as the number of infalling satellites with stellar mass ratio above $1/4$ living within 5 and 30 kpc from the center of the central galaxy. We assume that the distance of the satellite galaxies scales proportionally to its dynamical friction timescale \citep[][]{guo_2011}. We make use of the projected two-dimensional distances computed following the recipe in \citet{mundy_2017} and \citet{simons_2019}, assigning stochastically a polar angle in spherical coordinates and projecting the three-dimensional distances onto the $z$ axis. Analogously to what inferred for the number of major mergers, Model 1 tends to produce more pair fractions with $\mu>1/4$ than Model 2. The available data on pairs is still sparse, and still subject to systematics in the determination of the stellar masses, but overall Model 2 tends to be more aligned with the data, at least at $z\geq0.5$. On the other hand, Models 1 and 2a, characterized by a flatter slope in the SMHM relation at the high-mass end, predict a higher pair fraction with respect to the data and to Model 2. This prediction is in line with the fact that a flatter SMHM relation produces a higher number of major mergers, as discussed above.
        
        Finally, in the lower panel of Figure \ref{fg_major_mergers_f_pair} we show a prediction of the major merger rates from \decode's Models 1, 2 and 2a. As we can see, the flatter the SMHM relation (Models 1 and 2a) the higher the rate of implied major mergers, in line with what shown in the upper panel. On the other hand, Model 2, characterized by a steeper SMHM high-mass end, predicts a much lower major merger rate, at least at lower redshifts. We do not show the comparison with observational data or any other model because the merging timescales that we use \citep[][]{mccavana_2012} are different to those adopted in other theoretical and observational works.
        
        In the next Sections we will show and discuss how the different shapes and evolution of the input SMHM relation play a crucial role in determining the satellite abundances, fraction of ellipticals, and B/T ratios.
        
        \begin{figure}
            \includegraphics[width=\columnwidth]{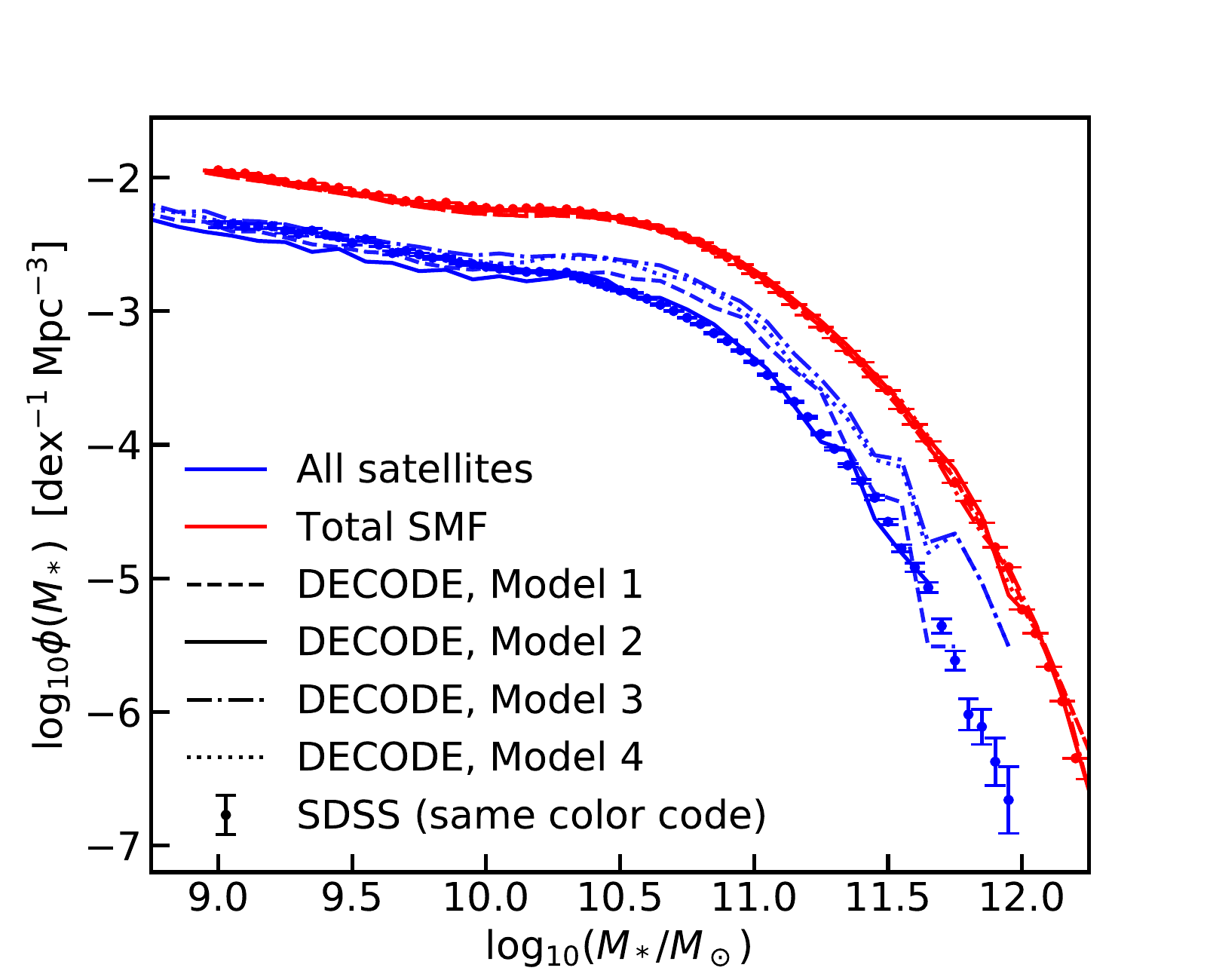}
            \caption{Total and satellite galaxies stellar mass function predicted by \decode compared to the data from SDSS. The solid, dashed, dash-dotted and dotted lines show the prediction for Model 1, 2, 3 and 4, described in Section \ref{sec_smhm_models}, respectively. The dots and error bars represent the data from SDSS.}
            \label{fg_SMF_sat}
        \end{figure}
        
        \subsection{Predicting the abundances of satellite galaxies}\label{sec_res_sate}
        
        Having defined the mapping between galaxy stellar mass and host DM halo mass, we can start to predict galaxy observables that can be used to validate \decode and the input SMHM relation. As a very first test, we compute the number density of surviving satellite galaxies in the local Universe and compare it with SDSS data. Satellites can be effectively considered as the other side of the same coin with respect to mergers. In fact, surviving satellites in a hierarchical DM-dominated Universe, can be interpreted as "failed mergers", i.e., all those infalling satellite galaxies that have not yet had the time to merge with their central galaxies at the time of observation. Therefore satellites, just like mergers, represent a pivotal test of hierarchical models and of the input SMHM relation. Although the total SMF is an input in \decode, the satellite SMF is an actual prediction of the model, as it depends both on the satellite evolution after infall and on the rate of galaxy mergers, which in turn depend on the high-z SMHM relation and on the dynamical friction timescales.
        
        \begin{figure*}
	        \includegraphics[width=0.98\textwidth]{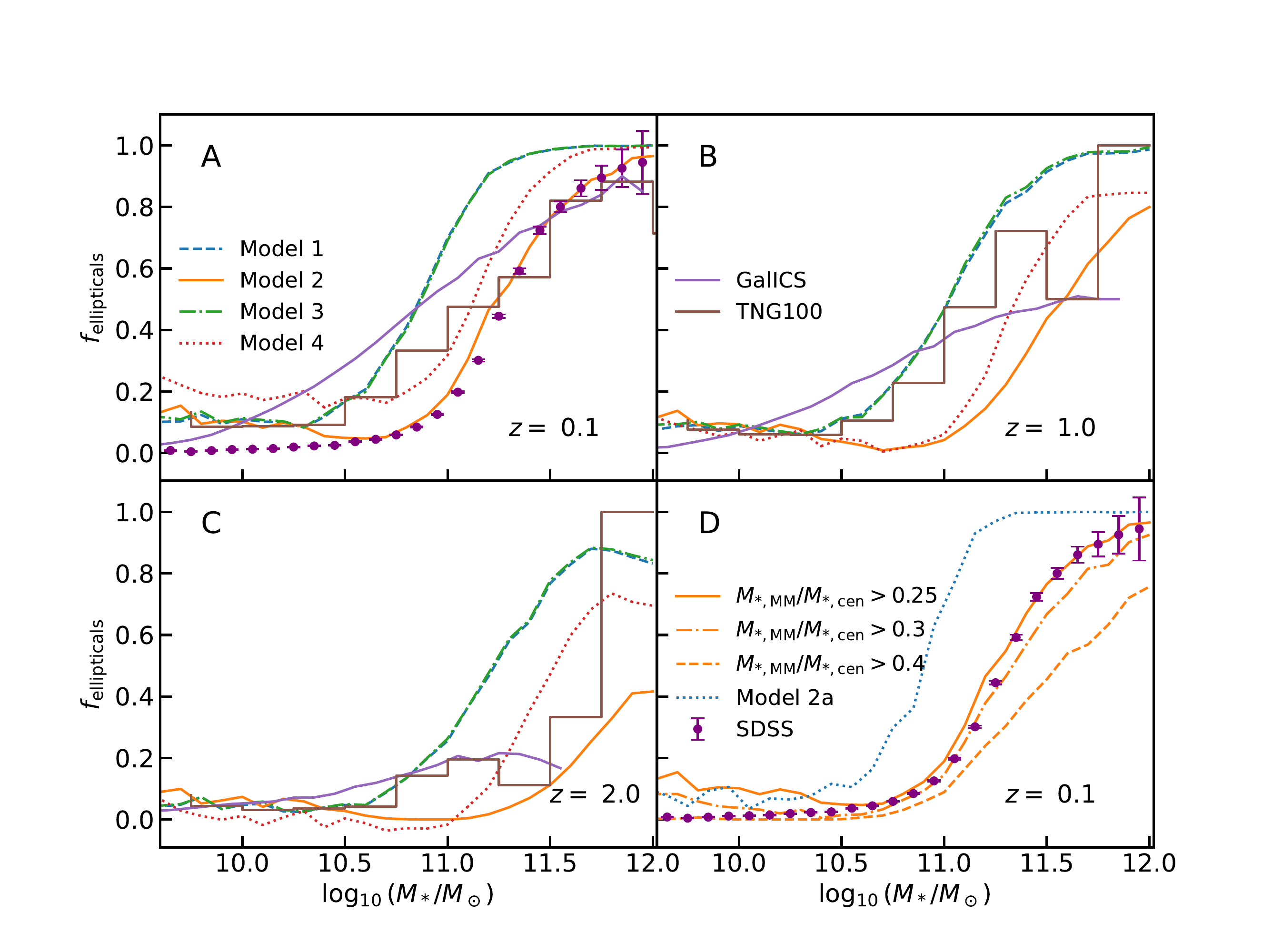}
	        \caption{Fraction of elliptical galaxies as a function of the galaxy stellar mass as predicted by \decode. Panels A, B and C show the predictions for the four models described in Section \ref{sec_smhm_models} at redshifts $z = 0.1$, $1$ and $2$, respectively. Panel D shows the predictions for Model 2 at redshift $z = 0.1$, along with the impact of assuming different mass ratio for the major mergers (denoted as MM in the legend), as well as the prediction for Model 2a. The data from the SDSS Survey, \textsc{GalICS} semi-analytic model and the TNG100 hydrodynamical simulation are included, as labelled, for comparison.}
	        \label{fg_frac_ellipt_models_comp}
	    \end{figure*}
        
        Figure \ref{fg_SMF_sat} shows the results of the SMF for all galaxies (centrals + satellites) and only satellites (red and blue lines, respectively). The different types of line distinguish the four SMHM models, and the blue and red dots with error bars are, respectively, the satellite\footnote{Centrals and satellites classification is from \citet{Yang+07}.} and total galaxy SMF as measured in SDSS using, for consistency with our Models, the data from \citet{bernardi_2017}. As one can see from the red lines, the total SMF is well reproduced by \decode, as expected by construction via the abundance matching relation given in Equation (\ref{eq_aversa_AM}). For all SMHM models reported in Figure \ref{fg_SMHM_AM_scatter_comparison}, we assume a ``frozen'' scenario in which satellite galaxies are assumed to retain the same stellar mass after infall with no further growth via star formation or loss via, e.g., stellar stripping. \citetalias{grylls_paper1} showed that the frozen model was able to reproduce the bulk of the observed satellite population in their SEM \textsc{steel}. We do find a similar result with \decode in Figure \ref{fg_SMF_sat}, despite using a significantly more accurate distributions of satellites of first- and second-order and dynamical friction timescales. We will explore in separate work the impact of star formation and stellar stripping on the satellite population and star formation histories. We anticipate here that including standard recipes for stellar stripping as given by, e.g., \citet{cattaneo_2011}, and for star formation after infall following the analytic recipes by \citetalias{grylls_paper2} (and references therein), we obtain very similar results to those shown in Figure \ref{fg_SMF_sat}. Figure \ref{fg_SMF_sat} shows that only Model 2, the one characterized by an evolving SMF and constant scatter, better lines up with the observational data of satellites, especially at stellar masses $M_* \gtrsim 3 \cdot 10^{10} M_\odot$. We will see below that Model 2 also performs better against other observational constraints.
        
        \begin{figure}
            \includegraphics[width=\columnwidth]{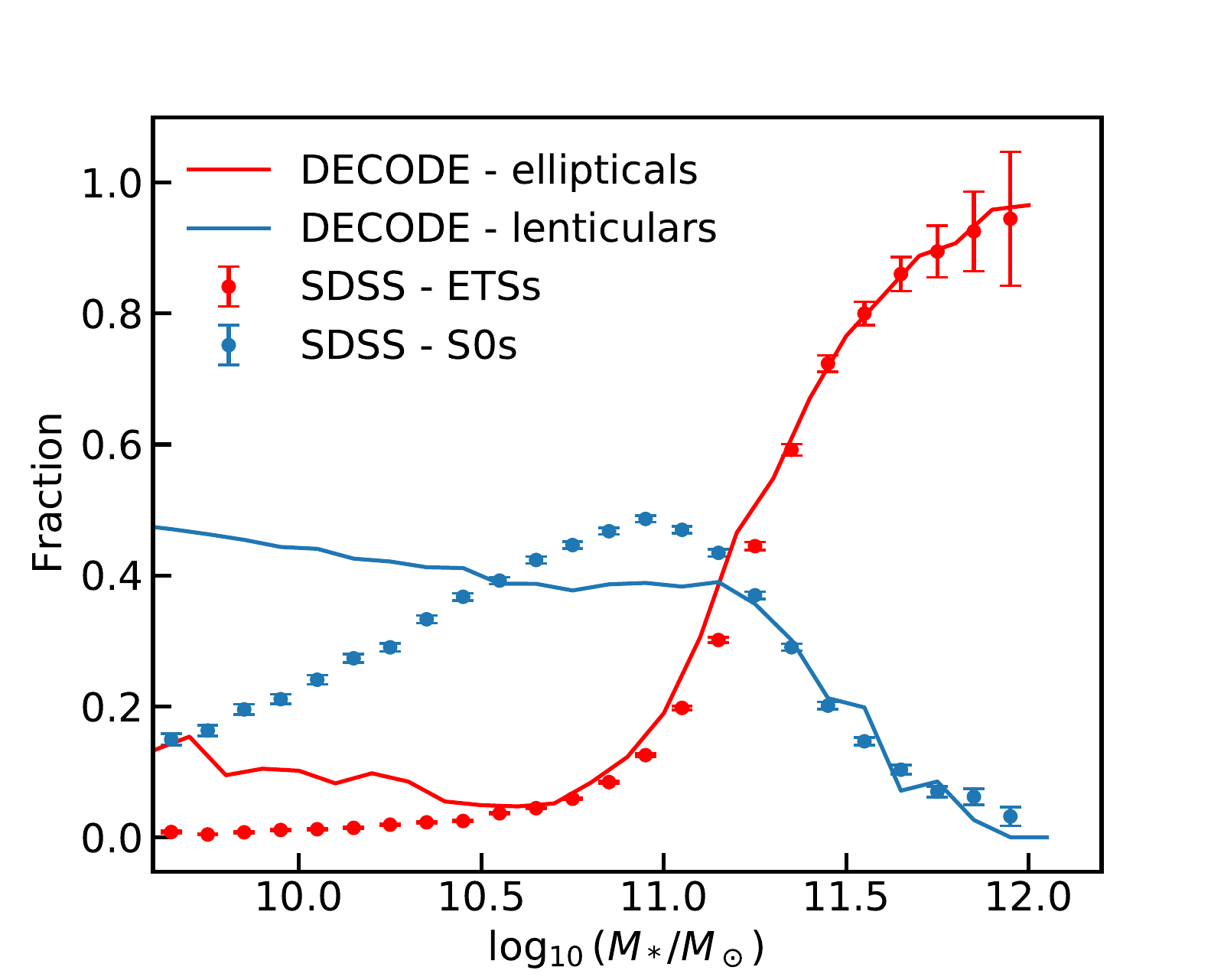}
            \caption{Fraction of lenticular and elliptical galaxies as predicted by \decode for Model 2 compared with the data from the SDSS Survey, as labelled.}
            \label{fg_frac_lenticulars}
        \end{figure}

	    \subsection{Morphology of central galaxies}\label{sec_res_morphology}   
	    
        In this Section we apply \decode to study another key aspect of galaxy evolution: the role of mergers in shaping galaxies, and in particular in generating elliptical type galaxies. Many works have in fact suggested a close link between the number of major mergers and the number of ellipticals \citep[e.g.,][]{bournaud_2007, hopkins_2009b, hopkins_2010, shankar_2013, fontanot_2015}. Major mergers (for which we assume $f = M_{\rm *, sat} / M_{\rm *,cen} \gtrsim 0.25$) may be capable of destroying pre-existing galactic disks and form stellar spheroids, as suggested by hydrodynamical simulations and analytic models \citep[see, e.g.,][]{baugh_2006, malbon_2007, bower_2010, tacchella_2019, lagos_2022}. However, as discussed by \citetalias{grylls_paper2} and above, as the number of major mergers is strongly dependent on the shape of the input SMHM relation, we expect the fraction of ellipticals to be similarly impacted by the choice of SMHM relation. We will investigate this possibility in this Section.

        For each SMHM Model, we follow the merger history of galaxies in the mock from redshift $z = 4$, and we label each galaxy that has undergone a major merger as an "elliptical", and then compute the fraction of ellipticals at different redshifts. We show the results in Figure \ref{fg_frac_ellipt_models_comp}. Panels A, B and C show the fraction of ellipticals predicted by \decode using the four Models of SMHM described in Section \ref{sec_smhm_models} for redshifts $z = 0.1$, $1$ and $2$ respectively, along with the predictions from the TNG simulation and from the semi-analytic model \textsc{GalICS}, as well as with the SDSS data (see Section \ref{sec_data_sdss}), only shown in the panels at $z=0.1$. We note that, as discussed in Section \ref{sec_data}, GalICS and TNG define elliptical galaxies in a somewhat different way than a simple cut in $\mu$, but we still include their predictions in Figure \ref{fg_frac_ellipt_models_comp} for completeness.
        
        It is immediately clear from Figure \ref{fg_frac_ellipt_models_comp} that Model 2 is the only one among our four chosen SEMs that can faithfully reproduce the SDSS data at $z=0.1$, on the assumption that ellipticals are strictly formed from major mergers above a mass ratio of $\mu > 0.25$, the standard limit adopted in state-of-the-art SAMs \citep[e.g.,][]{guo_2011, fontanot_2015, lacey_2016}. Model 4, which only differs from Model 2 on the assumed scatter around the SMHM relation, is moderately close, but still higher than the data for the same cut in merger ratio $\mu > 0.25$, further proving that even a modest variation in the scatter of the SMHM relation can significantly impact galactic outputs. In particular, an increase of the scatter at fixed slope at high stellar masses tends to increase the number of major mergers. On the other hand, Models 1 and 3 produce a significantly higher fraction of ellipticals. The main reason why models including evolution in the SMF predict less elliptical galaxies than models with constant SMF is a direct consequence of the number of mergers that they predict, as discussed in Section \ref{sec_res_mergers_rate}.
        Figure \ref{fg_frac_ellipt_models_comp} also shows that the TNG100 simulation provides a decent match to the SDSS at $z=0.1$, especially at galaxy masses $M_* > 2 \cdot 10^{11} \, M_\odot$, although we should caution that some internal self-inconsistencies naturally arise when comparing with the TNG100 outputs with these data as the TNG100 simulation does not exactly reproduce the same SMF of \citet{bernardi_2017} \citep[see, e.g.,][]{pillepich_2018_Mstar_content}, on which the elliptical fractions are based, and thus has a SMHM relation that slightly differs from the one in Model 2. The SAM \textsc{GalICS}\footnote{We remind the reader that elliptical galaxies in \textsc{GalICS} are identified as those galaxies with $B/T > 0.7$.} also well matches the SDSS elliptical fractions at high stellar masses, but it overproduces them at lower stellar masses.
        Lastly, we show in panel D of Figure \ref{fg_frac_ellipt_models_comp} the fraction of ellipticals predicted by Model 2a. This model, characterized by a flatter high-mass end slope, produces a much larger number of mergers, as expected from the discussion in Section \ref{sec_res_mergers_rate}, and thus it increases the implied fraction of ellipticals at all stellar masses.
        
        In conclusion, the question whether major mergers are the triggers for the formation of local ellipticals, is strongly dependent on the SMHM relation in input or generated by the model in use, and, at fixed SMHM relation, on the exact threshold chosen for being classified as a major merger, as shown in panel D.
        
        Finally, we show for completeness the fraction of lenticular type galaxies as a function of stellar mass in Figure \ref{fg_frac_lenticulars} predicted from \decode for Model 2, compared with the observational data from SDSS. Several works have suggested that lenticular galaxies might be created by mergers as well \citep[e.g.,][]{christlein_2004, laurikainen_2005, blanton_2009}. Here, we label the lenticulars as the galaxies that have had at least one merger with mass ratio $0.05 < \mu < 0.25$, whilst above $\mu=0.25$ they would end up as ellipticals. This simple merger recipe is capable of reproducing the data for stellar masses $M_* \gtrsim 10^{11} \, M_\odot$, but it fails at lower stellar masses, suggesting that additional processes may be at work in forming less massive lenticulars, such as disc instabilities and/or disc regrowth. %We note that models based on mergers can reproduce the fraction of lenticulars for stellar masses $M_* \gtrsim 10^{11} \, M_\odot$. At low stellar masses, \decode overpredicts the lenticular fraction.

        \subsection{Bulge-to-total ratios}\label{sec_res_bt_ratio}
        
        In the previous Sections we showed that the shape of the SMHM relation drives the number and thus the rate of mergers galaxies undergo through cosmic time (at fixed dynamical friction timescale). Only specific SMHM relations, such as the one defined in Model 2, which is characterized by a larger number density of massive galaxies and a significant evolution in normalization at $z > 0.5 - 1$, are able to simultaneously reproduce the number of local satellites and the number of local ellipticals, on the assumptions that the latter are formed out of mergers between galaxies with a mass ratio $M_{*,1} / M_{*,2} > 0.25$. We now move a step forward in our modelling and test how well our Models 1 and 2 reproduce the B/T ratios of local galaxies, as measured in MaNGA (see Section \ref{sec_data}).
        
        To perform a meaningful and instructive comparison between models and data, we make use of two simple but theoretically well-motivated toy models for the formation of bulges in hierarchical models:
        \begin{itemize}
            \item In Model BT1, we assume that when a major merger occurs, with $M_{*,1} / M_{*,2} > 0.25$, the descendant galaxy is strictly an elliptical with $B/T = 1$. This is a common assumption made in semi-analytic models of galaxy evolution \citep[e.g.,][]{cole_2000, hatton_2003, bower_2006, de_lucia_2007, guo_2011, croton_2016, lacey_2016, cattaneo_2017}. We then assume that in minor mergers the mass of the satellite can be accreted either onto the bulge or onto the disc component.
            \item In Model BT2, we instead assume that the remnant galaxy has a surviving disc with $B/T = 0.5$. In other words, we assume that the disc is not entirely disrupted in a major merger, irrespective of the gas fraction in the progenitor galaxies, but in fact a significant fraction of it survives and/or is rapidly reaccreted \citep[e.g.,][]{hopkins_2009a, hammer_2012}. We then explore the impact on the final $B/T$ of assuming the satellite mass in minor mergers to be added systematically to the disc or to the bulge component. We note that, in what follows, we consider the B/T at fixed galactic stellar mass averaged over \textit{all} central galaxies that enter that bin in stellar mass.
        \end{itemize}
        
        \begin{figure*}
            \includegraphics[width=\textwidth]{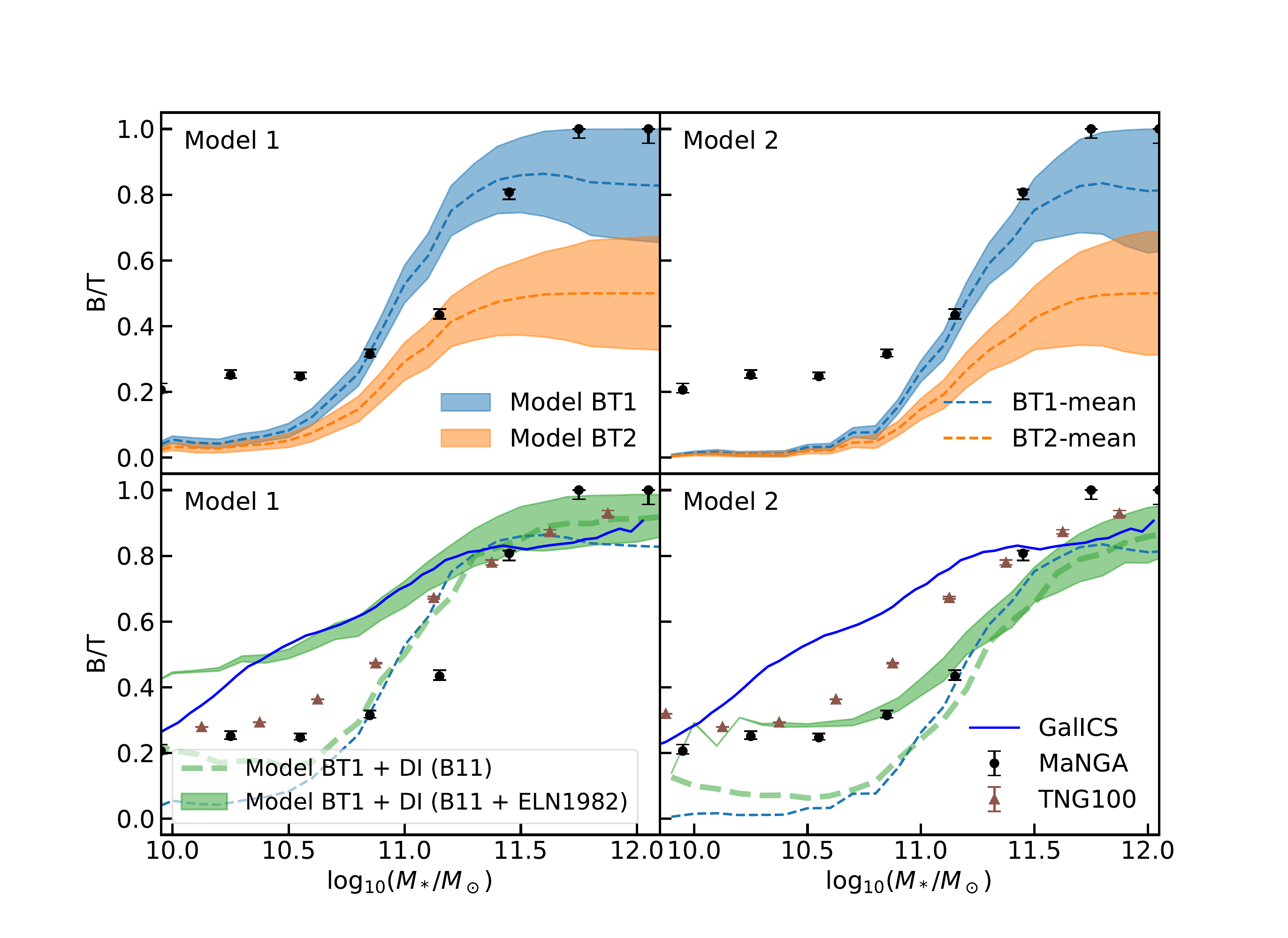}
    	    \caption{Mean bulge-to-total ratios for the two different toy models described in Section \ref{sec_res_bt_ratio} and SMHM relationships of Models 1 and 2. The blue and orange shaded areas in the upper panels show the results for Models BT1 and BT2 (for Models 1 and 2 respectively), while the green areas in the lower panels show the results for Model BT1 but by including disc instabilities of \citet{efstathiou_1982} (Equation \ref{eq_discInst_ELN}). The dashed tick green lines show the B/T ratios for Model BT1 by including only \citet{bournaud_2011} disc instabilities. The shaded areas are constrained by the two limit cases, where all the mass from minor mergers goes into the bulge and disc, respectively, and the thin dashed lines show the mean. The black error bars represent the $1\sigma$ bound from the MaNGA survey. Finally the blue solid line shows the prediction from \textsc{GalICS} semi-analytic model and the brown triangles with error bars the results from the TNG simulation. The uncertainties for the MaNGA survey and the TNG simulation are estimated using the standard deviation on the median.}
    	    \label{fg_BT_ratio_multi_models}
        \end{figure*}
        
        Another popular route to form stellar bulges in galaxies is via disc instabilities, which are usually implemented in SAMs in broadly two ways. A first type envisions that when the circular velocity of the disc becomes larger than a given reference circular velocity, then the disc is considered unstable and a mass is transferred from the disc to the bulge. The amount of mass transferred from the disc to the bulge varies significantly from model to model \citep[e.g.,][]{cole_2000, bower_2006, monaco_2007, guo_2011, lacey_2016, henriques_2020, izquierdo_2019}. The disc instabilities of the second type are triggered by high redshift cold flows of gas, which favour the formation of (possibly) long-lived gas clumps that migrate towards the centre via dynamical friction in the gaseous disc \citep[e.g.,][]{dekel_2006, dekel_2009, bournaud_2011, dimatteo_2012, oklopcic_2017, dekel_2019}. To include an example of the second type of disc instabilities in \decode to generate stellar bulges, we adopt the parameterization of the baryonic inflow rate from \cite{bournaud_2011}
        \begin{equation}\label{eq_bournaud}
            \dot M_{\rm b} = 25 \frac{M_{\rm disc}}{10^{11} M_\odot} \bigg( \frac{1+z}{3} \bigg)^{3/2} \; M_\odot / {\rm yr} \; ,
        \end{equation}
        with $M_{\rm disc}$ the mass of the disc at redshift $z$. Equation (\ref{eq_bournaud}) assumes that most of the mass inflow rate, which is in gaseous form, will form clumps that via dynamical friction will end up forming a stellar bulge at the galaxy centre. We apply this recipe only to galaxies which have a gas fraction $f_{\rm gas} \geq 0.5$ which are more likely to have undergone disc instabilities, since large amount of gas and mass densities inevitably lead to disc fragmentation \citep[see][]{lang_2014}. Gas masses are not present in \decode. To this purpose, following other SEMs \citep[e.g.,][]{hopkins_2009a, shankar_2014}, we assign gas fractions to any galaxy in the mock via the empirical mean relations derived by \citet{stewart_2009} from a number of galaxy samples out to $z \sim 3$. We find that Equation (\ref{eq_bournaud}) does not generate enough large stellar bulges at $z = 0$, as expected by simple direct integration of Equation (\ref{eq_bournaud})\footnote{We note that the growth of stellar bulges via clumpy accretion can be further increased by lowering the $f_{\rm gas}$ threshold, though it tends to still be moderate, reaching a value of up to $B/T \sim 0.2$ at low stellar masses, $M_\star \lesssim  3 \cdot 10^5 M_\odot$.}. Therefore, some slightly stronger disc instabilities need to be implemented to better match local data. To this purpose, we follow the recipe from \citet{efstathiou_1982} building on the analytic modelling of many previous works \citep[e.g.,][]{cole_2000, monaco_2007}
        \begin{equation}\label{eq_discInst_ELN}
            \epsilon \sqrt{G M_{\rm disc} / R_{\rm disc}} > V_{\rm ref} \; ,
        \end{equation}
        where $M_{\rm disc}$ is the mass of the disc, $R_{\rm disc}$ is the half mass radius calculated via the redshift-dependent analytic fit by \citet{shen_2003}, $V_{\rm ref}$ is the reference velocity calculated assuming an exponential profile \citep[see, e.g.,][]{tonini_2006}, and $\epsilon$ is a factor of order unity \citep[see, e.g.,][and references therein]{shankar_2014}. When the condition in Equation (\ref{eq_discInst_ELN}) is verified in galaxies that still have a disc-dominated structure, we assume that the disc transfers a sufficient stellar mass to the bulge to reestablish dynamical equilibrium \citep[e.g.,][]{hatton_2003, shen_2003}.
        
        The top panels of Figure \ref{fg_BT_ratio_multi_models} report our results for the merger Models BT1 and BT2, for the SMHM relation in Model 1 (left panel) and Model 2 (right panel), shown with cyan and orange shaded areas, respectively, where the upper and lower bounds mark the limiting cases where all the stellar mass of the minor merger is transferred to the bulge and to the disc, respectively, and the dashed lines represent the mean of the two cases. We compare our predictions with data on the mean B/T as a function of galaxy stellar mass from MaNGA (black dots with error bars), as detailed in Section \ref{sec_data_sdss}. The fact that the shaded areas in SMHM Model 2 are slightly broader at high stellar masses than those in Model 1 is an artefact of the number of mergers predicted by these Models, as shown in Figures \ref{fg_num_mergers} and \ref{fg_num_mergers_Mstar}. In particular, at fixed number of DM mergers, Model 2 predicts less major mergers, i.e., more minor mergers than Model 1. This leads to a higher bulge/disc mass in the cases where all the mass from minor mergers goes to the bulge/disc, leading, therefore, to a larger bound.
        
        \begin{figure*}
            \includegraphics[width=0.9\textwidth]{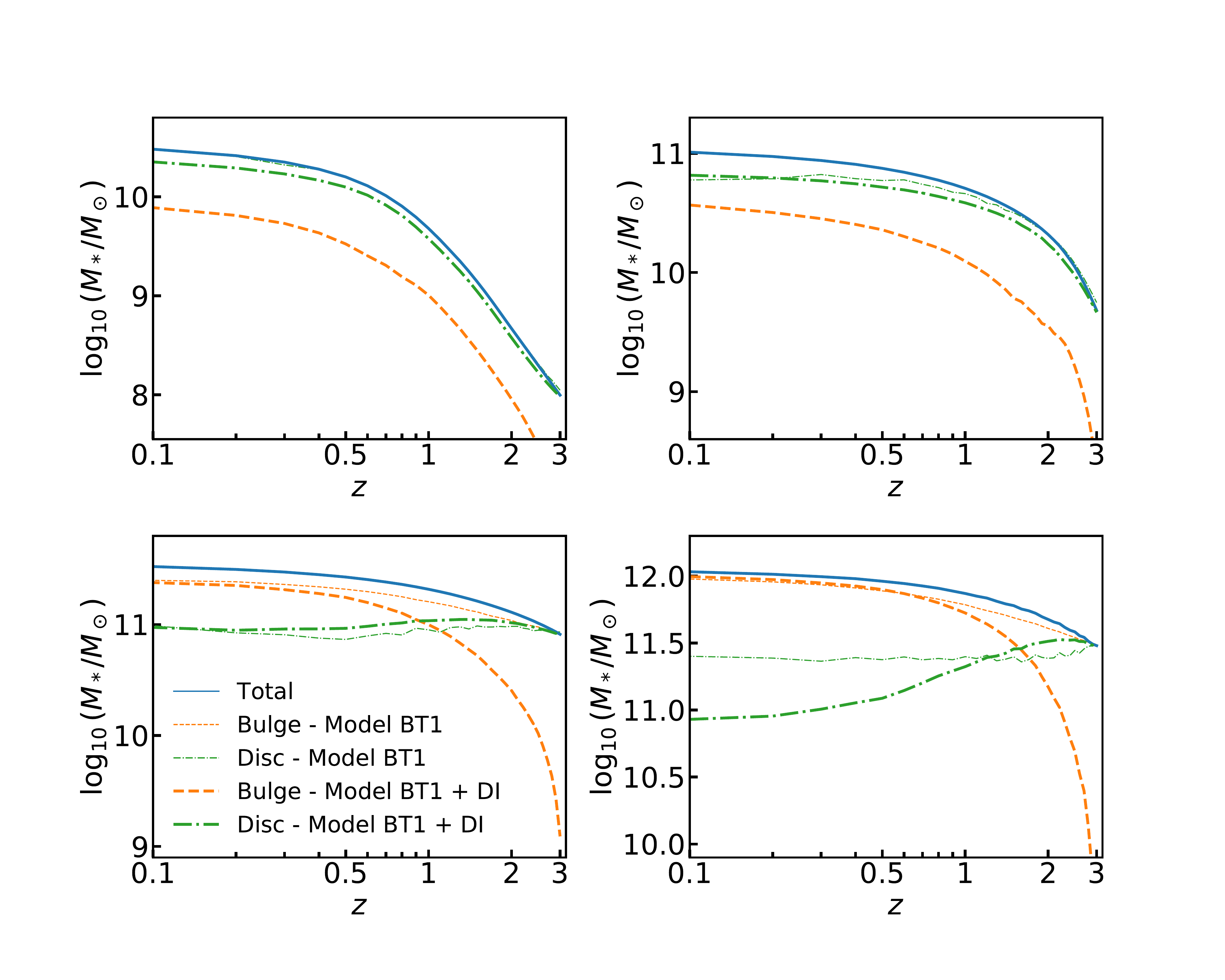}
    	    \caption{Average growth of total, bulge and disc stellar mass for four different galaxy mass bins at $z=0$. The solid blue lines show the total stellar mass growth. The thin dashed and dash-dotted lines show the mass growth of the bulges and discs, respectively, for Model BT1, while the thick lines show the mass growths for Model BT1 including disc instabilities.}
    	    \label{fg_bulge_growth}
        \end{figure*}
        
        The MaNGA data clearly indicate that all local galaxies have a B/T ratio that is confined within $0.2 - 1$, with a mean value slightly rising with increasing stellar mass, from $\sim 0.2$ reaching an average value of $B/T \sim 1$ only in the most massive galaxies with $M_* \gtrsim 10^{11} \, M_\odot$. Models that, like our BT1, assume a strict $B/T=1$ during major mergers, can better reproduce the data at high stellar masses, especially when assuming that all the minor mergers contribute to the bulge component. On the other hand, models that, like BT2, assume that a significant fraction of the disc survives and/or is rapidly reaccreted after the merger, fall short in matching the data at high stellar masses, suggesting that these kinds of model are not extremely suitable to describe the average B/T ratios, at least at high stellar masses.

        Irrespectively of what we assume for the B/T in major (or even minor) mergers, both our BT1 and BT2 Models drastically fail in reproducing the observed B/T in galaxies with $M_* \lesssim 2 \times 10^{11} M_\odot$ (top panels of Figure \ref{fg_BT_ratio_multi_models}). An \emph{additional} component/process should be included in \decode to reproduce the data. Here again we can witness the usefulness of a semi-empirical approach that, as discussed in the Introduction, can include additional layers of complexities wherever necessary, as guided by the data, in a transparent and efficient way. The bottom panels of Figure \ref{fg_BT_ratio_multi_models} show the prediction of Model BT1 by including disc instabilities as discussed above. We do not show Model BT2 as it fails to model the B/T ratios even with the addition of disc instabilities. The main effect of including disc instabilities in our Model BT1 is that it tends to increase the bulge masses at lower, but not at higher, galaxy stellar masses, where the condition in Equation (\ref{eq_discInst_ELN}) is more easily met mainly because lower mass galaxies retain their disc morphologies for a longer time. In particular, we find that, irrespective of the exact input SMHM relation, to broadly match the data we need to assume that a fraction of the disc mass is transferred to the bulge at each disc instability event, in line with some previous cosmological models \citep[e.g.,][]{bower_2006}. %In particular, we show the maximal cases for the two Models where 100\% of the disc mass is transferred to the bulge.
        In the bottom panels of Figure \ref{fg_BT_ratio_multi_models}, we show both the cases where clumpy accretion \citep[following][]{bournaud_2011} and "classical" disc instabilities \citep[][]{efstathiou_1982} are implemented. We find that Model BT1\footnote{We note that the narrower shaded areas of the Model with disc instabilities are a direct consequence of the transfer of mass from the disc to the bulge and tends to shrink into a single line in the case where disc instabilities take place at any time step since the formation epoch of the galaxy, which explains also why the bounds are even thinner at low masses where the condition of Equation \ref{eq_discInst_ELN} is more easily satisfied.}, the one without disc regrowth, with SMHM Model 2 broadly matches the SDSS local average B/T ratios when some level of disc instabilities is included in the model, while it fails to match the data at low masses with SMHM Model 1. In particular, disc instabilities implemented following \citet{bournaud_2011} appear not to be sufficient to provide enough boost to the average B/T at the low-mass end to match the data, even if the baryonic inflow rate in Equation (\ref{eq_bournaud}) is doubled. On the other hand, disc instabilities as in \citet{efstathiou_1982} allow to broadly reproduce the observational data of SDSS, when the factor $\epsilon$ in Equation (\ref{eq_discInst_ELN}) is roughly $0.5$. Intriguingly, one could argue that strong disc instabilities, and not necessarily major mergers, are responsible for forming most stellar bulges, even at the highest stellar masses. In fact, the strength of a disc instability in forming a bulge closely depends on the D/T ratio, the more disc mass there is, the more potential there is for the bulge to grow in mass after a disc instability. We however checked that even on the assumption of ineffective major mergers preserving a $B/T \sim 0.2$, the disc instabilities would still fall short in boosting the B/T up to unity at high stellar masses.
        
        In the bottom panels of Figure \ref{fg_BT_ratio_multi_models}, for completeness, we compare our predictions with the outputs from the TNG100 simulation and the \textsc{GalICS} SAM (brown triangles with error bars and blue solid line, respectively). Interestingly, both TNG100 and GalICS predict large mean $B/T \gtrsim 0.9$ at high stellar masses in reasonable agreement with our predictions and the data. In addition, they also include bulge formation via disc instabilities \cite[see, e.g.,][]{tacchella_2019, cattaneo_2020}, predicting indeed $B/T > 0.2$ at low masses in line with the data, though the mean B/T predicted by \textsc{GalICS} tends to be larger than the one measured in MaNGA towards lower stellar masses (we provide in Appendix \ref{app_bt_ratios} a detailed comparison between the B/T ratio found in MaNGA and those from other samples/studies). Our Model 1 is roughly consistent with the predictions of \textsc{GalICS} at all masses but not with the observational data and the TNG simulation at low masses, while Model 2 behaves in nearly the opposite fashion, highlighting once again the dependence of galactic properties, this time the mean B/T ratios, on the input SMHM relation.
        
        Finally, in Figure \ref{fg_bulge_growth} we show the mean mass growth of bulges and discs for four galaxy masses at $z=0$ for Model BT1 with and without disc instabilities, as labelled. We note that, when disc instabilities are not included, at low stellar masses (e.g., $M_*(z=0) \sim 10^{10.5} \, M_\odot$) the disc dominates the overall mass growth of the galaxies and the bulge component is almost negligible, unless disc instabilities are included (long-dashed orange lines). Moving towards higher stellar masses, the bulge begins to be gradually more dominant, as a direct consequence of the increasing number of mergers, and adding disc instabilities does alter this trend noticeably.

        \subsection{Brightest cluster galaxies history}\label{sec_res_bcgs}
        
        As a final application of \decode, we study the stellar mass growth history of BCGs, which are massive elliptical galaxies that constitutes an additional source of information for understanding the evolution of galaxies and large-scale structure. Several studies have addressed already this issue, both from observations \citep[e.g.,][]{whiley_2008, collins_2009, stott_2010, lidman_2012, bellstedt_2016, lin_2017, zhang_2017} and numerical works, such as SAMs \citep[e.g.,][]{de_lucia_2007, contini_2014}, SEMs \citep{shankar_2015} and hydrodynamic simulations \citep[e.g.,][]{pillepich_2018_Mstar_content, ragone_figueroa_2018}. Here, we show the predictions of \decode for the stellar mass assembly of BCGs and how these compare to the results from other works.
        
\begin{figure}
            \includegraphics[width=\columnwidth]{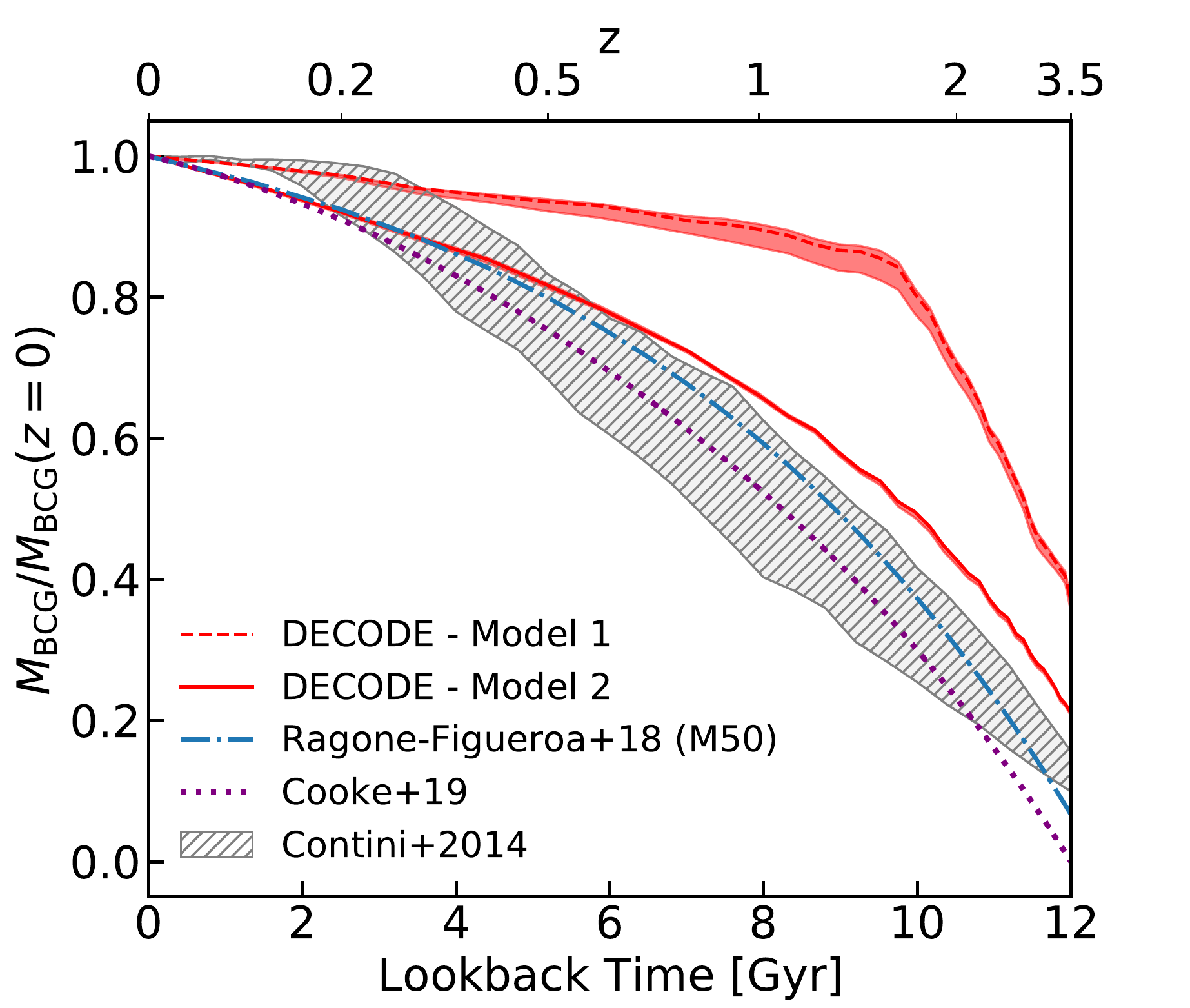}
    	    \caption{Fractional stellar mass growth of BCGs predicted by \decode for Models 1 and 2 as a function of lookback time with their $1\sigma$ bounds (red lines and shaded areas). The blue dash-dotted line corresponds to the fit M50 selected BCGs according to Equation (2) in \citet{ragone_figueroa_2018}. The purple dotted line shows the observed median growth from COSMOS} as presented in \citet{cooke_2019}. The grey dashed area includes the evolutionary histories of the models in the \citet{contini_2014} semi-analytic model.
    	    \label{fg_Mstar_BCGs}
        \end{figure}

        The results are shown in Figure \ref{fg_Mstar_BCGs}. The red dashed and solid lines show the total stellar mass fractional growth of BCGs, selected in haloes with present day mass $M_{\rm h} (z=0) > 8 \times 10^{14} M_\odot$, predicted by \decode with Models 1 and 2, along with their $1\sigma$ uncertainties (shaded areas). The data are compared with the COSMOS data \citep{cooke_2019}, results from the hydrodynamic simulations of \citet{ragone_figueroa_2018} and SAM of \citet{contini_2014} (as labelled in the Figure), selected in the same mass region. According to Model 1 BCGs have already formed most of their mass at redshifts $z > 1.5$, because of the assumed constant SMF up to that redshift which maps the DM halo mass accretion history into a higher average stellar mass growth. On the other hand, assuming Model 2 BCGs have only formed roughly $50 \%$ of their mass by $z = 1.5$ and have grown the remaining mass at later epochs.
        
        The BCG stellar masses in \citet{ragone_figueroa_2018} are computed with the mass within the spherical radius of 50 kpc (M50). Our Model 2 is in relatively good agreement on the mass formation history of BCGs with M50 selected galaxies, predicting a factor of $\sim 1.5$ in the mass increase between $z=1$ and $z=0$ against the factor $1.4$ of \citet{ragone_figueroa_2018}.
        
        Figure \ref{fg_Mstar_BCGs} shows that models characterized by a SMHM relation with a significantly evolving underlying SMF as in Model 2, are once again better tuned to reproduce the current data, this time data on BCGs, in line with predictions from hydrodynamic simulations and SAMs. We note that in all our Models satellites are frozen. Increasing their masses via residual star formation after infall would possibly steepen the evolution for both Models, possibly slightly improving the match with the data for Model 1.

    \section{Discussion}\label{sec_discussion}
        
        The discrete statistical and semi-empirical model, \decode, presented in this work constitutes an invaluable complementary tool to the existing cosmological models for modelling galaxy evolution. These models, either analytic or numerical, are affected by significant volume/mass resolution limitations and/or a large amount of input assumptions and parameters. Also other existing works on SEMs, such as \citet{behroozi_2019} and \citet{moster_2018}, based on abundance matching between galaxy-halo properties, still need large computational resources to run. Instead our model, based on statistical input distributions, allows to rapidly simulate a large volume box and investigate the mean properties of galaxies without relying on ab-initio analytic models or simulations.
        
        In the last decade several SEMs have been developed. We mention here some examples and discuss how \decode differs from and complements them. \citet{moster_2018} showed an empirical relation between the mass growth of DM haloes and the galaxy star formation rate, from which they retrieve the mergers history and other properties. \citet{moster_2013} and \citet{behroozi_2019} proposed SEMs where they populate DM haloes via SMHM relations with galaxies in vast catalogues. These models connect galaxy to DM halo properties via several analytic relations, which could involve a non-negligible number of free parameters. Our present model, instead, although more restricted in scope as it only focuses on mean galaxy assembly histories, essentially relies on only one input parameter, the scatter in the SMHM relation. From this single parameter and abundance matching, \decode can then generate robust predictions on mean galaxy growth histories, merger rates, satellite abundances, and star formation histories, thus providing a flexible and transparent tool to probe galaxy evolution in a full cosmological context. \citetalias{grylls_paper1} recently developed a SEM where they apply abundance matching to statistical distributions of DM haloes and galaxies, which is limited by the non-discreteness since it is based on statistical weights over continuous probability densities. Our model instead applies the SMHM relation from abundance matching directly to the mock universe generated stochastically by making a realization of the distributions themselves.
        
        The fact that \decode starts from the SMF to compute the SMHM relation, makes \decode a powerful tool to set more stringent constraints on the SMF. As discussed in Section \ref{sec_smhm_models}, the galaxy SMF is far from being well known, especially at high redshifts, and many works suggested different (and sometimes contrasting) results, in terms of shape and/or evolution in time \citep[e.g.,][]{tomczak_2014, bernardi_2013, bernardi_2016, bernardi_2017, davidzon_2017, huang_2018, kawinwanichakij_2020, leja_2020}. Our current tests show a preference for SMFs characterised by a larger number of massive galaxies and a significant evolution in time. These SMFs generate SMHM relations that, in turn, produce a sufficient number of mergers to match the local fraction of ellipticals, satellite abundances, and BCG growth. Our study thus highlights the strong dependence of galaxy stellar mass assembly histories on the input SMHM relation. \citet{stewart_2009a} and \citet{hopkins_2010a} also found that galaxy merger rates depend on the input SMHM relation. Their results, along with ours, imply that, for a fixed DM merger tree, the major merger rates and other quantities strongly depend on the mapping between stellar mass and halo mass, which in turns depends on the systematics, shape and evolution of the SMF. Following their path, a fraction of $~ 40\%$ of mass loss during mergers can alter significantly the merger rates and, therefore, also the fraction of elliptical galaxies and B/T ratios, up to more than a factor of 2, which would allow also flatter SMHM relations to perform well.
        
        In this work, we also found evidence for the need of disc instabilities to boost the formation of bulges at lower stellar masses. This result is in line with the general notion of fast and slow rotators \citep[e.g.,][]{bernardi_2019, dominguez_sanchez_2020}, which suggest that the former dominates at $M_* \lesssim 10^{11.5} M_\odot$ and the latter at $M_* \gtrsim 10^{11.5} M_\odot$, or also with the distinction between pseudo and classical bulges \citep[see discussions in][and references therein]{gadotti2009, fisher_2010, shankar_2012, shankar_2013}.
        Similar findings are retrieved in SAMs. \citet{guo_2011} found that at stellar masses $M_* \gtrsim 10^{11} \, M_\odot$ mostly all galaxies have a $B/T > 0.7$, and mostly $B/T < 0.7$ below the same stellar mass threshold, in line with the results of this work. Similar conclusions are derived from \textsc{GalICS} SAM and the TNG100 simulation (Figure \ref{fg_BT_ratio_multi_models}), as well as from other works in the literature \citep[e.g.,][]{rodriguez_gomez_2015, tacchella_2019, kannan_2015, fontanot_2015}. \citet{shankar_2014} found that if strong disc instabilities are included in the models, then these could also contribute to a stronger dependence of the mean galaxy size on environment (halo mass), which is in line with our results that point to relatively mild disc instabilities.

	\section{Conclusions}\label{sec_conclusions}
    
    In this paper we have presented \decode, the Discrete statistical sEmi-empiriCal mODEl, built to accurately simulate \textit{mean} DM halo and galaxy growth and merger histories, for any input SMHM relation. \decode generates discrete populations of DM haloes and assigns an average mass accretion history to each of them via an input mass function. It subsequently generates their merger trees via an input subhalo distribution function, and assigns to each subhalo the infall redshift and dynamical friction timescale using statistical density functions that we fit and accurately test against the Millennium simulation. We then populate the (sub)haloes with central/satellite galaxies via diverse and observationally motivated SMHM relations, computed via numerical abundance matching techniques which are very sensitive to the shape of the input SMF. Thanks to its statistical nature, \decode is flexible, rapid, and not affected by limitation in volume or mass resolution. In this work we provide useful analytic recipes for the infall redshift distributions of subhaloes of first- to third-order (Section \ref{sec_valid_millennium}), along with the correction to the halo mass function for unmerged and unstripped subhaloes (Appendix \ref{app_corr_HMF}).
    
    We apply \decode to predict the galaxy-galaxy merger rates, satellite abundances (which can be considered as unmerged satellites) and BCG growth. We also explore how merging pairs can impact on the fraction of ellipticals and mean B/T ratios of local galaxies, by assuming that the former are formed in major mergers, and the latter are shaped by both major/minor mergers and disc instabilities.
    
    Our main results on the galaxy evolution probed via \decode can be summarized as follows:
    \begin{itemize}
        \item \decode can generate accurate galaxy stellar mass assembly and merger histories starting from an input SMHM relation with only one input parameter, the scatter around the SMHM relation. \decode can reproduce the average galaxy growth histories of hydrodynamic simulations and SAMs when inputting their SMHM relations.
        \item Via \decode, we showed how sensitive many galaxy observables are on the input SMHM relation, and thus on the input SMF, in particular galaxy merger rates, satellite abundances, and BCG growths.
        \item A SMHM relation implied by a SMF characterized by a larger number of massive galaxies and a normalization significantly decreasing at high redshift, is more suitable to reproduce the correct abundances of satellite galaxies in the local Universe and the stellar mass growth of BCGs at $z<1$, as well as the combined major merger pair fractions as inferred from GAMA, UDS, VIDEO and COSMOS.
        \item Our reference SMHM relation is also able to reproduce the fraction of local elliptical galaxies on the assumption that these are formed from major mergers with $\mu>0.25$, as often assumed in cosmological SAMs. In other words, the validity of the $\mu>0.25$ threshold is strongly dependent on the input SMHM relation.
        \item The same SMHM relation is also able to reproduce the mean B/T ratio of local MaNGA galaxies, with a contribution from disc instabilities at stellar masses below $M_* \lesssim 10^{11} \, M_\odot$.
    \end{itemize}
    
    In conclusion, \decode is a valuable, complementary tool for probing galaxy evolution and the relevant physical processes involved therein. It can indeed rapidly probe galaxy merger rates, satellites abundances, morphologies, star formation histories for any given input SMHM relation and with minimal input parameters. \decode will also constitute a very precious instrument for generating robust galaxy mock catalogues for the upcoming large-scale extra-galactic surveys such as Euclid \citep[e.g.,][]{euclid_prepI, euclid_prepIII, euclid_prepXIII} and LSST \citep[e.g.,][]{bridge_2009, ptak_2011, covey_2010, gawiser_2013, riccio_2021}.

    \section*{Acknowledgements}

    We warmly thank Philip J. Grylls for reading the manuscript, and for useful discussion and comments. We thank the referee for a careful reading of the manuscript and for useful inputs. We also thank Sergio Contreras for useful discussions. This work received funding from the European Union’s Horizon 2020 research and innovation programme under the Marie Sk\l odowska-Curie grant agreement No. 860744. HF acknowledges partial support from the "Torno Subito" programme.
	MA acknowledges funding from the Deutsche Forschungsgemeinschaft (DFG) through an Emmy Noether Research Group (grant number NE 2441/1-1). YRG acknowledges the support of the “Juan de la Cierva Incorporation” fellowship (IC2019-041131-I) 

    %%%%%%%%%%%%%%%%%%%%%%%%%%%%%%%%%%%%%%%%%%%%%%%%%%
    \section*{Data Availability}
    
    The data underlying this article will be shared on reasonable request to the corresponding author. \decode is available for download at \href{https://github.com/haofu94/DECODE}{https://github.com/haofu94/DECODE}.



\bibliographystyle{mnras}
\bibliography{main}

%\begin{thebibliography}{99}
%\end{thebibliography}

%%%%%%%%%%%%%%%%%%%%%%%%%%%%%%%%%%%%%%%%%%%%%%%%%%

%%%%%%%%%%%%%%%%% APPENDICES %%%%%%%%%%%%%%%%%%%%%

\appendix

    \section{Testing the self-consistency of DECODE}\label{app_weighted_discrete_consistency}
    
    We develop a variant of our model, which we refer to as \textit{weighted} method, to check the self-consistency of our approach in \decode (referred to as \textit{discrete} method) in reproducing the average properties of dark matter haloes and galaxy evolution. In the main body of the paper we make use of the discrete method only. We describe in this section the details of the weighted method. We reiterate here that the weighted method is more difficult to generalise to all model variants, for example when multiple galaxy properties are included in the evolution, which makes the discrete method more flexible.
    
    \begin{figure}
        \includegraphics[width=\columnwidth]{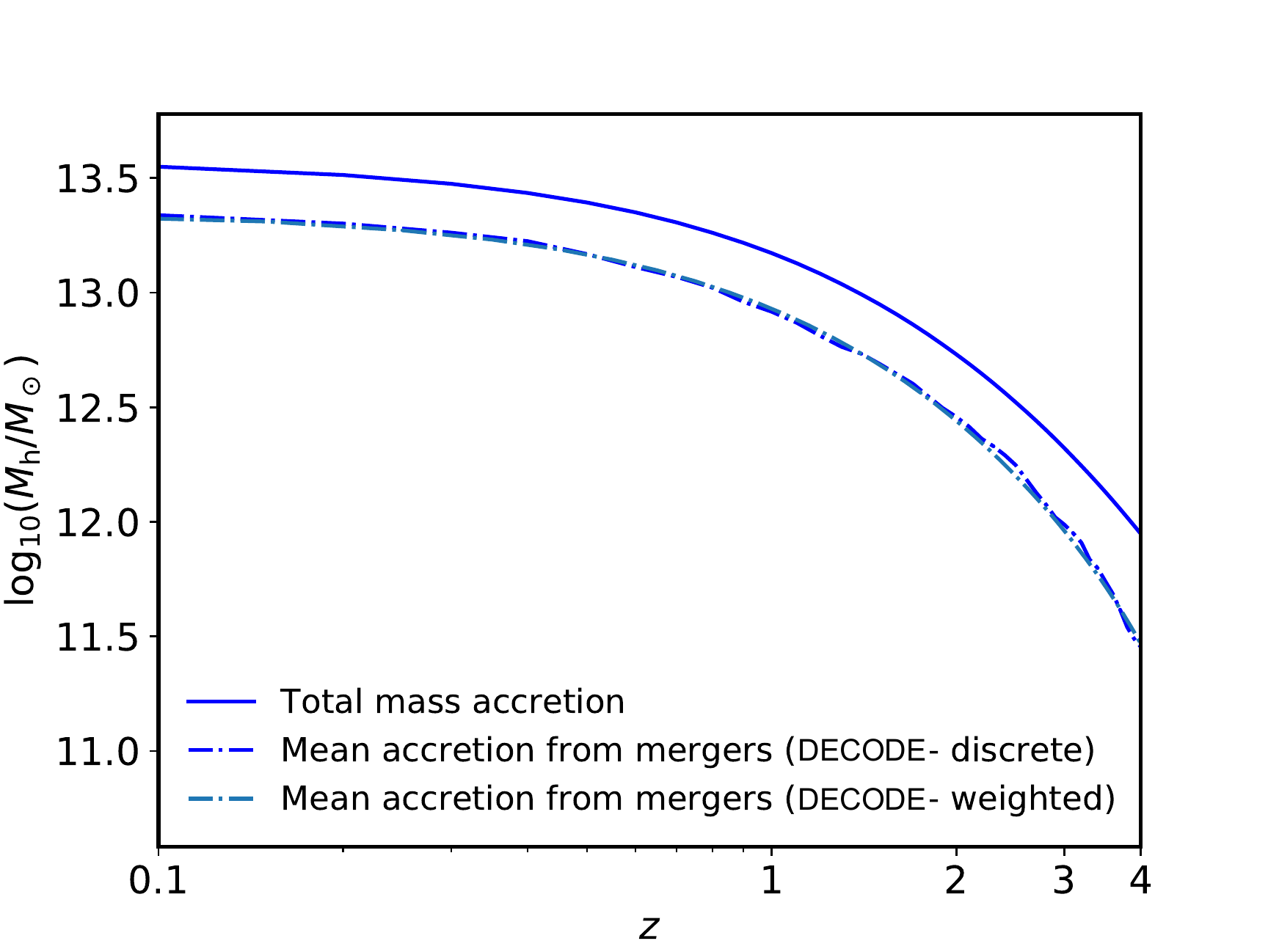}
	    \includegraphics[width=\columnwidth]{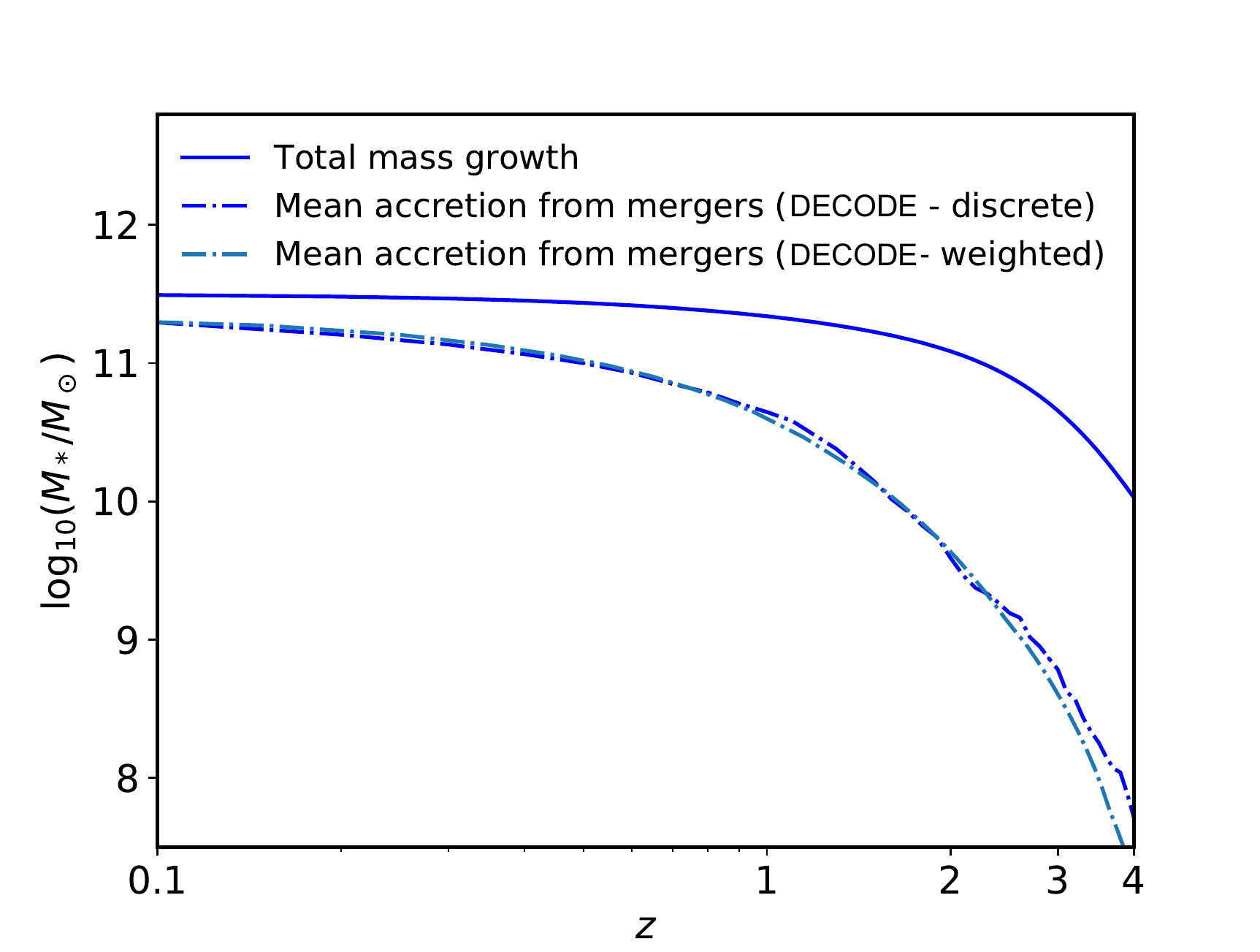}
	    \caption{Upper panel: total halo mass accretion history (solid line) along with the comparison between the mergers history of from the discrete and weighted methods (dashed lines). Lower panel: same as upper panel but for galaxy stellar mass.}
	    \label{fg_mergers_comparison}
    \end{figure}
    
    \begin{figure*}
        \centering
        \includegraphics[width=\linewidth]{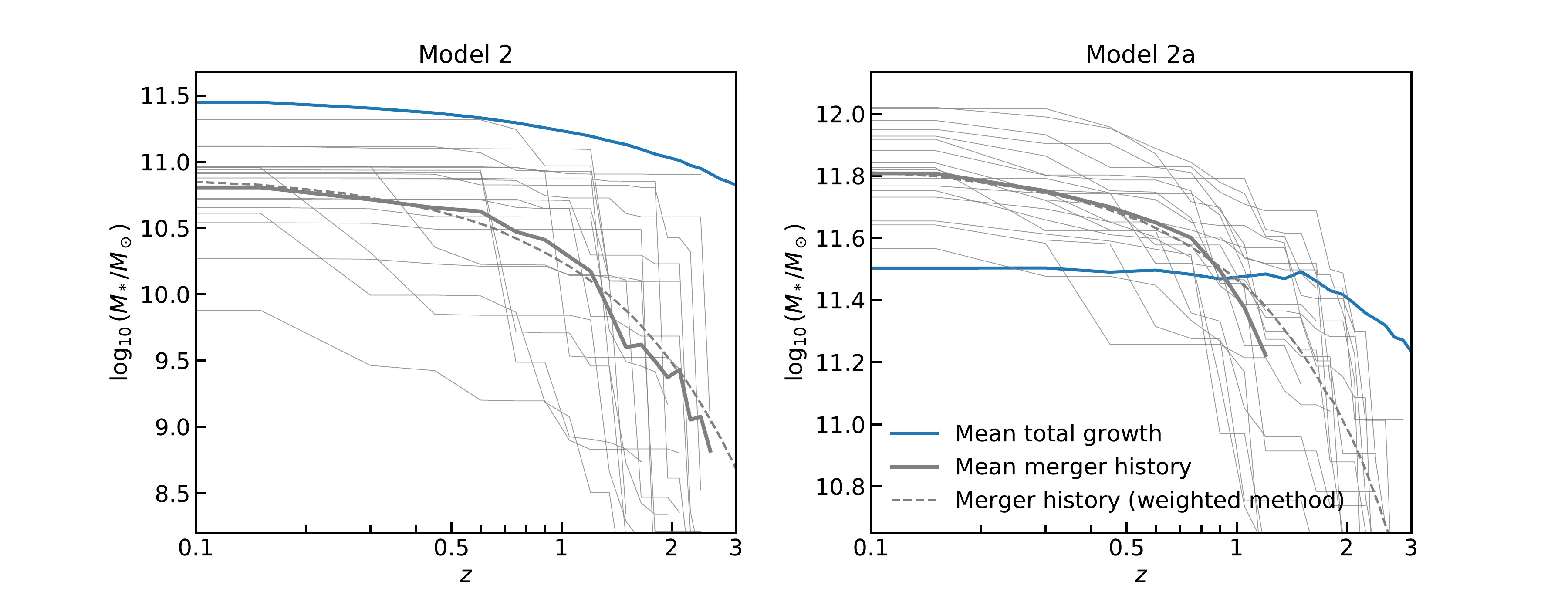}
        \caption{Merger histories of a sample of galaxies with stellar mass at $z=0$ of $\sim 10^{11.5} \, M_\odot$, for Models 2 and 2a, as labelled by the panels. The solid blue and grey lines how the average total stellar mass growth and merger history, respectively. The dashed grey lines show the average merger history from the weighted model.}
        \label{fg_merger_history_comp}
    \end{figure*}
    
    First of all, we focus our attention on the mergers history of DM haloes. To compute the latter, in the weighted method we employ the recipe from Section 3.1.3 of \citetalias{grylls_paper1}, where we interpret the difference of the SHMF between redshifts $z$ and $z+\mathrm dz$ as the weight (or probability) of the infalling subhaloes in each mass bin. In other words the weight, at given redshift step and mass bin, is the fractional average number of subhaloes of that mass which cross the virial radius of the parent halo at that redshift. The comparison between the weighted and discrete methods for one parent halo mass bin is shown in the upper panel of Figure \ref{fg_mergers_comparison}, where the total mass assembly history of \cite{vdb_2014} is also shown for completeness. Despite the fact that in the discrete method we perform our analysis on the assumption of identical mean accretion for all haloes competing to the same bin of host halo mass at $z=0$, the resulting mean contribution from subhalo mergers in the discrete method appears to be in very good agreement with the one computed from the weighted method, further supporting the validity of our discrete approach.
    
    \begin{figure}
        \includegraphics[width=\columnwidth]{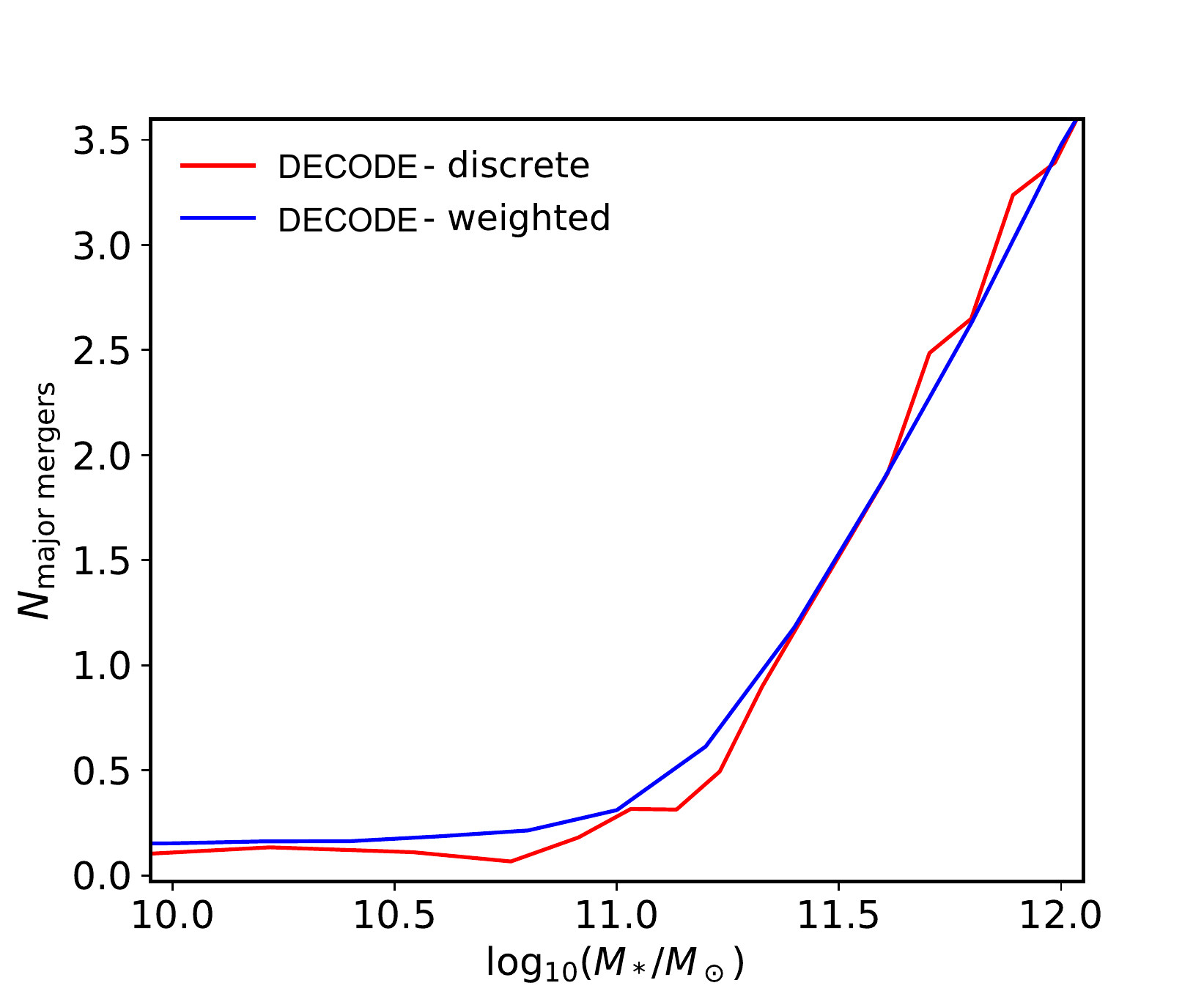}
	    \caption{Number of major mergers predicted by the discrete version of \decode compared to that computed with the weighted method.}
	    \label{fg_num_major_steel_vs_dream}
    \end{figure}
    
    Similarly to the host dark matter haloes, the bottom panel of Figure \ref{fg_mergers_comparison} compares the merger contributions to the central galaxy growth computed via the discrete and weighted methods, as labelled, showing again very good agreement between the two methods. We note that this agreement is, as expected, independent of the choice of the input SMHM relation or dynamical friction timescales, as long as the same parameters are adopted in both methods. We specify once again that, as already noted by \citetalias{grylls_paper2}, each merger tree generated in the \emph{discrete} \decode can show sometimes a merger history that goes beyond the total stellar mass growth of the galaxy, which might seem not physical. This is a direct consequence of the fact that each merger tree in \decode is a stochastic realization of the mass functions and probability distributions used as input. However, we test that in the SMHM models that we adopt in this work, the average merger history always lives below the total mass growth and is also fully consistent with what the \emph{weighted} method predicts, as already shown in Figure \ref{fg_mergers_comparison}. We show in the left panel of Figure \ref{fg_merger_history_comp} the merger history for our fiducial Model 2 for a galaxy mass bin of $\sim 10^{11.5} \, M_\odot$, where we see that all the single merger histories lie below the total mean mass assembly of the galaxy. On the other hand, in the right panel we show the same results but for Model 2a, where we can clearly see the impact of a lower high-mass end in the SMF leading to a much higher number of mergers, and in many cases not physical as it goes beyond the total growth. The effect of generating an unphysical merger history that, on average, is larger than the total mean stellar mass growth, could be, at least in part, be alleviated by including strong stellar stripping \citep[see e.g.,][]{cattaneo_2011, smith_2016} and/or stellar mass loss in mergers \citep[e.g.,][]{moster_2018}. We will further explore these interesting variants to the model in future work, in combination with the amount of associated star formation our stellar mass accretion tracks predict.
    
\begin{figure}
        \includegraphics[width=\columnwidth]{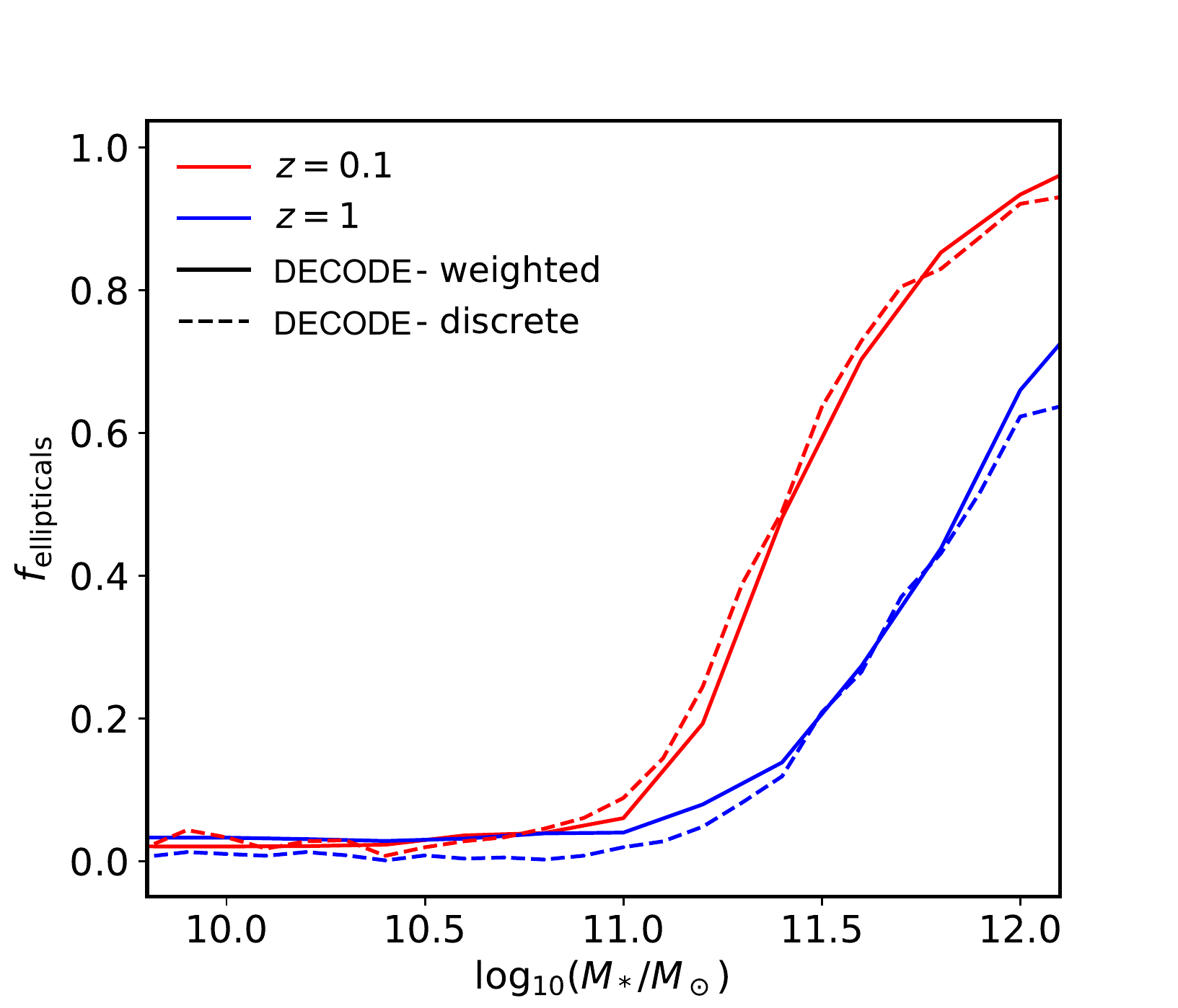}
	    \caption{Fraction of ellipticals predicted by the discrete version of \decode compared to those computed with the weighted method, for redshifts 0.1 and 1.}
	    \label{fg_frac_ell_steel_vs_dream}
    \end{figure}

\begin{figure}
        \includegraphics[width=\columnwidth]{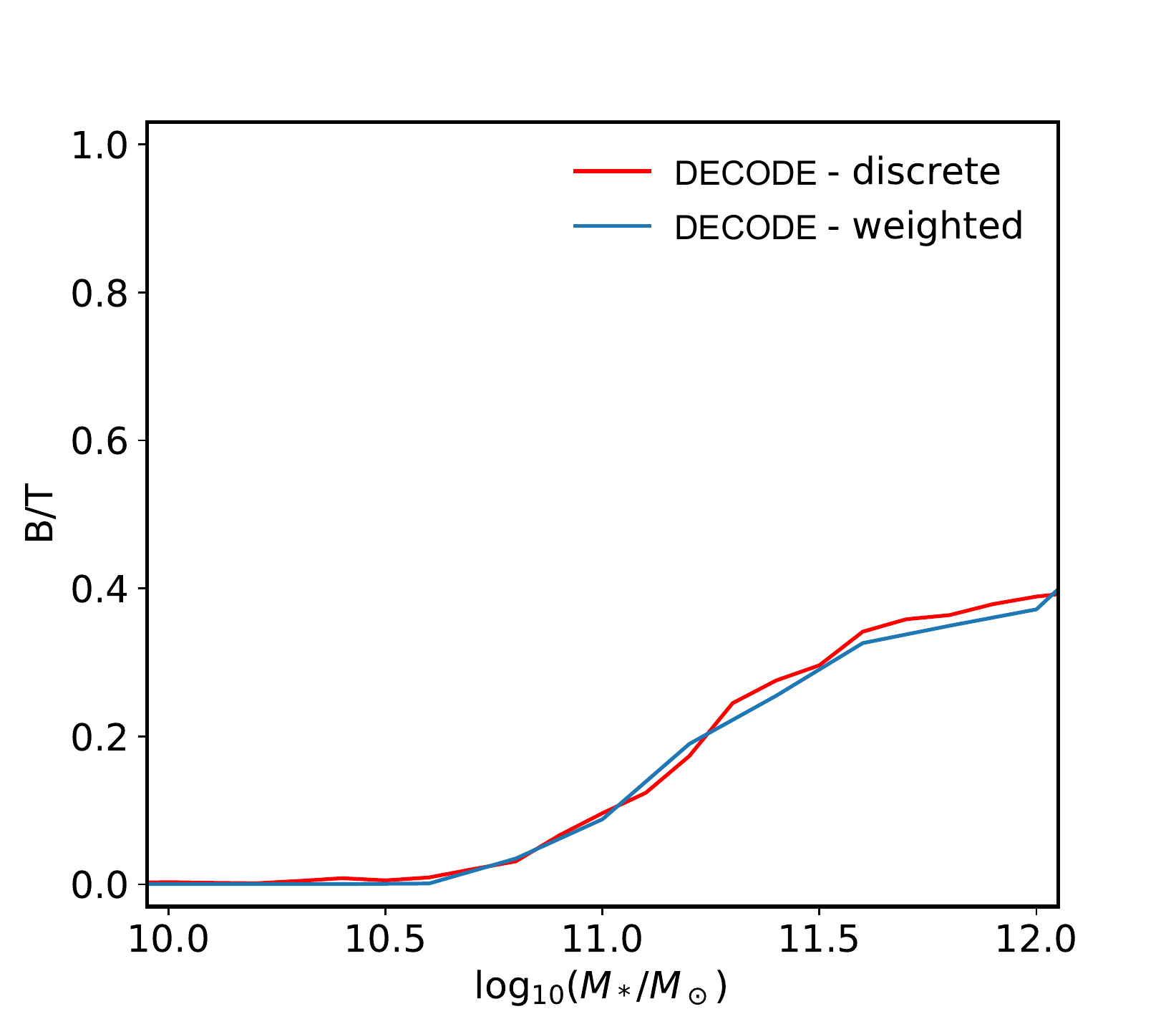}
	    \caption{Bulge-to-total ratio predicted by the discrete version of \decode compared to that computed with the weighted method for Model BT2.}
	    \label{fg_BT_steel_vs_dream}
    \end{figure}

    We also provide the number of major mergers, implied fraction of ellipticals and mean B/T ratios as predicted from the two methods in Figures \ref{fg_num_major_steel_vs_dream}, \ref{fg_frac_ell_steel_vs_dream} and \ref{fg_BT_steel_vs_dream}, respectively. The selection of the major mergers and the ellipticals in the discrete method is already described in Section \ref{sec_res_morphology}. In the weighted method the number of major mergers is computed by directly integrating the merging satellites SMF at each redshift over the range of mass $M_{\rm *,sat} / M_{\rm *,cen} > \mu$, with $\mu$ being the major mergers mass ratio threshold. Concerning the fraction of ellipticals, we also label the galaxies that had at least one major merger as ellipticals, similarly as we do in the discrete method. To this purpose, we initialize the fraction of ellipticals at redshift $z_{\rm ini} = 4$ to 0, assuming that all galaxies are disc-like at that time. From that epoch, we proceed forward in time and at each redshift we analytically compute the probability of a galaxy to have had at least one major merger, which we interpret as the fraction of ellipticals itself, according to the following formula
    \begin{equation}\label{eq_P_MM}
        \mathcal{P}_{\rm 1MM} = 1 - \mathcal{P}_{\rm MM}^{W} \; ,
    \end{equation}
    where $\mathcal{P}_{\rm 1MM}$ is the probability of having at least one major merger, $\mathcal{P}_{\rm MM}$ is the probability of a generic merger to be a major one, and the exponent $W$ is the weight integrated over the major mergers stellar mass range. At each time step, we update the fraction of spirals and ellipticals according to Equation (\ref{eq_P_MM}). Finally, we also provide the comparison of the B/T ratio for Model BT2 (Section \ref{sec_res_bt_ratio}), which we compute in the weighted model as the cumulative sum of the probabilities of having at least one major merger in each time bin. It is clear from the results reported in Figures \ref{fg_num_major_steel_vs_dream}-\ref{fg_BT_steel_vs_dream} that the both methods provide extremely consistent predictions on the aforementioned quantities, further validating the use of the discrete method to predict mean galactic properties.
    
	\section{Correction to the halo mass function}\label{app_corr_HMF}
	
\begin{figure}
        \includegraphics[width=\columnwidth]{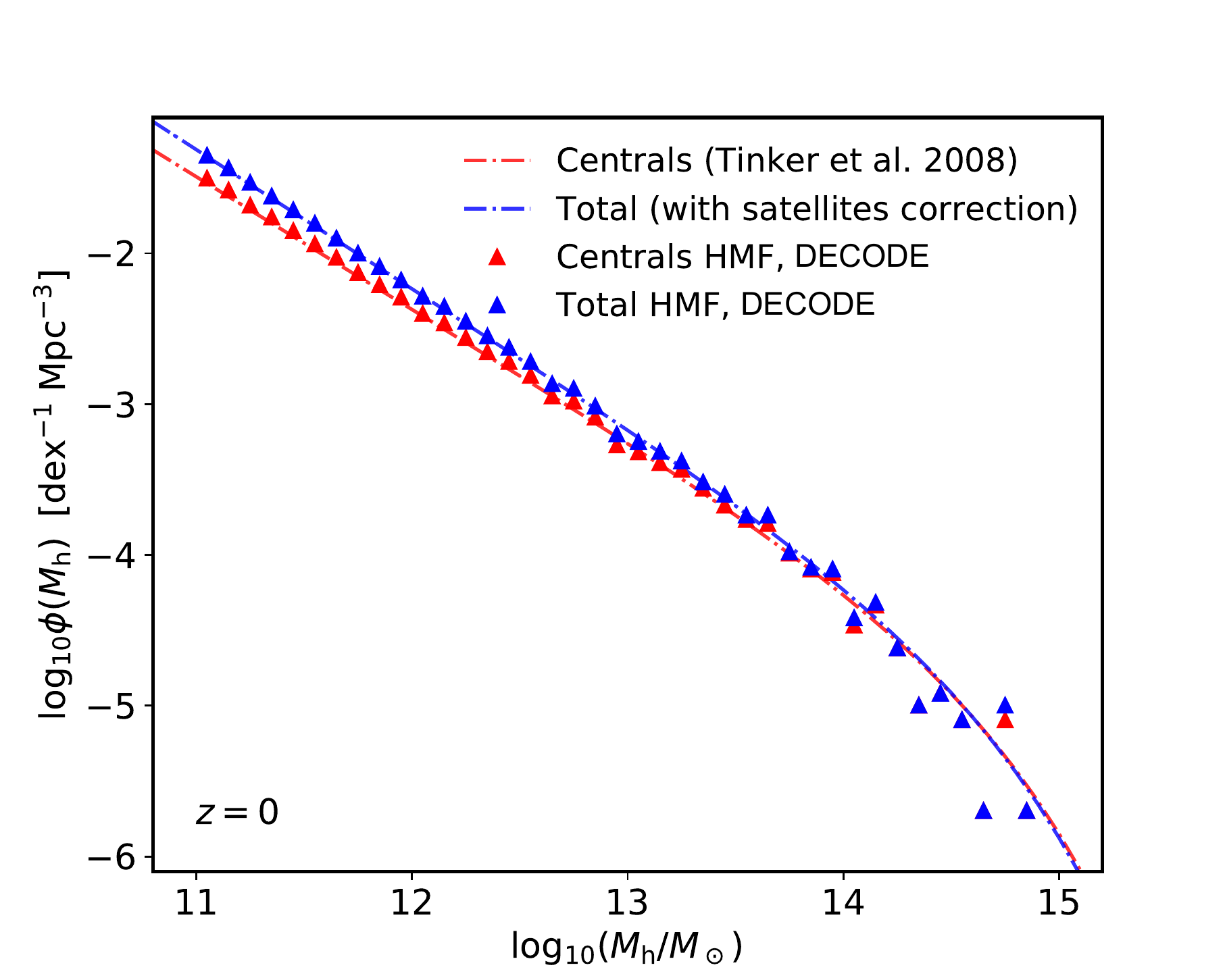}
        \caption{Halo mass function for parent dark matter haloes of \citet{tinker_2008} (red dash-dotted line) and the total HMF obtained by applying the correction with satellites (blue dash-dotted line), compared to the mass functions calculated from \decode (triangles).}
        \label{fg_hmf_corr_comparison}
    \end{figure}

	The abundance matching procedure of Equation (\ref{eq_aversa_AM}) between the galaxy SMF and the dark matter HMF also includes galactic satellites in the former, and dark matter subhaloes in the latter. However, the HMF is usually given as a fit to the number densities of only the parent haloes existing in any given simulation snapshot, and thus it must be corrected by the abundances of the unstripped subhaloes surviving at any given epoch. To determine this correction, we first compute the number densities of unstripped and surviving subhaloes in \decode at any redshift of interest, and then fit it following, for convenience and for ease of comparison, the same analytic expression adopted by \cite{behroozi_2013}
	\begin{equation}\label{eq_behroozi_corr}
	    \frac{\phi_{\rm satellites} (M_{\rm h})}{\phi_{\rm centrals} (M_{\rm h})} \sim C(a) \log \bigg( \frac{M_{\rm cutoff} (a)}{M_{\rm h}} \bigg) \; ,
	\end{equation}
	where $a = 1/(1+z)$ is the scale factor. Our \decode Monte Carlo simulations provide the following fitting formulae for the two free parameters in Equation (\ref{eq_behroozi_corr})
	\begin{equation}
	    \log (C(a)) = -2.42 + 11.68a - 28.88a^2 + 29.33a^3 - 10.56a^4 \; ,
	\end{equation}
	\begin{equation}
	    \log (M_{\rm cutoff} (a)) = 10.94 + 8.34a - 0.36a^2 - 5.08a^3 + 0.75a^4 \; ,
	\end{equation}
    We show in Figure \ref{fg_hmf_corr_comparison} the \cite{tinker_2008} HMF for centrals along with the aforementioned correction for including unstripped, surviving (i.e., unmerged) satellites, and we compare them with the numerical mass functions calculated from \decode.

    \section{Halo and stellar mass growths}\label{app_mass_growth}
    
    \begin{figure*}
        \centering
        \includegraphics[width=0.495\linewidth]{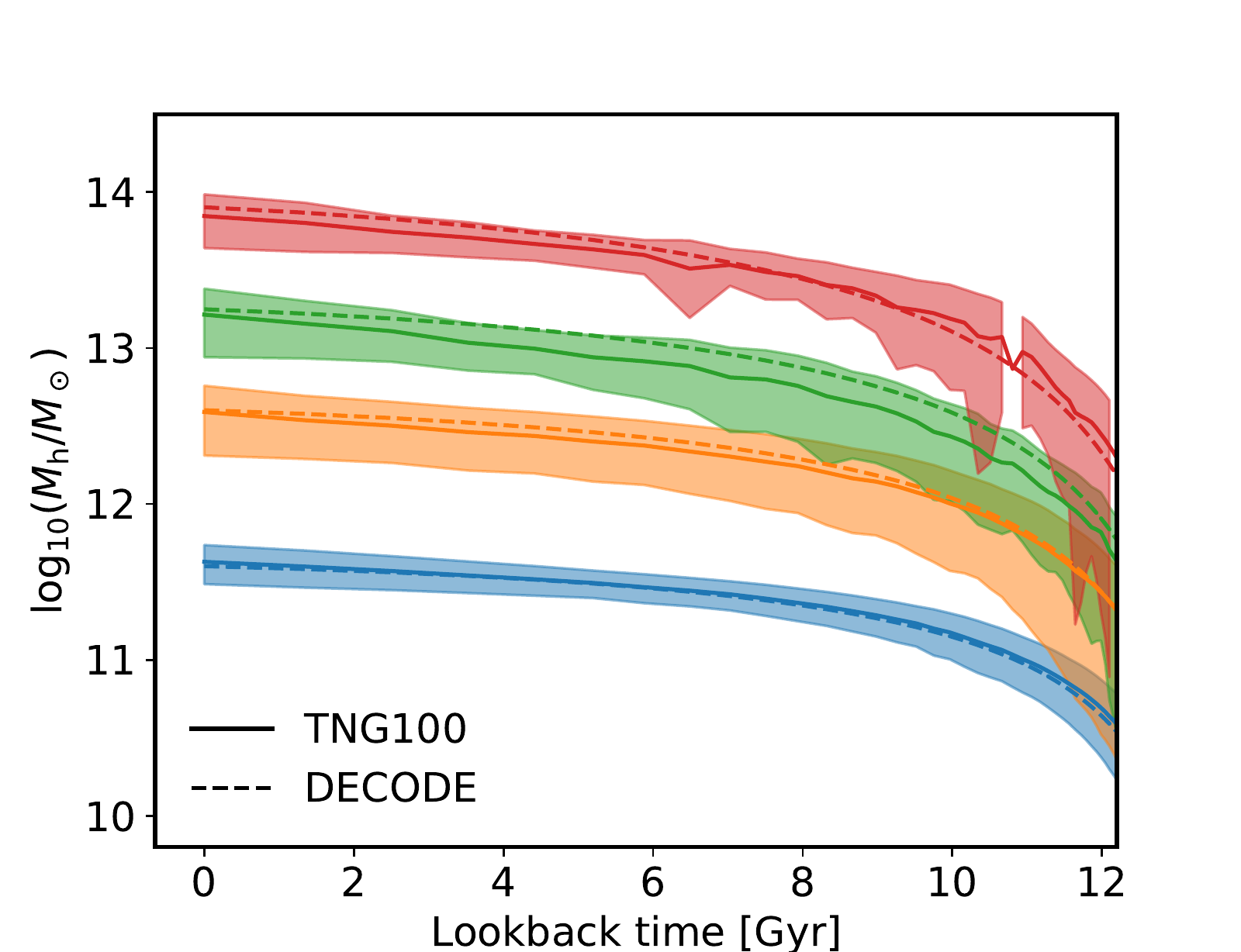}
        \includegraphics[width=0.49\linewidth]{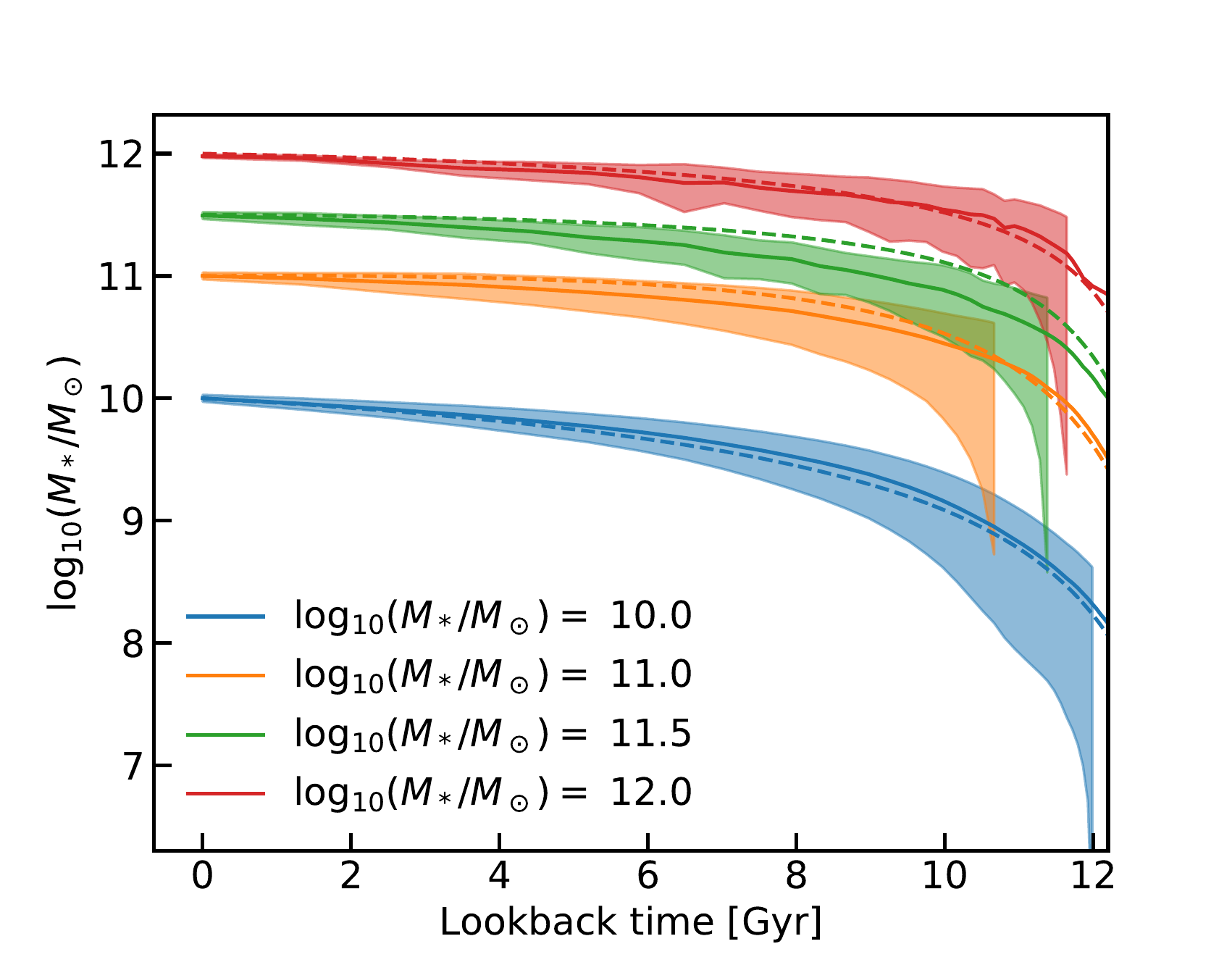}
        \caption{Left panel: halo mass assembly history from the TNG100-1 simulation, for four stellar mass bins at $z=0$. Solid lines and shaded areas show the mean and $1\sigma$ error from the simulation, and the dashed lines show the accretion history from \citep{vdb_2014} that we adopt in this work. Right panel: total stellar mass growth for the same four mass bins. The dashed lines show the mean galaxy growth computed using the SMHM relation of TNG as input in \decode, while the solid lines show the results extracted directly from the TNG database.}
        \label{fg_growths_tng}
    \end{figure*}
    
    \begin{figure*}
        \centering
        \includegraphics[width=0.495\linewidth]{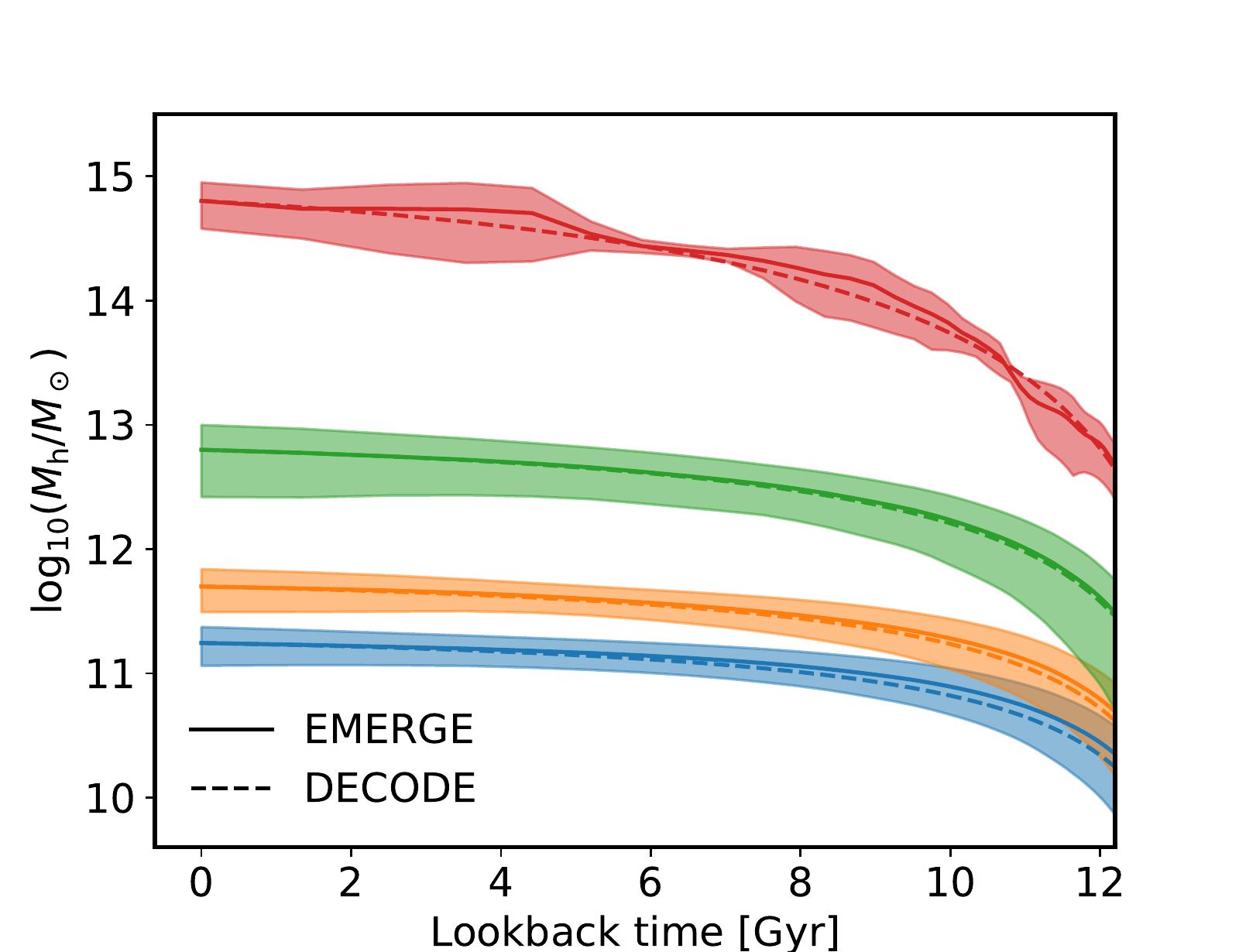}
        \includegraphics[width=0.49\linewidth]{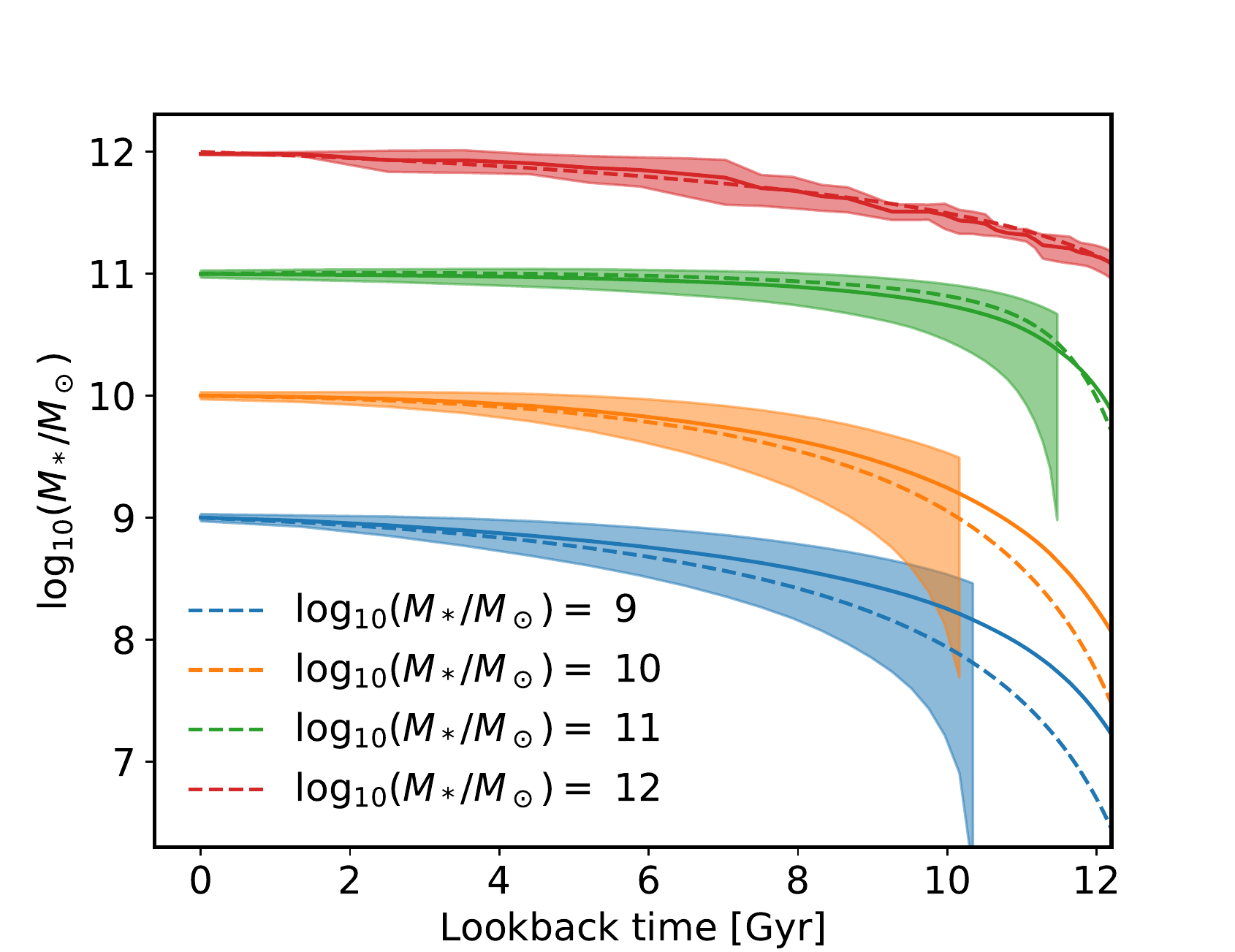}
        \caption{Same as Figure \ref{fg_growths_tng} but for EMERGE.}
        \label{fg_growths_emerge}
    \end{figure*}
    
    As already described in Section \ref{sec_num_density_haloes}, we assign the parent halo with a mean mass assembly history. For the details of how we numerically calculate the latter, we readdress the interested reader to the work of \cite{vdb_2014}. We show the mean DM accretion for four different values of stellar mass bin (as labelled in the legends) in the left panels of Figures \ref{fg_growths_tng} and \ref{fg_growths_emerge}, where we compare with the data from the TNG simulation and EMERGE respectively. This is a cross check that the predictions of the TNG simulation and EMERGE are consistent with the mean assembly history that we are employing.
    
    Similar considerations are valid for the total mean galaxy stellar mass growths. We show in the right panels of Figures \ref{fg_growths_tng} and \ref{fg_growths_emerge} the results predicted by our model \decode using as input the SMHM computed from the two simulations, and we compare with the data from the simulations themselves. The stellar mass growth histories for the four masses shown comes out to be consistent within $\sim 1 \, {\rm dex}$ with TNG and EMERGE.

    \section{Bulge-to-total ratios modelling}\label{app_bt_ratios}
    
    \begin{figure}
        \includegraphics[width=\columnwidth]{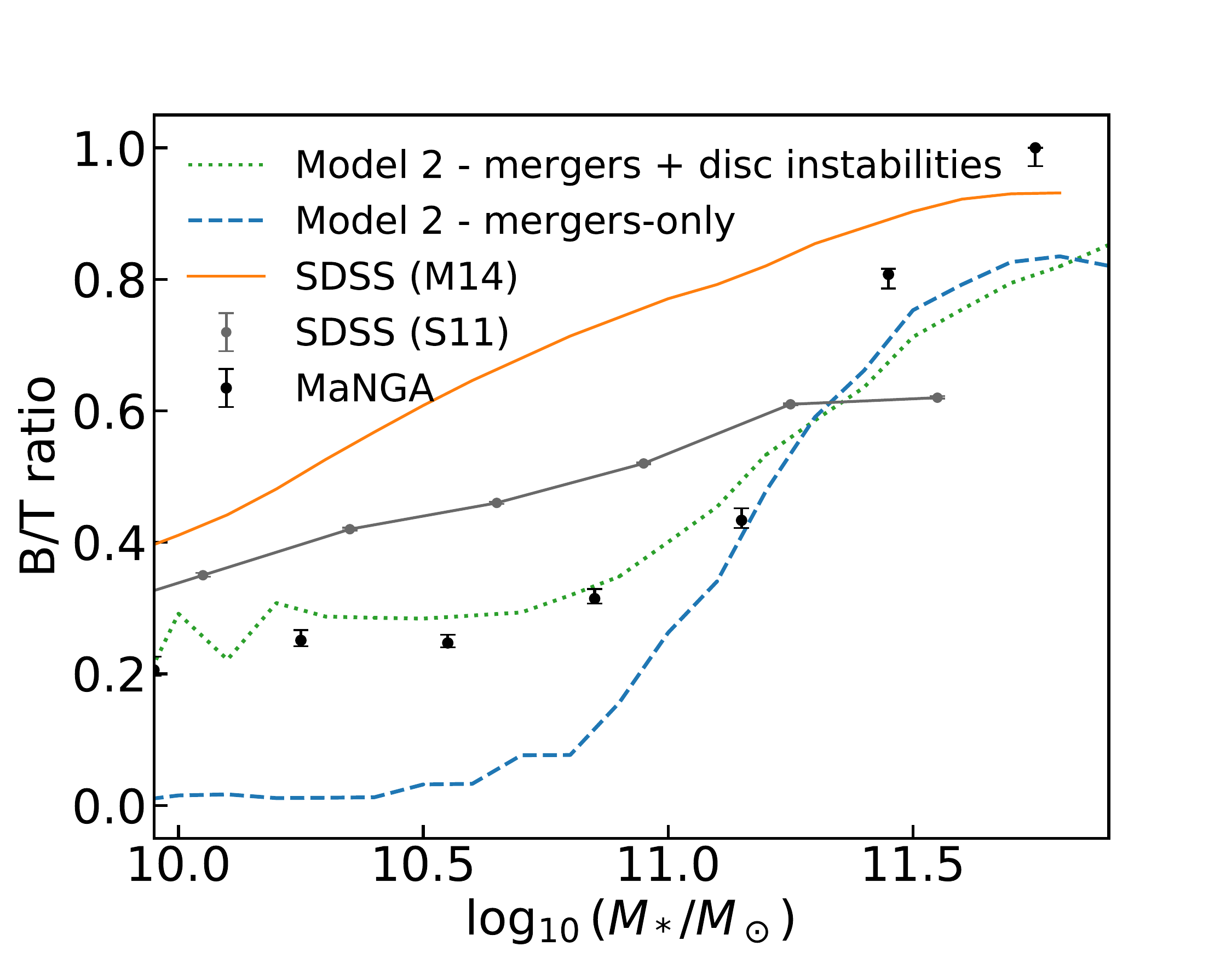}
        \caption{Bulge-to-total ratios as a function of stellar mass from this work compared to different observational samples. The green dotted and blue dashed lines show the Model 2 predictions for the mergers+disc instabilities and mergers-only toy models, respectively. The orange line shows the mean B/T of SDSS using the sample of \citet{Mendel+14}, as shown in \citet{devergne_2020}. The grey error bars show the SDSS B/T computed using the \citet{simard_2011} catalogue and the black error bars the MaNGA data.}
        \label{fg_bt_comp_mendel}
    \end{figure}
    
    We show in Figure \ref{fg_bt_comp_mendel} a comparison of the B/T ratios from different observational datasets. In particular, we compare the MaNGA data (black error bars), described in Section \ref{sec_data_sdss} and used as a reference for the models in this work, with the SDSS data from \citet{Mendel+14} who selected a subsample of the \citet{simard_2011} catalogue. We also show the B/T ratio of SDSS that we have computed directly from the \citet{simard_2011} catalogue (grey error bars), as well as the predictions of Models 2 from this work (green dotted and blue dashed lines). Interestingly, our results for SDSS are not consistent with those from \citet{Mendel+14}. Nevertheless, our results discussed in the main text are still valid, irrespective of the exact data set chosen for computing mean B/T ratios. All the three observational B/T ratios show in fact that models based only on mergers, such as our BT1 and BT2 described in Section \ref{sec_res_bt_ratio}, are not sufficient to reproduce the measure B/T ratios, at least at low masses. On the other hand, models that include also disc instabilities perform much better in reproducing the observational data. In summary, all observational B/T data suggest that at low masses some level of disc instabilities is still expected in addition to mergers in order to well describe the evolution of galactic bulges.

    \section{Subhalo abundance matching}\label{app_sham}
    
    For completeness, we compare our SMHM relations computed from direct abundance matching between the SMF and the HMF, with the SMHM relation derived from the stellar mass-peak velocity (SMPV) relation \citep[e.g.,][]{guo_2014, chaves_montero_2016, contreras_2021, favole_2022}. To this purpose, we first compute the SMPV relation using Equation (\ref{eq_aversa_AM}), where we input the SMFs of Models 1 and 2, as described in Section \ref{sec_smhm_models}, and we replace the HMF with the peak velocity function. We extract the peak velocity function from the MultiDark simulation \citep[][]{klypin_2016} at different redshift snapshots. Once the SMPV relation is computed, we calculate the implied SMHM relation and its dispersion at fixed halo mass by using the halo masses competing to each $V_{\rm peak}$ in the simulation.
    
    \begin{figure*}
        \centering
        \includegraphics[width=0.9\textwidth]{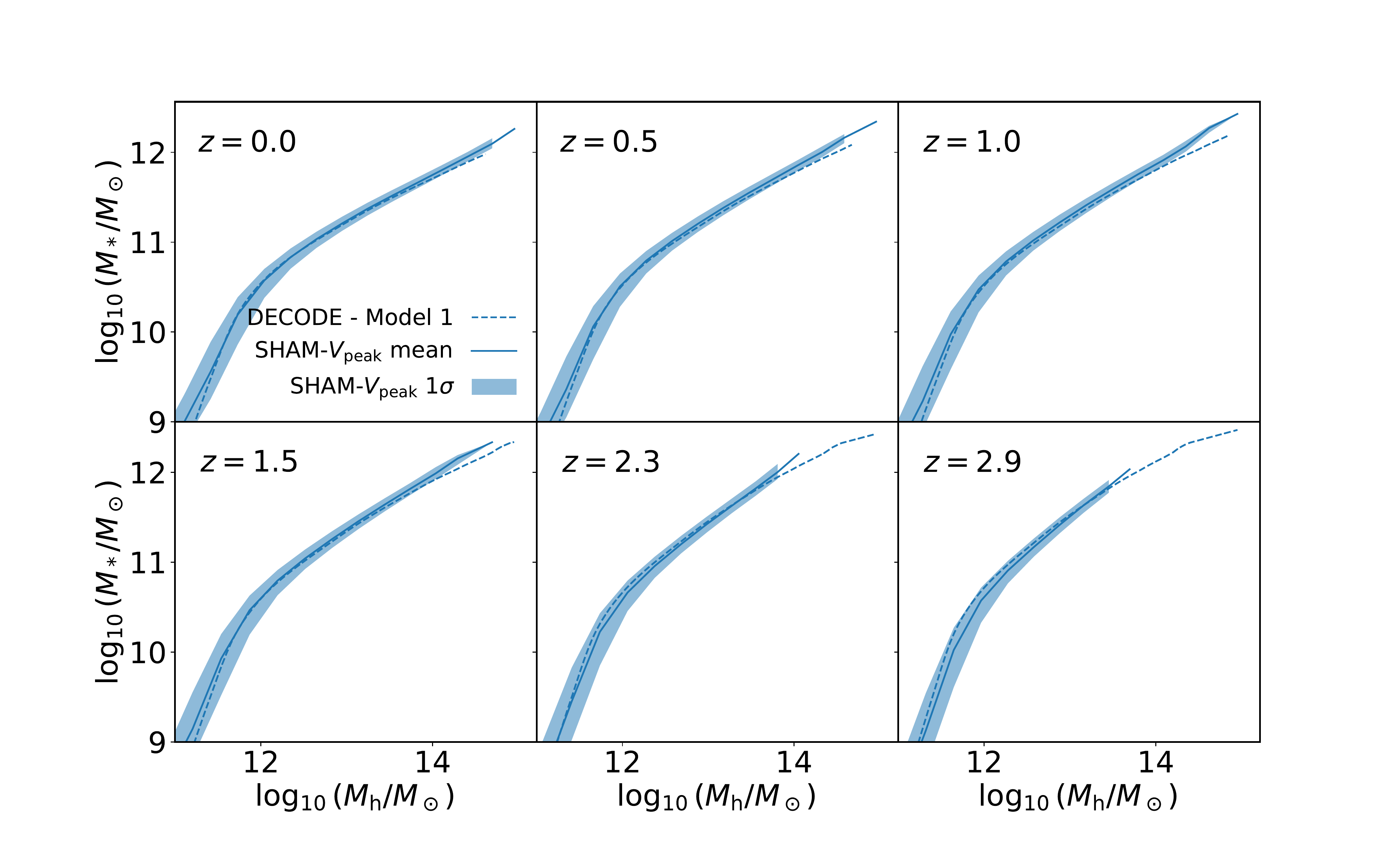}
        \caption{SMHM relation computed via the SHAM technique using the DM $V_{\rm peak}$ data from the MultiDark compared to \decode at different redshifts. The dashed lines show Model 1 SMHM relation from \decode, at different redshifts, as described in Section \ref{sec_smhm_models}. The solid lines and shaded areas show the SMHM relation computed via SHAM with the velocity function using the Model 1 SMF as input and the $1\sigma$ dispersion.}
        \label{fg_smhm_sham_scatter000_Model_1}
    \end{figure*}
    
    \begin{figure*}
        \centering
        \includegraphics[width=0.9\textwidth]{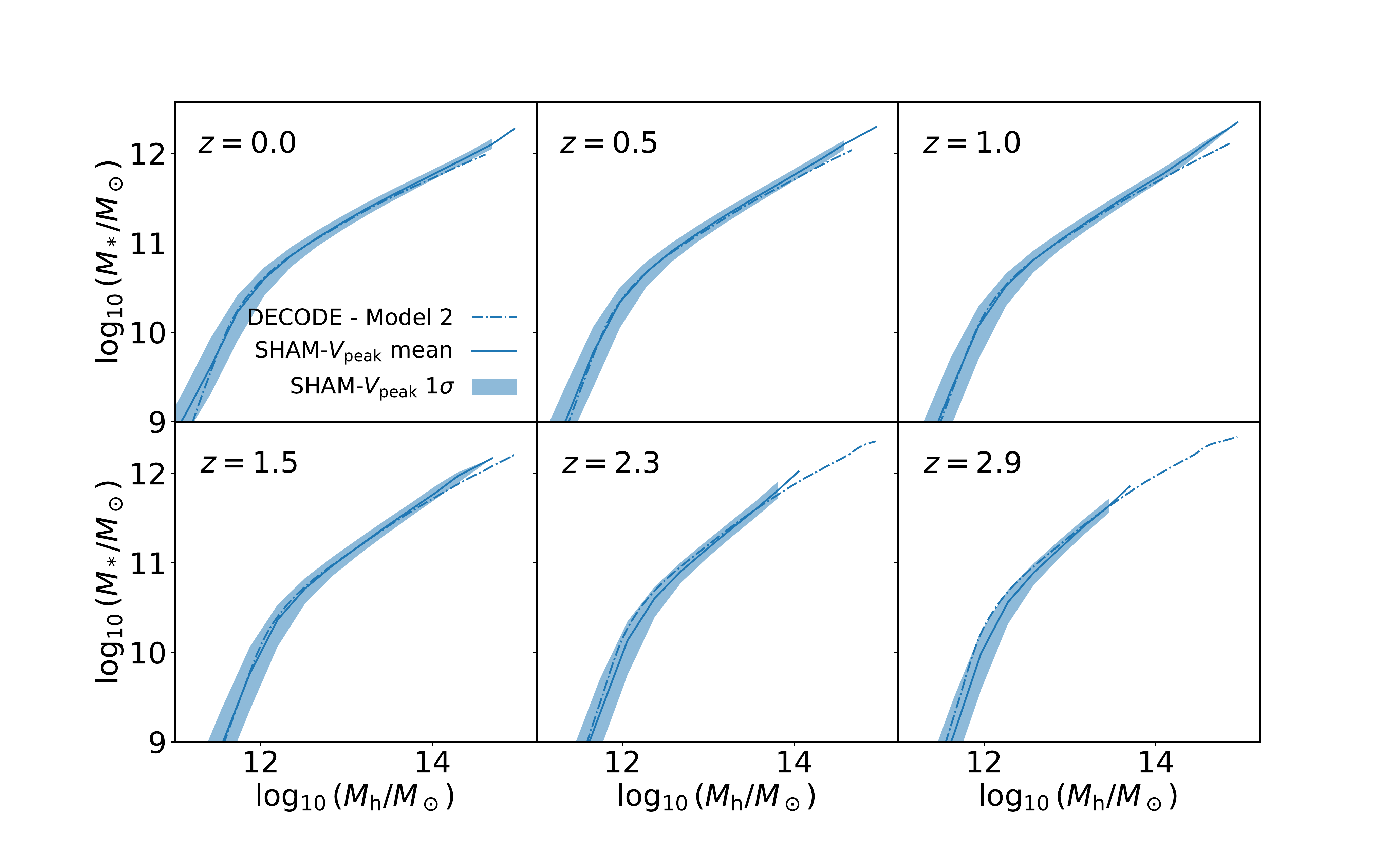}
        \caption{Same as Figure \ref{fg_smhm_sham_scatter000_Model_1} but for Model 2.}
        \label{fg_smhm_sham_scatter000_Model_2}
    \end{figure*}

    We show the resulting SMHM relation in Figures \ref{fg_smhm_sham_scatter000_Model_1} and \ref{fg_smhm_sham_scatter000_Model_2} at different redshifts (solid blue lines and blue regions), compared with the SMHM relations from \decode's Models 1 and 2, respectively. We see that the relations obtained via the SMPV relation match very well with the SMHM relations from Models 1 and 2. This agreement, on one hand, provides a further validation of our unstripped, surviving subhalo abundance matching technique presented in Section \ref{sec_dream_gal_halo_connect} and, on the other hand, further highlights that the systematics in the shape and/or redshift evolution of the observed SMF have a strong impact on the resulting mapping between stellar mass and halo mass.

%%%%%%%%%%%%%%%%%%%%%%%%%%%%%%%%%%%%%%%%%%%%%%%%%%

% Don't change these lines
\bsp	% typesetting comment
\label{lastpage}
\end{document}